\documentclass[%
 reprint,
 floatfix,
 amsmath,amssymb,
 aip
]{revtex4-2}

\makeatletter

\renewcommand*{\p@subsection}{}

\renewcommand*{\p@subsubsection}{}
\makeatother

\usepackage{bm}
\usepackage[scaled]{helvet}
\usepackage{tabularx,booktabs}
\newcolumntype{Y}{>{\centering\arraybackslash}X}
\usepackage{graphicx}
\usepackage{cleveref}
\usepackage{dcolumn}
\usepackage{setspace}

\usepackage{float}
\usepackage[caption = false]{subfig}

\usepackage[utf8]{inputenc}
\usepackage[T1]{fontenc}
\usepackage{mathptmx}
\usepackage{etoolbox}

\usepackage[dvipsnames]{xcolor}
\usepackage{soul}
\usepackage{cancel,siunitx}

\usepackage{comment}
\usepackage{soul}

\usepackage{braket} 

\usepackage{natbib}
\setcitestyle{super}

\begin{document}
\title{Polaritonic response theory for exact and approximate wave functions}

\author{Matteo Castagnola}
\author{Rosario Roberto Riso}
\affiliation{Department of Chemistry, Norwegian University of Science and Technology, 7491 Trondheim, Norway}
\author{Alberto Barlini}
\affiliation{Scuola Normale Superiore, Piazza dei Cavalieri 7, 56126 Pisa, Italy}

\author{Enrico Ronca}
\email{enrico.ronca@unipg.it}
\affiliation{Dipartimento di Chimica, Biologia e Biotecnologie, Università degli Studi di Perugia, Via Elce di Sotto, 8,06123, Perugia, Italy}
\author{Henrik Koch}
\email{henrik.koch@ntnu.no}
\affiliation{Department of Chemistry, Norwegian University of Science and Technology, 7491 Trondheim, Norway}
\affiliation{Scuola Normale Superiore, Piazza dei Cavalieri 7, 56126 Pisa, Italy}

\begin{abstract}
    \textit{Polaritonic chemistry} is an interdisciplinary emerging field that presents several challenges and opportunities in chemistry, physics and engineering. 
    A systematic review of polaritonic response theory is presented, following a chemical perspective based on molecular response theory. 
    We provide the reader with a general strategy for developing response theory for \textit{ab initio} cavity quantum electrodynamics (QED) methods and critically emphasize details that still need clarification and require cooperation between the physical and chemistry communities.
    We show that several well-established results can be applied to strong coupling light-matter systems, leading to novel perspectives on the computation of matter and photonic properties. 
    The application of the Pauli-Fierz Hamiltonian to polaritons is discussed, focusing on the effects of describing operators in different mathematical representations.
    We thoroughly examine the most common approximations employed in \textit{ab initio} QED, such as the dipole approximation.
    We introduce the polaritonic response equations for recently developed \textit{ab initio} QED Hartree-Fock and QED coupled cluster methods. 
    The discussion focuses on the similarities and differences from standard quantum chemistry methods, providing practical equations for computing the polaritonic properties. 
 \end{abstract}

\maketitle


\section{Introduction}
Hybrid light-matter states, referred to as \textit{polaritons}, were described in the 1950s from the interaction of optical modes and lattice excitons in crystals.\cite{huang1951lattice, hopfield1958theory, tolpygo1950physical}
In 1963, Jaynes and Cummings described with a phenomenological quantum picture the formation of polaritons from the interaction of photons and atomic states.\cite{jaynes1963comparison}
Their predictions were later confirmed experimentally for Rydberg atoms in a quantum microcavity.\cite{rempe1987observation, brune1996quantum} 
The use of quantum optical devices to confine the fields, such as Fabry-Pérot cavities or metal nanostructures,\cite{long2015coherent, gordon1964equivalence, fox1961resonant, kojima2002laser, schmidt2016quantum, santhosh2016vacuum} increase the coupling between matter and light, and polaritons have been observed since the 1970s for inorganic semiconductors\cite{weisbuch1992observation, yakovlev1975surface} and from the 1990s for organic molecules in optical cavities.\cite{lidzey1998strong, fujita1998tunable} 
Polaritonics has recently attracted the interest of chemists since the pioneering work of Ebbesen and coworkers, who showed that the formation of polaritons can influence  photochemical\cite{hutchison2012modifying} and ground state reactivity. \cite{thomas2016ground, lather2019cavity, thomas2019tilting}
Several experiments have now shown that electromagnetic confinement leads to essential modifications of processes such as chemical reactions,\cite{hutchison2012modifying, thomas2016ground, lather2019cavity, thomas2019tilting, canaguier2013thermodynamics, sau2021modifying} singlet fission,\cite{eizner2019inverting, takahashi2019singlet, martinez2018polariton} intersystem crossing,\cite{stranius2018selective, yu2021barrier, ulusoy2019modifying} crystallization and assembly,\cite{joseph2021supramolecular, hirai2021selective} as well as optical properties like absorption, scattering and emission,\cite{garcia2021manipulating, chervy2018vibro, george2015ultra, xue2018ultrastrong, del2015signatures, baranov2020circular, guo2021optical, itoh2018reproduction, herrera2017absorption, wang2020coherent, takele2021scouting, barachati2018tunable, mund2020optical, ebadian2017extending, wang2021large, wang2014quantum} although the reproducibility of the chemical reactivity modifications is not always straightforward.\cite{imperatore2021reproducibility}
Crucial experimental aspects are still to be clarified, and cooperation between experimentalists and theoreticians can be of great assistance in unravelling the conditions and the consequences of vibrational and electronic strong coupling.
Since both the photon and the molecular degrees of freedom are important in such systems, their theoretical modelling is a challenging interdisciplinary task.\cite{ruggenthaler2022understanding, fregoni2022theoretical, hirai2020recent, feist2018polaritonic, sidler2022perspective, hertzog2019strong, nagarajan2021chemistry, schafer2022shining} 
Following well-established electronic structure theory, several \textit{ab initio} quantum electrodynamics (QED) approaches have been developed recently. 
These include QED density functional theory (QEDFT),\cite{ruggenthaler2014quantum} reduced density matrix approaches for matter-photon systems,\cite{buchholz2019reduced, mallory2022reduced} polaritonic coupled cluster,\cite{mordovina2020polaritonic} QED Hartree-Fock (QED-HF),\cite{haugland2020coupled, haugland2021intermolecular} QED coupled cluster (QED-CC),\cite{haugland2020coupled, haugland2021intermolecular} QED full configuration interaction (QED-FCI),\cite{haugland2020coupled, haugland2021intermolecular} polarized Fock states approach (PFSs),\cite{mandal2020polarized} polaritonic unitary coupled cluster (QED-UCC),\cite{pavosevic2021polaritonic} strong coupling (SC)-QED-HF,\cite{riso2022molecular} QED Møller-Plesset perturbation
theory of second-order (QED-MP2),\cite{bauer2023perturbation} and second-order QED algebraic diagrammatic construction scheme for the polarization propagator (QED-ADC(2)).\cite{bauer2023perturbation}
In addition, several approaches have been investigated combining phenomenological models from the quantum optics community with computational chemistry.\cite{salmon2022gauge, jaynes1963comparison, dicke1954coherence, tavis1968exact, frisk2019ultrastrong, imamouglu2009cavity, knight1978super, vukics2012adequacy, grynberg2010introduction, sidler2022perspective, f2018theory, gonzalez2016uncoupled, galego2015cavity, luk2017multiscale, galego2019cavity, li2021cavity, li2020cavity, fregoni2020photochemistry, fregoni2018manipulating, hopfield1958theory, todorov2012intersubband}\\

\noindent 
The response theory formalism is a very successful and well-known framework employed in quantum chemistry to calculate approximate ground and excited state properties.\cite{olsen1985linear, norman2018principles, casida2012progress, helgaker1999, christiansen1998response, sasagane1993higher, langhoff1972aspects, cammi1999linear, helgaker2012recent, lazzeretti2004assessment} 
While several of these results are still valid for QED systems, essential and subtle differences arise due to the explicit modelling of the photon field, highlighting additional challenges and opportunities.  
Although some response schemes for QED methods have been proposed,\cite{flick2019light, yang2021quantum, welakuh2023tunable, welakuh2022frequency, bonini2022ab, flick2020ab, welakuh2022transition, fregoni2021strong} a systematic discussion of polaritonic response theory and its connections to molecular response theory is still not available.\\

\noindent This paper aims to fill this gap by proposing a systematic discussion of wave function based methods for computing polaritonic properties. 
We will review several results from molecular response theory that can be applied to polaritonic wave functions and critically discuss the limitations and future challenges such methods have to tackle. 
In section \eqref{sec:cavities}, we briefly introduce the most common optical devices employed to achieve strong coupling. 
In section \eqref{PH_chapt}, we present the Pauli-Fierz Hamiltonian and focus on how this framework is used to model the strong coupling regime. 
We review the most common approximations for \textit{ab initio} QED in section \eqref{sec:approximations_qed}. 
In section~\eqref{exact_resp_chapt}, we discuss polaritonic response theory for exact states, focusing on similarities and differences with molecular response theory, and provide several examples of polaritonic response properties.  
The derived equations are equivalent to QED-FCI response theory.
In section~\eqref {Approx_resp_chapt}, we discuss how to develop response theory for approximate models and derive linear and quadratic response equations for QED-HF and QED-CC. 
Finally, section~\eqref{conclusion_chapt} recaps the main results described in this paper and contains some concluding remarks and future perspectives.

\section{Optical cavities}\label{sec:cavities}
This section provides a brief introduction to quantum optical devices. 
The discussion, although far from being exhaustive, aims at reviewing the most common experimental setups employed in polaritonic chemistry and  provides the reader with phenomenological basics of light-matter strong coupling.\\

The use of quantum optical devices provides a non-invasive way to engineer a material's properties by exploiting the confinement of the electromagnetic field. 
The geometry and the material of the device define the resonator eigenmodes.
In a real cavity, the confinement is always defective and depends specifically on the frequency of the field and the cavity material. 
The so-called $Q$-factor measures the quality of a cavity\cite{jackson1977classical}
\begin{equation}
    Q(\omega_0) = \omega_0\frac{\textit{energy stored}}{\textit{power loss}},
\end{equation}
where $\omega_0$ is the resonance frequency of the device.
This equation implies that the electric field will exhibit an exponential decay
\begin{equation}
    E(t)=E_0 e^{-i\omega_0 t}e^{-\frac{\omega_0 }{2Q}t},
\end{equation}
such that the energy spectrum will show a Lorenzian lineshape
\begin{equation}\label{real_cavity_lineshape_Q}
    |E(\omega)|^2\propto\frac{1}{(\omega-\omega_0^2)+(\omega_0/2Q)^2}.
\end{equation}
From Eq. \eqref{real_cavity_lineshape_Q}, we have a more practical definition of the $Q$-factor as the ratio between the frequency of the hosted field mode $\omega_0$ and the full width at half maximum (FWHM) $\Delta\omega$ of the cavity spectrum
\begin{equation}
    Q(\omega_0) = \frac{\omega_0}{\Delta\omega}.
\end{equation}
The coupling $\lambda$ between matter and the confined cavity modes is proportional to the inverse square root of the effective confinement volume $V_{eff}$
\begin{equation}
    \lambda\propto \frac{1}{\sqrt{V_{eff}}}.
\end{equation}
The efficiency of a device in enhancing the interaction between light and matter can then be measured by the $Q/V_{eff}$ ratio.
Strong coupling conditions are obtained once the coupling strength exceeds the material and cavity losses (i.e. the coherent energy exchange rate between the electromagnetic field and the molecule exceeds the dissipation processes).
Decoherence interactions with the environment are suppressed, and hybrid states between the cavity and the molecular eigenstates naturally describe the system.
In this regime, the molecular properties are therefore affected even in the absence of an external photon pumping of the device.\\

\begin{figure}[ht]
    \centering
    \includegraphics[width=.45\textwidth]{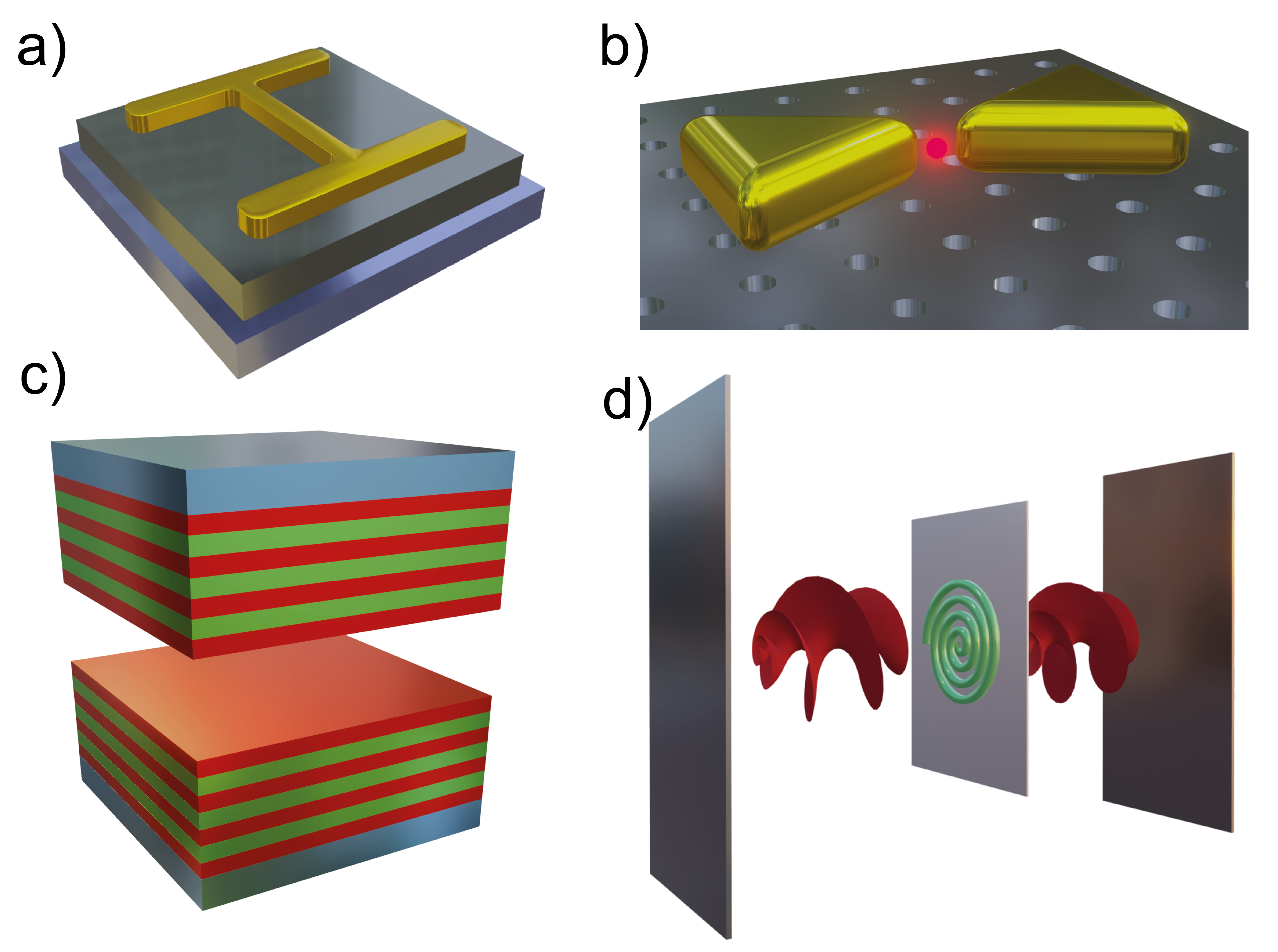}
    \caption{(a) Metallic nanostructures can be used to confine electromagnetic fields through surface plasmon resonances. A "dogbone" metamaterial can be employed as a subwavelength resonator, confining the electromagnetic field on a scale of a tenth of the mode wavelength.\cite{benz2013strong}
    Subwavelength optical devices can overcome the diffraction limit by employing suitable geometries and metamaterials, facilitating the fulfilment of strong coupling conditions.\cite{maier2006plasmonics, lee2010review, benz2013strong, dintinger2005strong, ballarini2019polaritonics, todorov2009strong}
    (b) Two metal nanoparticles are coupled on a crystal substrate, increasing the electric field's intensity in a hotspot between them.\cite{maier2006effective, hugall2018plasmonic, mondal2022strong, liang2020fine, zhang2020hybrid}  Metal nanoparticles can be used to achieve very low-volume confinement due to surface plasmon resonances, although dissipation processes are often significant.
    (c) Distributed Bragg reflectors (DBRs) are composed of alternate slabs of dielectric materials. Carefully choosing the number of slabs and their material, it is possible to achieve efficient confinement of the electromagnetic fields between two DBRs.\cite{zhang2019distributed, emsley2002silicon, menghrajani2020strong, tao2015strong, butte2005recent} 
    (d) A 2D chiral polymer film, obtained by torsional sheer stress, inserted inside a Fabry-Pérot resonator endows the system with chiral eigenmodes.\cite{gautier2022planar} }
    \label{fig:strong_coupling_devices}
\end{figure}

The most straightforward optical device to confine light is a Fabry-Pérot cavity, composed of two planar parallel metal plates kept at a fixed distance.\cite{fabry1899theorie, pfeifer2022achievements, muller2010ultrahigh, steinmetz2006stable, rakhmanov2002dynamic, schlawin2022cavity} 
These resonators are often produced by sputtering gold or silver on a substrate (often $SiO_2$, $BaF_2$  or $ZnSe$), and direct contact with the sample is prevented by using a thin polymer film. 
The sample is then coated by this multilayer structure or injected into the hollow cavity.
The cavity eigenmodes are linearly-polarized standing waves whose wavelength $\lambda_n$ depends on the distance $L$ between the mirrors
\begin{equation}
    \lambda_n=\;\frac{2L}{n} \qquad n=1,2,\;\dots
\end{equation}
The spatial distribution of the optical mode can be observed by changing the position of a thin sample slab inside the cavity.
The strong coupling features of this system are a function of the slab position, which gives us the field strength distribution.\cite{wang2014quantum, schouwink2002dependence}
Changing the form of the plates, e.g. using curved mirrors similar to lenses, can lead to modifications of the field shape and possibly an increase in the coupling constant.\cite{wang2022manipulating, mckeever2003experimental, culver2016collective, herskind2009realization, favero2009optomechanics, plum2015chiral, viviescas2003field}
The coupling strength $\lambda$ in Fabry-Pérot cavities is limited by the diffraction limit, which restrains the field confinement.
Devices which bound the fields on a scale smaller than the diffraction limit are called \textit{subwavelength} cavities and can facilitate ultrastrong coupling.
To this end, using nanostructured metamaterials in a suitable geometric arrangement can allow efficient confinement and overcome the diffraction limit, leading to efficient subwavelength cavities.
In Fig. (\ref{fig:strong_coupling_devices}a), we show an example of a subwavelength "dogbone" metamaterial resonator, which can confine light on a scale of about one-tenth of the mode wavelength.\cite{maier2006plasmonics, lee2010review, benz2013strong, dintinger2005strong, ballarini2019polaritonics, todorov2009strong}
Metallic nanostructures can be used to confine electromagnetic fields through surface plasmon resonances, allowing for remarkably low-volume confinement. 
It is also possible to employ non-metallic plasmonic structures, such as graphene.\cite{xiao2018theoretical, li2022strong, qing2022strong, li2017graphene, gan2012strong, koppens2011graphene} 
Coupling two or more plasmonic nanoparticles can lead to extreme field localization in hot spots between them, as schematically shown in Fig. (\ref{fig:strong_coupling_devices}b) for two nano prisms on a silicon substrate.\cite{maier2006effective, hugall2018plasmonic, mondal2022strong, liang2020fine, zhang2020hybrid}
However, these systems can be very lossy, i.e. they have a low $Q$-factor, which effectively limits their use for strong coupling.
Alternatively, dielectric materials can also be used to confine the electromagnetic field.
In Fig. (\ref{fig:strong_coupling_devices}c), we show a cavity obtained by two distributed Bragg reflectors (DBRs) composed of several alternate dielectric slabs. 
By carefully choosing the number of slabs and their material, DBRs can achieve low-volume confinement and high $Q$-factors.\cite{zhang2019distributed, emsley2002silicon, menghrajani2020strong, tao2015strong, butte2005recent}
Transparent high-reflection index dielectrics in curved or polygonal shapes can confine electromagnetic waves by total internal reflection.
The optical modes of such resonators, called whispering-gallery modes (WGMs), can achieve extreme field localization and quality factors.\cite{hu2021strong, farr2014strong, gupta1995strong, o2011all, matsko2006optical, strekalov2016nonlinear, matsko2005review, kaliteevski2007whispering}\\
The devices described so far cannot, in general, host any chiral eigenmodes. 
The helicity of a circularly polarized wave changes sign when reflected by a mirror surface, creating a mode with zero global helicity.
In recent years, different helicity-preserving optical devices (chiral cavities) have been proposed based on the introduction of a chiral element in the resonator.\cite{gautier2022planar, voronin2022single, feis2020helicity, beutel2021enhancing, plum2015chiral, scott2020enhanced, yoo2015chiral, liu2020switchable, sofikitis2014evanescent, hodgkinson2000vacuum, graf2019achiral,  hentschel2017chiral, zheng2021discrete, wang2023excitation, govorov2012theory, lan2016self} 
In Fig. (\ref{fig:strong_coupling_devices}d), we show a chiral Fabry-Pérot cavity formed by the insertion of a 2D chiral polymer slab.\cite{gautier2022planar}  
The 2D chirality is achieved by applying torsional sheer stress, and chiral eigenmodes are sustained in the device by a circular conversion dichroism mechanism.\cite{gautier2022planar} 
Other chiral optical resonators have been proposed using chiral metasurfaces\cite{plum2015chiral, scott2020enhanced, feis2020helicity, beutel2021enhancing} and chiral plasmonics.\cite{lan2016self, govorov2012theory, wang2023excitation, zheng2021discrete, hentschel2017chiral}

\section{Pauli-Fierz Hamiltonian}\label{PH_chapt}

The starting point of molecular modelling is the nonrelativistic molecular Hamiltonian
\begin{align}\label{std_ham}
    &H = \sum_M\frac{1}{2m_M}\mathbf{p}_M^2 + \sum_i\frac{1}{2}\mathbf{p}_i^2 \nonumber\\
    &+ \frac{1}{2}\sum_{i\neq j} \frac{1}{|\mathbf{r}_i-\mathbf{r}_j|}-\sum_{i,M}\frac{Z_M}{|\mathbf{r}_i-\mathbf{R}_M|}+\frac{1}{2}\sum_{M\neq N}\frac{Z_MZ_N}{|\mathbf{R}_N-\mathbf{R}_M|}.
\end{align}
In Eq.~\eqref{std_ham}, the capital indices $M$ and $N$ refer to nuclei of mass and charge $m_{M}$ and $Z_{N}$, $i$ and $j$ to electrons with mass $m=1$ and charge $q=-1$. 
The electronic and nuclear positions are labelled  $\mathbf{r}$ and $\mathbf{R}$ respectively, and $\mathbf{p}$ is the linear momentum of the particles. 
The first line of Eq.~\eqref{std_ham} contains the nonrelativistic kinetic energy operators for nuclei and electrons, while the second line describes the Coulomb interaction. \\

\noindent Although it is often assumed that the Hamiltonian in Eq.~\eqref{std_ham} is sufficient to model a molecular system, this is not the case. 
This is apparent if we realise that the excited states of the system have an infinite lifetime, i.e. they do not radiate. 
In Eq.~\eqref{std_ham}, we disregard the electromagnetic degrees of freedom, retaining only the electrostatic interaction between particles while neglecting retardation effects. 
The motion of electric charges generates radiation fields which in turn influence the particles' motion: this back-reaction mechanism is the origin of spontaneous decay. 
The inclusion of the electromagnetic degrees of freedom is also fundamental when optical devices are employed to modify the photon environment. 
It is then necessary to go beyond the Coulomb interaction terms and resort to Maxwell's electrodynamics to describe self-consistently the interaction between light and matter.\cite{cohen1997photons, craig1998molecular, jackson1977classical, landau2013electrodynamics}
The particle motion is determined by the Lorenz force $\mathbf{F}_L$
\begin{equation}\label{Lorenz_Force}
    \mathbf{F}_L = m {\mathbf{a}} = q\bigg(\mathbf{E}(\mathbf{r},t) + \frac{{\bm{v}}}{c} \times \mathbf{B}(\mathbf{r},t)\bigg),
\end{equation}
where $c$ is the speed of light, $\bm{v}$ and $\mathbf{a}$ are the velocity and the acceleration of the particle, $q$ is the charge and $m$ is the mass. 
The electric $\mathbf{E}$ and magnetic $\mathbf{B}$ fields are determined by Maxwell's equations, here expressed in cgs units
\begin{equation}\label{Maxwell_Equations}
    \begin{cases}
    \nabla \cdot\mathbf{E} = 4 \pi \rho(\mathbf{r},t)\\
    \nabla \cdot \mathbf{B} = 0\\
    \nabla \times\mathbf{E} = - \frac{1}{c}\frac{\partial \mathbf{B}}{\partial t}\\
     \nabla \times \mathbf{B} = \frac{4\pi}{c}\mathbf{j}(\mathbf{r},t) +\frac{1}{c}\frac{\partial \mathbf{E}}{\partial t}
    \end{cases},
\end{equation}
where $\mathbf{r}$ is a vector defining a point in space, $\rho(\mathbf{r},t)$ is the charge density and $\mathbf{j}(\mathbf{r},t)$ is the current density. 
The fields are naturally described using \textit{potentials} $(\phi, \mathbf{A})$, such that
\begin{align}
    \mathbf{B} &= \nabla\times\,\mathbf{A} \label{eq:Curl_A}\\
    \mathbf{E} &= -\nabla\, \mathbf{\phi} - \frac{1}{c}\frac{\partial \mathbf{A}}{\partial t} \label{eq:div_phi_minus_A'}.
\end{align}
We notice that, unlike the fields $\mathbf{B}$ and $\mathbf{E}$, the vector potential $\mathbf{A}$ and the scalar potential $\phi$ are not measurable quantities and must, therefore, only be seen as mathematical tools. 
In Eqs.~\eqref{eq:Curl_A} and \eqref{eq:div_phi_minus_A'}, the potentials are identified only through their derivatives. 
This implies that the same fields, i.e. the same system, can be modelled using different potentials. 
The transformation
\begin{align}\label{pot-field}
    \mathbf{A}&\to\mathbf{A} + \nabla\, \chi\\
    \mathbf{\phi}&\to \mathbf{\phi} - \frac{1}{c}\frac{\partial \chi}{\partial t},
\end{align}
known as \textit{gauge transformation}, leads to unchanged electromagnetic fields for any well-behaved scalar function $\chi(\mathbf{r},t)$. 
This gauge freedom can be exploited to simplify the equations. 
We will make use of the so-called \textit{Coulomb gauge}
\begin{equation}\label{CoulombGauge}
    \nabla \cdot \mathbf{A} = 0
\end{equation}
which allows us to account for the Coulomb interactions among the particles explicitly. 
The scalar potential is, in this gauge, determined by the Poisson equation as in the electrostatic case
\begin{equation}\label{Poisson_equation}
    \nabla^2 \phi = - 4\pi \rho,
\end{equation}
while the vector potential is obtained from the following equation:
\begin{equation}
    \nabla^2\mathbf{A}-\frac{1}{c^2}\frac{\partial^2 \mathbf{A}}{\partial t ^2}=-\frac{4\pi}{c}\mathbf{j}^\perp,
\end{equation}
where $\mathbf{j}^\perp$ is the solenoidal (divergence-free) part of the density current. \\

\noindent The Hamiltonian for the field-particle system is described in terms of the position of the particles, the vector potential and their conjugate momenta. The self-consistent interaction between molecules and light is described by the Pauli-Fierz Hamiltonian, for which Hamilton's equations can be shown to be equivalent to Eqs. \eqref{Lorenz_Force} and \eqref{Maxwell_Equations}\cite{cohen1997photons, craig1998molecular} 
\begin{widetext}
    \begin{align}\label{eq:Minimal_coupling}
    H_{PF}&=\sum_M\frac{1}{2m_M}\bigg(\mathbf{p}_M-\frac{Z_M}{c}\mathbf{A}(\mathbf{R}_M)\bigg)^2 + \sum_i\frac{1}{2}\bigg(\mathbf{p}_i+\frac{1}{c}\mathbf{A}(\mathbf{r}_i)\bigg)^2 \nonumber\\
    & +\frac{1}{2} \sum_{i\neq j} \frac{1}{|\mathbf{r}_i-\mathbf{r}_j|}-\sum_{i,M}\frac{Z_M}{|\mathbf{r}_i-\mathbf{R}_M|}+\frac{1}{2}\sum_{M\neq N}\frac{Z_MZ_N}{|\mathbf{R}_N-\mathbf{R}_M|} \nonumber \\
    &+\frac{1}{2}\int \left[\mathbf{E}_\perp\left(\mathbf{r}\right)^{2}+\mathbf{B}\left(\mathbf{r}\right)^{2}\right]d\mathbf{r}.
\end{align}
\end{widetext}
In the first line of Eq. \eqref{eq:Minimal_coupling}, we have the vector potential $\mathbf{A}$ computed at the positions of nuclei and electrons, and the second line describes the Coulomb interactions between the particles. 
The explicit separation of Coulomb interactions provided by the Coulomb gauge allows for the separation of Hamiltonian in Eq. \eqref{std_ham}, providing a natural chemical description of the system. 
The last line accounts for the electromagnetic field energy. 
Expanding the first terms, we obtain an interaction term between the particle Hamiltonian in Eq. \eqref{std_ham} and the radiation fields mediated by the vector potential $\mathbf{A}$
\begin{equation}
    H_{int}=-\sum_{x}\frac{q_x}{2m_xc}\big(\mathbf{p}_x\cdot\mathbf{A}(\mathbf{r}_x)+\mathbf{A}(\mathbf{r}_x)\cdot\mathbf{p}_x\big)+\sum_{x}\frac{q^2_x}{2m_xc^2}\mathbf{A}^2(\mathbf{r}_x),
\end{equation}
where $x$ runs on both electrons and nuclei. Following the QED prescription, the vector potential in Eq. \eqref{eq:Minimal_coupling} is promoted to an operator
\begin{equation}\label{eq:promotion}
\mathbf{A}\longrightarrow \hat{\mathbf{A}} = \sum_{\mathbf{k},\tau}\frac{c\lambda}{\sqrt{2\omega_{k}}}\left(\boldsymbol{\epsilon}_{\mathbf{k}\tau}e^{i\mathbf{k}\cdot\mathbf{r}}b_{\mathbf{k}\tau}+\boldsymbol{\epsilon}^{*}_{\mathbf{k}\tau}e^{-i\mathbf{k}\cdot\mathbf{r}}b^{\dagger}_{\mathbf{k}\tau}\right),    
\end{equation}
where $b^{\dagger}_{\mathbf{k}\tau}$ and $b_{\mathbf{k}\tau}$ respectively create and annihilate a photon with wave vector $\mathbf{k}$ and frequency $\omega_{k}=\frac{|\mathbf{k}|}{c}$. 
The field polarization $\boldsymbol{\epsilon}_{\mathbf{k}\tau}$ with $\tau=1,2$ spans the 2D plane perpendicular to $\mathbf{k}$. 
If we take the polarization $\boldsymbol{\epsilon}_{\mathbf{k}\tau}$ to be real, the fields are described by a superposition of linearly polarized plane waves. 
In an equivalent way, we can take the complex orthogonal unit vectors
\begin{equation}\label{eq:complex_polarization}
    \boldsymbol{\epsilon}_{\mathbf{k}\pm}= \frac{1}{\sqrt{2}}(\boldsymbol{\epsilon}_{\mathbf{k}1}\pm i\boldsymbol{\epsilon}_{\mathbf{k}2})
\end{equation}
which describe left and right circularly polarized waves.

\subsubsection*{Remarks}
There are now some subtle points to mention. Choosing the Coulomb gauge, the vector potential only has (two) transverse components $\mathbf{A}=\mathbf{A}_\perp$.
     At the same time, the longitudinal component of the electric field (Coulomb field) is described by the scalar potential. 
     In this gauge, the equations are not manifestly covariant.
     However, this does not imply a loss of relativistic invariance since the electromagnetic field prediction agrees with special relativity. 
     In fact, the retardation of the interaction between particles is obtained by a cancellation of the instantaneous interactions from the Coulomb term and the transverse field.
     For a different gauge, the Coulomb interactions would also be mediated by the longitudinal component of the vector potential.\cite{cohen1997photons, craig1998molecular}

    \noindent The canonical momentum $\bm{p}_i$, contrary to the standard quantum mechanical formulation, is not the linear momentum of the particle but here has a field-dependent component\cite{craig1998molecular, cohen1997photons}
    \begin{equation}\label{velocity_momentum}
        \bm{p}_i=m_i\bm{v}_i + \frac{q_i}{c}\mathbf{A}(\bm{r}_i).
    \end{equation}
    This means that, although in Eq.~\eqref{eq:Minimal_coupling} we can identify the original particle Hamiltonian in Eq.~\eqref{std_ham}, the physical meaning of these two operators is different.

    \noindent The Pauli-Fierz Hamiltonian is self-adjoint and bounded from below\cite{griesemer2001ground, hiroshima2002self} such that a stable ground state exists. However, the original excited states of the particle Hamiltonian Eq. \eqref{std_ham} become metastable (i.e. with a finite lifetime) because of their coupling to the continuous photonic spectrum.\cite{golenia, derezinski, Bach1995, Bach1999}

    \noindent In Eq.~\eqref{eq:Minimal_coupling}, we have disregarded the nuclear and electronic spin, which could be introduced in Eq.~\eqref{eq:Minimal_coupling} with the addition of the following contribution
    \begin{equation}\label{spinPF}
        H_{spin}=\frac{1}{2}\sum_i\bm{\sigma}_i\cdot\mathbf{B}(\bm{r}_i)-\sum_N\gamma_N\bm{I}_N\cdot\mathbf{B}(\bm{R}_N),
    \end{equation}
    where the first term of Eq.~\eqref{spinPF} accounts for the interaction of the electronic spin, described by the Pauli matrices $\bm{\sigma}_i$, with the (internal) magnetic field $\mathbf{B}$, while the second term describes the energy associated with the nuclear spin $\bm{I}_N$ with magnetogyric ratio $\gamma_{N}$.
    
    \noindent Finally, in Eq. \eqref{eq:Minimal_coupling} particles can in principle interact with infinitely high-frequency modes.
    However, the description of the kinetic energy is nonrelativistic.
    When the field frequency is comparable with the rest energy of the particles $\omega\sim mc^2$, relativistic effects (e.g. creation of electron-positron pairs) not included in Eq. \ref{eq:Minimal_coupling} become relevant.\cite{cohen1997photons} 
    It is then necessary to introduce a cutoff as the field's momentum $\mathbf{k}$ increases (ultraviolet cutoff). 
    Moreover, the masses in Eq.~\eqref{eq:Minimal_coupling} are not the \textit{physical masses} usually employed in quantum mechanics, which include the contribution of the electromagnetic energy created by the particle, but they are their \textit{bare masses}.\cite{craig1998molecular}

\section{Approximations for \textit{ab initio} polaritonic chemistry}\label{sec:approximations_qed}
 While the Hamiltonian in Eq.~\eqref{eq:Minimal_coupling} allows for a consistent treatment of light and matter, approximations are needed to perform computational studies. 
The problem of the continuum of photonic modes could be overcome by employing a fine discretization of the spectrum.
Moreover, different approaches can be developed depending on the treatment of the photonic degrees of freedom. 
In the Cavity Born-Oppenheimer approach (CBO),\cite{flick2017cavity, flick2018strong, flick2017atoms} the photon coordinates $q_\alpha$ are embedded in the nuclear wave function $\chi(\bm{R},\bm{q},t)$ and separated from the electronic degrees of freedom, described by and electronic wave function $\phi(\bm{r};\bm{R},\bm{q})$. The complete wave function can then be expanded as
\begin{equation}
    \Psi(\bm{r},\bm{R},\bm{q},t)=\sum_k\phi_k(\bm{r};\bm{R},\bm{q})\chi_k(\bm{R},\bm{q},t).
\end{equation}
In a polaritonic approach, the electronic and photon coordinates are treated on the same footing and described by a polaritonic wave function which depends parametrically on the nuclear coordinates only\cite{fregoni2020photochemistry, kowalewski2016cavity, fregoni2018manipulating, ribeiro2018polariton, haugland2020coupled, ruggenthaler2014quantum, bennett2016novel}
\begin{equation}
    \Psi(\bm{r},\bm{R},\bm{q},t)=\sum_k\phi_k(\bm{r},\bm{q};\bm{R})\chi_k(\bm{R},t).
\end{equation}
While these Born-Huang expansions are in principle equivalent, approximations will lead to different results.
In the simplest case, we take into account only a single term of these expansions as in standard electronic Born-Oppenheimer approximation.
In the discussion of exact response theory in Section \eqref{exact_resp_chapt} we do not refer to any explicit form of the Hamiltonian.
In Section \eqref{Approx_resp_chapt}, we focus on the polaritonic approach and describe the QED-CC and QED-HF response theory.
The interaction between the nuclear motion and the electromagnetic degrees of freedom can lead to novel non-adiabatic coupling terms that could, in some situations, jeopardize the validity of the BO approximation.\cite{feist2018polaritonic, vendrell2018collective, ulusoy2019modifying, schafer2018ab, fregoni2020photochemistry, fabri2021born}
Nevertheless, we will not explore such effects, and we will work in a fixed-nuclei framework. 
Although these approximations already reduce the complexity of the problem, further simplifications are needed to deal with the electromagnetic environment and treat the molecule and its interaction with the fields with chemical accuracy.

\subsection{Cavity QED}
The Hamiltonian  in Eq.~\eqref{eq:Minimal_coupling} allows for simultaneous treatment of light and matter, where we could, in principle, include any optical device that contributes to a modification of the electromagnetic environment. 
Nevertheless, the molecular complexity and the presence of the optical apparatus make the problem impractical from a computational {point of} view.
 
 The explicit modelling of the photonic device is avoided by defining an effective change in the structure of the electromagnetic modes. 
 To this end, we impose boundary conditions on the fields (e.g. vanish on the surface of the optical device) to model the electromagnetic confinement induced by the cavity. 
 The vector potential is expanded in terms of a complete orthonormal set of suitable electromagnetic normal mode functions $\{\bm{S}_\alpha(\bm{r})\}$.\cite{jackson1977classical, power1982quantum} 
 For a rectangular parallelepiped with perfectly conducting walls of sides $L_x$, $L_y$ and $L_z$, the transverse mode functions are standing waves\cite{power1982quantum}
\begin{align*}
    S_x(\bm{r})&=\sqrt{\frac{8}{V_c}}\cos \big(k_xx\big)\sin\big(k_yy\big)\sin\big(k_zz\big)\\
    S_y(\bm{r})&=\sqrt{\frac{8}{V_c}}\sin\big(k_xx\big)\cos\big(k_yy\big)\sin\big(k_zz\big)\\
    S_z(\bm{r})&=\sqrt{\frac{8}{V_c}}\sin\big(k_xx\big)\sin\big(k_yy\big)\cos\big(k_zz\big),
\end{align*}
where $V_c=L_x \times L_y \times L_z$ is the total volume of the parallelepiped and $\mathbf{k}$ is the wave vector which is now quantized:
\begin{equation}
    k_i =  n\frac{\pi}{L_i}\qquad n=1,2,\,\dots
\end{equation}
In the absence of external charges, if we assume that the walls of the parallelepiped are grounded, the scalar potential vanishes identically ${\phi}=0$. 
The quantization is then performed by promoting the expansion coefficients to quantum operators, and the radiation Hamiltonian for the free fields inside the perfect optical device is discretized in terms of the cavity eigenmodes
\begin{equation}
    H_{rad}=\sum_{\mathbf{\alpha}}\omega_{\mathbf{\alpha}}b^\dagger_{\mathbf{\alpha}} b_{\mathbf{\alpha}},
\end{equation}
where $b^\dagger_{\mathbf{\alpha}}$ is the creation operator for the $\alpha$-mode of frequency $\omega_{\mathbf{\alpha}}$. \\

\noindent In the presence of molecules, which can be considered as a source of electromagnetic fields, we must also ensure that the scalar potential in Eq.~\eqref{Poisson_equation} is consistent with the boundary conditions. 
We can enforce them by adding an auxiliary scalar field ${F}$ to the particles' Coulomb potential. This potential can be thought of as a potential generated by \textit{image charges}\cite{jackson1977classical, power1982quantum} placed outside the volume of quantization (i.e. outside the volume of interest of the system). 
This potential can be expressed in terms of a response kernel $\Theta(\bm{s};\bm{r})$ and the real charge distribution $\rho$ inside the cavity\cite{jackson1977classical, power1982quantum}
\begin{equation}\label{image_charge_potential}
    {F}(\bm{r})=\int_{\bar{V}} d^3s\;\frac{\int_{V_c} d^3r'\;\rho(\bm{r}')\Theta(\bm{s};\bm{r}')}{|\bm{r}-\bm{s}|}
\end{equation}
where $\bm{s}$ is a point outside the cavity volume. 
The Hamiltonian describing the molecular system then reads\cite{power1982quantum}
\begin{widetext}
\begin{align}\label{velocity_image_charges}
    H_{c}=&\sum_i\frac{1}{2}\mathbf{p}_i^2 + \frac{1}{2}\sum_{i\neq j} \frac{1}{|\mathbf{r}_i-\mathbf{r}_j|}-\sum_{i,M}\frac{Z_M}{|\mathbf{r}_i-\mathbf{R}_M|}+\frac{1}{2}\sum_{M\neq N}\frac{Z_MZ_N}{|\mathbf{R}_N-\mathbf{R}_M|}\nonumber\\
    &+\sum_i\frac{1}{2c}(\mathbf{p}_i\cdot\mathbf{A}(\bm{r}_i)+\mathbf{A}(\bm{r}_i)\cdot\mathbf{p}_i)+\sum_i\frac{1}{2c^2}\mathbf{A}^2(\bm{r}_i)\nonumber\\
    &+\sum_{\alpha}\omega_{\alpha}{b}^\dagger_{\alpha}{b}_{\alpha}+\frac{1}{2}\sum_MZ_M{F}(\bm{R}_M)-\frac{1}{2}\sum_i{F}(\bm{r}_i).
\end{align}  
\end{widetext}
As seen from \eqref{image_charge_potential}, the last two terms introduced by ${F}$ correspond to a modified Coulomb interaction kernel due to the boundary conditions, i.e. they lead to a modified longitudinal interaction among the particles. 
The factor $1/2$ appears because it describes an interaction with an \textit{image} charge distribution.\cite{jackson1977classical}  
While some authors suggest that this contribution is the major difference from free space,\cite{schuler2020vacua, de2018cavity, barut1987quantum} the discussion of these terms for \textit{ab initio} QED is often neglected. 
Nevertheless, we will see that they can be handled in a more practical way by a suitable unitary transformation.\cite{power1982quantum, vukics2012adequacy} 

\subsubsection*{Remarks}
\noindent    The Hamiltonian in Eq. \eqref{velocity_image_charges} now acts on a Hilbert space $\mathcal{V}$ which is the direct product of the full-CI Fock space for the electrons $\mathcal{V}_{FCI}$ and the photon space spanned by the cavity eigenmodes $\mathcal{V}_{C}$
\begin{equation}\label{radiation-matter-hilbert-space}
    \mathcal{V}=\mathcal{V}_{FCI}\otimes \mathcal{V}_{C}.
\end{equation}
    However, each cavity setup corresponds to specific boundary requirements, leading to different quantum fields: different boundary conditions could substantially modify both the shape of the eigenmodes $\bm{S}_\alpha(\bm{r})$ and the Coulomb kernel $\Theta(\bm{s};\bm{r})$.
    
\noindent    The photonic space $\mathcal{V}_{C}$ is usually truncated, considering only a few eigenmodes relevant to the system. Therefore, the excited states of the Hamiltonian in Eq. \eqref{velocity_image_charges} are again true eigenstates with an infinite lifetime. Notice that, in the limit of an infinitely large cavity, if we retain a considerable number of modes we obtain {again} a fine energy discretization of the photon bath that recovers the decay channels.\cite{ruggenthaler2022understanding} 

\noindent    The masses of the particles in ~\eqref{velocity_image_charges} are dependent on the electromagnetic mode structure and, in general, are different from their physical and free-space bare masses.\cite{craig1998molecular, rokaj2022free} However, in QED \textit{ab initio} calculations, the physical masses are usually employed, implicitly considering the interaction energy with the continuum photon bath\cite{ruggenthaler2022understanding}. The effects of the modification of the masses on chemical and physical properties from an \textit{ab initio} perspective, to the best of our knowledge, is still to explore.
    
\noindent    Finally, because of the states of the optical device $\mathcal{V}_{C}$, the system's symmetries now differ from the bare molecule ones due to the cavity's modified eigenmodes and potential in Eq.~\eqref{image_charge_potential}. This leads to modifications of selection rules and other symmetry-related properties.\\

\subsection{Dipole approximation}\label{chapter_dipole_approx}
\begin{figure}[ht]

        \includegraphics[width=.45\textwidth]{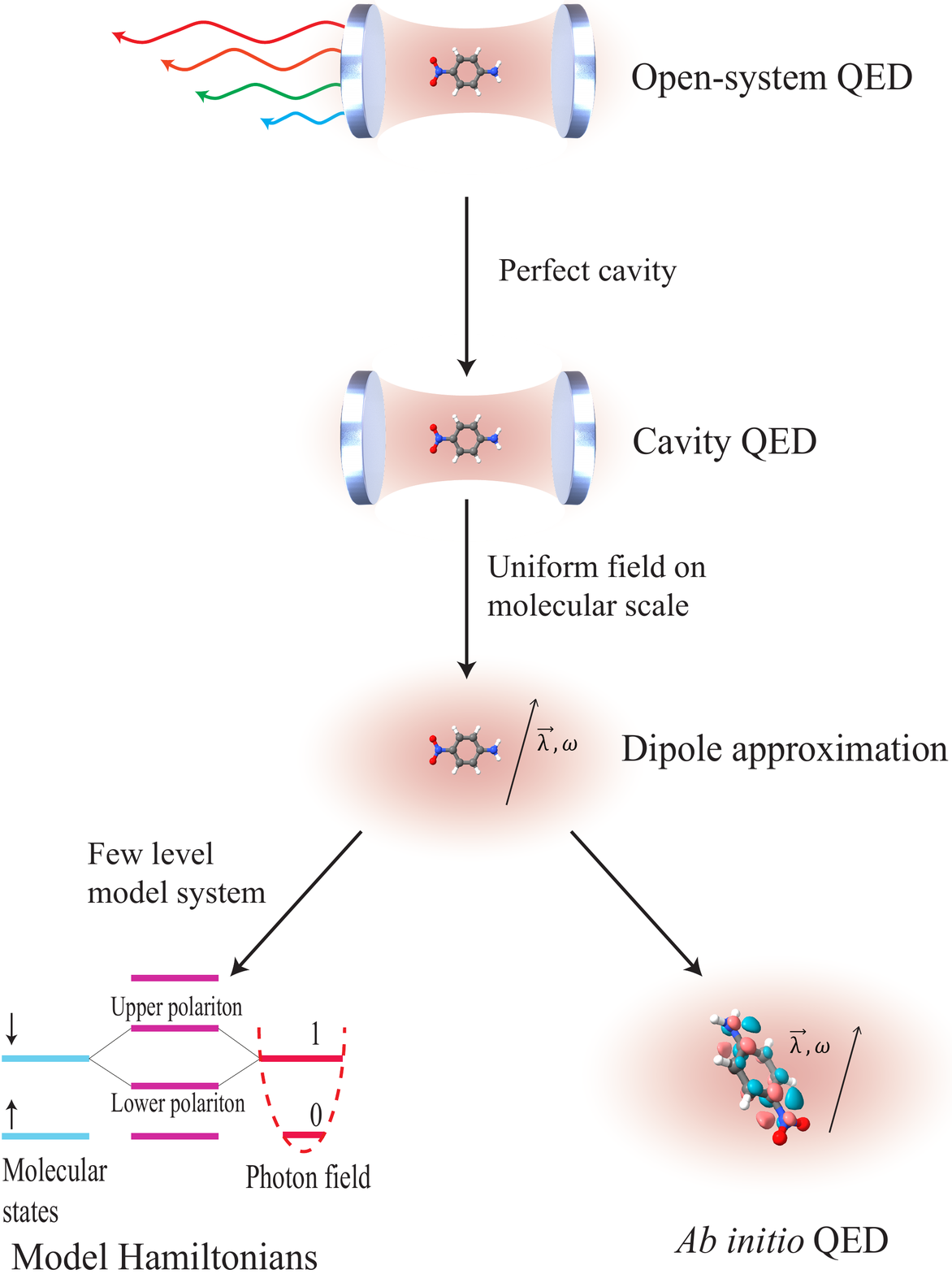}
        \caption{Graphic summary of the hierarchy of approximations for computational polaritonic chemistry. The starting point is the open system described by the nonrelativistic Pauli-Fierz Hamiltonian, which includes the molecular system and the optical device immersed in the photon continuum. As a first approximation, the device is assumed to be perfect, and cavity QED formalism with a limited number of effective photon states is employed. Then, in the dipole approximation, the field is assumed to be uniform over the molecular scale. 
        The exact shape of the photon states is irrelevant, and the system is described through an effective coupling. At the \textit{ab initio} level, the polaritonic problem is approximated with a suitable parametrization.\cite{haugland2020coupled, ruggenthaler2014quantum, riso2022molecular, mallory2022reduced, bauer2023perturbation, mordovina2020polaritonic, mandal2020polarized, haugland2021intermolecular, pavosevic2021polaritonic}
        Alternatively, there are phenomenological models where the molecular complexity is usually reduced to a few selected reference states, as, for instance, in the Jaynes-Cummings model.\cite{jaynes1963comparison}}
        \label{fig:approx_ladder}
    \end{figure}

The dipole approximation is commonly employed in \textit{ab initio} polaritonic chemistry. It assumes that the (relevant) electromagnetic modes have a wavelength much larger than the characteristic lengths of the molecules. 
The complete shape of the modes $\bm{S}_{\bm{\alpha}}(\bm{r})$ is therefore irrelevant, and we only need to evaluate the fields at a point internal to the molecular structure. 
In Eq.~\eqref{eq:Minimal_coupling}, we then set $\mathbf{A}(\bm{r})\to\mathbf{A}(0)$, and obtain the Pauli-Fierz Hamiltonian $H_{PF}^v$ in dipole approximation and velocity representation
\begin{widetext}
\begin{align}\label{PF_velocity}
    H_{PF}^v&=\sum_i\frac{1}{2}\mathbf{p}_i^2 + \frac{1}{2}\sum_{i\neq j} \frac{1}{|\mathbf{r}_i-\mathbf{r}_j|}-\sum_{i,M}\frac{Z_M}{|\mathbf{r}_i-\mathbf{R}_M|}+\frac{1}{2}\sum_{M\neq N}\frac{Z_MZ_N}{|\mathbf{R}_N-\mathbf{R}_M|}\nonumber\\
    &+\sum_i\frac{1}{2c}(\mathbf{p}_i\cdot\mathbf{A}(0)+\mathbf{A}(0)\cdot\mathbf{p}_i)+\sum_i\frac{1}{2c^2}\mathbf{A}^2(0)\nonumber\\
    &+\sum_{\alpha}\omega_{\alpha}{b}^\dagger_{\alpha}{b}_{\alpha}+\frac{1}{2}\sum_MZ_M\tilde{F}(\bm{R}_M)-\frac{1}{2}\sum_i\tilde{F}(\bm{r}_i),
\end{align}  
\end{widetext}
where we applied the BO approximation. 
The dipole approximation allows for the modelling of the photons only by adequately tuning the frequency $\omega_\alpha$ and the coupling strength ${\lambda}_\alpha$ for the cavity modes.\cite{ruggenthaler2022understanding} 
The problem is eventually reduced to a single effective cavity mode. Assuming to work with real field polarization vectors, the final Hamiltonian for our system is
\begin{align}\label{dipole}
    H&=\sum_{pq}h_{pq}E_{pq}+\frac{1}{2}\sum_{pqrs}g_{pqrs}e_{pqrs}+h_{nuc}\nonumber\\
    &+\sqrt{\frac{\omega_\alpha}{2}}(\bm{\lambda}_\alpha\cdot \bm{d})(b^\dagger_\alpha+b_\alpha)+\frac{1}{2}(\bm{\lambda}_\alpha\cdot \bm{d})^2\nonumber\\
    &+\omega_{\alpha}{b}^\dagger_{\alpha}{b}_{\alpha}+\frac{1}{2}\sum_MZ_M\tilde{F}(\bm{R}_M)-\frac{1}{2}\sum_i\tilde{F}(\bm{r}_i),
\end{align}
where $\bm{\lambda}_\alpha =\lambda_\alpha \bm{\epsilon}_\alpha$ is the light-matter coupling vector of the $\alpha$ electromagnetic mode, and we used standard second quantization notation for the electronic Hamiltonian.\cite{helgaker2014molecular} 
This formulation is suitable for nonchiral cavities, while for helicity-preserving devices, we necessarily need to work with complex polarization vectors (see Eqs \eqref{eq:promotion} and \eqref{eq:complex_polarization}).
The Hamiltonian in Eq. \eqref{dipole} can be recast in a more convenient form, called length representation $H_{PF}^l$, by employing a unitary transformation $U$ such that
\begin{equation}
    U\mathbf{p}U^\dagger=\mathbf{p}-\frac{1}{c}\mathbf{A}(0),
\end{equation}
which will allow us to write the interaction between the molecule and the fields in a manifestly dipolar fashion. 
The transformation is
\begin{equation}\label{dipole_transformation}
    U=\text{exp}\bigg[i\sum_i\mathbf{r}_i\cdot\frac{1}{c}\mathbf{A}(0)\bigg]=\text{exp}\bigg[-i\mathbf{d}\cdot\frac{1}{c}\mathbf{A}(0)\bigg]
\end{equation}
 where $\mathbf{d}$ is the electronic dipole operator. 
 By using Eq.~\eqref{dipole_transformation} and a transformation which changes the phases of the creation/annihilation operators\cite{schafer2020relevance, cohen1997photons, tokatly2013time, rokaj2018light}
\begin{equation}\label{eq:operator_phase_photon}
    V=\text{exp}\bigg[-i\frac{\pi}{2}\sum_\alpha b^\dagger_\alpha b_\alpha\bigg],
\end{equation} 
we finally obtain
   \begin{align}\label{PF_dipole}
    &H_{PF}^l=VUH_{PF}^vU^\dagger V^\dagger\nonumber\\
    &=\sum_i\frac{1}{2}\mathbf{p}_i^2+\frac{1}{2} \sum_{i\neq j} \frac{1}{|\mathbf{r}_i-\mathbf{r}_j|}-\sum_{i,M}\frac{Z_M}{|\mathbf{r}_i-\mathbf{R}_M|}+\frac{1}{2}\sum_{M\neq N}\frac{Z_MZ_N}{|\mathbf{R}_N-\mathbf{R}_M|}\nonumber\\
    &+\sum_\alpha\sqrt{\frac{\omega_\alpha}{2}}(\bm{\lambda}_\alpha\cdot \bm{d})(b^\dagger_\alpha+b_\alpha)+\frac{1}{2}\sum_\alpha(\bm{\lambda}_\alpha\cdot \bm{d})^2\nonumber\\ 
    &+\sum_{\alpha}\omega_{\alpha}{b}^\dagger_{\alpha}{b}_{\alpha}
\end{align}   
The bilinear light-matter interaction term of this Hamiltonian can be interpreted as the interaction of the molecular dipole with the displacement field.\cite{cohen1997photons} 
The quadratic term is called dipole self-energy and ensures the Hamiltonian leads to well-defined states and properties.\cite{schafer2020relevance} 
Moreover, transformation Eq. \eqref{dipole_transformation} cancels the image charge contribution from $\tilde{F}$ so that no reference to the image charges appears in Eq. \eqref{PF_dipole}.\cite{power1982quantum, vukics2012adequacy, craig1998molecular} 
Notice that these Hamiltonians are also often referred to as length or velocity \textit{gauge}, although in a QED framework this can be misleading, and the term \textit{form} or \textit{representation} is more appropriate. 
The hierarchy of approximations that are employed to study quantum light-matter systems is pictorially summarized in Fig. \ref{fig:approx_ladder}. 

\subsubsection*{Remarks}
\noindent There are several advantages to employing the Pauli-Fierz Hamiltonian in the length representation. Contrary to the velocity representation Eq.~\eqref{PF_velocity}, Eq.~\eqref{PF_dipole} has a light-matter interaction term linear in the field with no reference to the image charge distribution.
    
     \noindent The transformation in Eq.~\eqref{dipole_transformation} implies that the momentum operator $\mathbf{p}$ now represents the kinetic momentum of particles
    \begin{equation}
        \mathbf{p}=m\bm{v},
    \end{equation}
    so that the first line of Eq.~\eqref{PF_dipole} truly represents the Coulomb and kinetic energy of the matter subsystem only.
    
     \noindent The operator $b^\dagger$ is no longer a purely photonic operator since $U$ in Eq.~\eqref{dipole_transformation} mixes electromagnetic and matter degrees of freedom. The original photon creation operator in the length representation after the tranformatoin is\cite{schafer2020relevance, rokaj2018light, cohen1997photons}
    \begin{equation}\label{length_form_creation}
        VUb^\dagger_\alpha U^\dagger V^\dagger = -ib_\alpha^\dagger -i \frac{1}{\sqrt{2\omega_\alpha}}(\bm{\lambda}_\alpha\cdot \bm{d}).
    \end{equation}
    These operators are now connected to the auxiliary fields of the macroscopic Maxwell's equations, where we can recognize the transverse polarization\cite{cohen1997photons, schafer2020relevance}
    \begin{equation}\label{eq:polarization_displacement_field}
        \mathbf{P}_\perp =\frac{1}{4\pi} \sum_\alpha (\bm{\lambda}_\alpha\cdot \bm{d})\bm{\lambda}_\alpha.
    \end{equation}
    The separation into "matter" and "photon" degrees of freedom is, therefore, blurred in this representation.\cite{cohen1997photons}

\subsection{Beyond the dipole approximation}
\textit{Ab initio} QED calculations often rely upon the dipole approximation, which provides a good compromise between accuracy and affordability. 
Nevertheless, achieving a better description of the light-matter interaction is an essential challenge for polaritonic systems, and a few attempts in this direction have already been reported.\cite{riso2022strong, doi:10.1021/acs.jpclett.3c00286, PhysRevA.107.L021501}
To go beyond the dipole approximation, we could replace the vector potential $\mathbf{A}(\bm{r})$ with its first-order expansion around the molecular origin.
For a sinusoidal field, this reduces to an expansion in terms of the wavevector $\bm{k}$
\begin{equation}
    e^{i\bm{k}\cdot\bm{r}}\approx 1 + i \bm{k}\cdot\bm{r} + \dots
\end{equation}
This expansion is often employed in the semiclassical description of the fields to model, for instance, X-ray absorption leading to the well-known electric quadrupole and magnetic dipole interaction terms.\cite{list2015beyond, bernadotte2012origin, list2020beyond}
The corresponding QED Hamiltonian can then be obtained in the velocity representation by the  following substitution in Hamiltonian Eq. \eqref{velocity_image_charges}
\begin{equation}\label{eq:beyond_dipole_vector_potential}
    \mathbf{A}(\bm{r})\to\mathbf{A}(0)+\bm{r}\cdot \frac{\partial\mathbf{A}}{\partial\bm{r}}\bigg|_{\bm{r}=0}+\dots
\end{equation}
The multipolar Hamiltonian can be obtained similarly to the length representation of the dipole Hamiltonian, applying the Power-Zienau-Woolley (PZW) transformation with the multipolar expansion of the vector potential,\cite{craig1998molecular} i.e. the transformation of Eq. \eqref{dipole_transformation} with the vector potential of Eq. \eqref{eq:beyond_dipole_vector_potential}. 
This transformation also leads to electric (dipole and quadrupole) and magnetic self-energy terms, which guarantee the boundedness of the Hamiltonian.\cite{craig1998molecular,schafer2020relevance, doi:10.1021/acs.jpclett.3c00286} 
Overcoming the dipole approximation is particularly relevant when the molecules interact with chiral fields, which leads to novel phenomena such as cavity-induced circular dichroism or cavity enantiomeric discrimination.\cite{riso2022molecular, doi:10.1021/acs.jpclett.3c00286} 
The description of magnetic interactions is then essential.
However, the multipolar expansion of the interaction beyond the electric dipole presents some difficulties, already in the semiclassical approximation,\cite{list2015beyond, bernadotte2012origin, list2020beyond, lestrange2015consequences} as the interaction terms are now origin dependent. 
The dipole operator also depends on the choice of the origin for charged systems, but this can be shown to be equivalent to a gauge transformation on Hamiltonian Eq. \eqref{dipole}.
Manifest origin invariance can then be recovered by a suitable coherent state transformation, even for a finite electronic basis set (see Eqs. \eqref{coherent-transf} and \eqref{QEDHF-transf-Ham}).\cite{haugland2020coupled, haugland2021intermolecular, riso2022molecular} 
At the same time, to the best of our knowledge, a similar solution has yet to be developed for higher-order Hamiltonians, and therefore using such an expansion could lead to unphysical results. 
This issue could be solved by retaining the complete shape of the field $\bm{S}_\alpha(\bm{r})$ instead of relying upon a multipolar expansion. 
This is considered computationally and theoretically challenging also because of the dependence of the field shape on the boundary conditions. 
Nevertheless, some progress has been made for a sinusoidal field in the semiclassical and QED frameworks. \cite{list2017rotationally, list2020beyond, rokaj2019quantum, rokaj2022polaritonic, riso2022strong}
Alternatively, effective reformulations of the QED problem in terms of reduced quantities, such as in QEDFT, can convey a different perspective for higher-order approximations.\cite{ruggenthaler2014quantum, tokatly2013time, schafer2021making, schafer2022shortcut, schafer2022polaritonic}

\section{Exact polaritonic response theory}\label{exact_resp_chapt}
In this section, we highlight the similarities and differences with standard molecular response theory following the derivation of J\o rgensen and Olsen,\cite{olsen1985linear} with particular emphasis on the quantities that describe the cross-talk between light and matter.\\

\noindent From the exact solution of the time-dependent Schr{\"o}dinger equation
\begin{equation}
    i \frac{d}{d t}\ket{\Psi}={H}\ket{\Psi}
\end{equation}
we derive, for an operator $\Omega$, the Ehrenfest theorem\cite{ehrenfest1927bemerkung}
\begin{equation}\label{ehrenfest}
    \frac{d}{d t}\braket{\Psi|\Omega|\Psi}=-i\braket{\Psi|[\Omega,H]|\Psi} +\braket{\Psi|\frac{\partial \Omega}{\partial t}|\Psi},
\end{equation}
where ${H}$ is here a polaritonic Hamiltonian and $\ket{\Psi}$ are polaritonic states of entangled light-matter character. 
The operator $\Omega$ can then be an electronic, photonic or mixed electron-photon operator.
The exact eigenstates of the Hamiltonian are defined by the eigenvalue equation
\begin{equation}\label{EigenvalueEquationH}
    H\ket{n}=E_n\;\ket{n} \quad n=0,1,2\hdots \;,
\end{equation}
and are assumed to form a complete basis for the radiation-matter Hilbert space $\mathcal{V}$. 
We further assume that at the initial time the system is in its ground polaritonic state $\ket{0}$.
We then apply the following perturbation operator
\begin{equation}\label{pert}
    V^t =e^{\eta t}\sum_n \bigg(V^{\omega_n} e^{-i\omega_n t}+\big(V^{\omega_n}\big)^\dagger e^{i\omega_n t}\bigg)
\end{equation}
where $\eta$ is a positive infinitesimal that ensures the perturbation to vanish for $t\to-\infty$.
The time evolution of the initial ground state is then determined by the time-dependent Hamiltonian
\begin{equation}
    H + V^t
\end{equation}
and can be modelled by a unitary transformation
\begin{equation}
    \ket{\tilde{0}}=\text{exp}[iP'(t)]\ket{0},
\end{equation}
where we used a tilde to indicate the time-evolved state and $P'(t)$ is a Hermitian operator:
\begin{equation}
    P'(t)=\sum_{n>0}\big(P'_n\ket{n}\bra{0}+P_n'^*\ket{0}\bra{n}\big)+(P'_0+P_0'^*)\ket{0}\bra{0}.
\end{equation}
The ground state phase evolution can be explicitly factorized\cite{norman2018principles}
\begin{equation}
    \ket{\tilde{0}}=e^{i\phi(t)}\;\text{exp}\bigg(i\sum_{n>0}\big(P_n\ket{n}\bra{0}+P_n^*\ket{0}\bra{n}\big)\bigg)\ket{0},
\end{equation}
and if the operator $\Omega$ in Eq.~\eqref{ehrenfest} does not involve time differentiation, it can be disregarded. 
We will then make use of the phase-isolated wave function
\begin{align}
    \ket{\Bar{0}}&=\text{exp}\bigg(i\sum_{n>0}\big(P_n\ket{n}\bra{0}+P_n^*\ket{0}\bra{n}\big)\bigg)\ket{0}
\end{align}
and perform an expansion of the state transfer coefficients in orders of the perturbation
\begin{widetext}
\begin{equation}\label{eq:pert_coeff_exp}
    P_n=P_n^{(0)}+P_n^{(1)}+P_n^{(2)}+P_n^{(3)}+\dots
\end{equation}
which leads to the perturbative expansion of the phase-isolated state
    \begin{align}
    \ket{\bar{0}}^{(0)}&=\ket{0}\label{zero_wave}\\
    \ket{\bar{0}}^{(1)}&=i\sum_{n>0}\ket{n}P_n^{(1)}\\
    \ket{\bar{0}}^{(2)}&=-\frac{1}{2}\ket{0}\sum_{j>0}P_j^{(1)}P_j^{*(1)}+i\sum_{n>0}\ket{n}P_n^{(2)}\\
    \ket{\bar{0}}^{(3)}&=-\frac{1}{2}\ket{0}\sum_{j>0}\big(P_j^{(2)}P_j^{*(1)}+P_j^{(1)}P_j^{*(2)}\big)+i\sum_{n>0}\ket{n}\big(P_n^{(3)}-\frac{1}{6}P_n^{(1)}\sum_{j>0}P_j^{(1)}P_j^{*(1)}\big).\label{third_wave}
\end{align}
By defining $P_{-n}\equiv P_n^*$ and by using in Eq.~\eqref{ehrenfest} the state transfer operators $\Omega \in\big\{\ket{0}\bra{n}\;;\,\ket{n}\bra{0}\big\}$, we obtain the following hierarchy of differential equations, for positive and negative indices\cite{olsen1985linear}
    \begin{align}
    i\;\text{\text{sgn}}(k)\dot{P}_k^{(1)}-\omega_kP_k^{(1)}&=-iV_k^t\\
    i\;\text{\text{sgn}}(k)\dot{P}_k^{(2)}-\omega_kP_k^{(2)}&=\sum_nV^t_{k-n}P_n^{(1)}\\
    i\;\text{\text{sgn}}(k)\dot{D}_k^{(3)}-\omega_kD_k^{(3)}&=\sum_n\bigg[V^t_{k-n}P_n^{(2)}+i\theta(kn)V^t_nP^{(1)}_{-n}P_k^{(1)}\bigg]+iV^t_k\sum_{n>0}P^{(1)}_{-n}P_n^{(1)},
\end{align}
where $\theta(x)$ is the Heaviside step function, while $\omega_{k}$, $D^{(n)}_{k}$ and $V^{t}_{n}$ are defined as
\begin{align}
    &\omega_k = \omega_{-k} = E_k - E_0\\
    &D_k^{(3)}=P^{(3)}_k-\frac{2}{3}P_n^{(1)}\sum_{j>0}P^{(1)}_jP^{(1)}_{-j}\\
   &V_k^t = \begin{pmatrix}
         \braket{k|V^t|0}  \\
          -\braket{0|V^t|-k}
    \end{pmatrix};\begin{pmatrix}
         k>0  \\
         k<0
    \end{pmatrix}\\
    & V_{kn}^t= \begin{pmatrix}
    0 & \braket{k|V^t|-n}-\delta_{k,-n}\braket{0|V^t|0}\\
    \braket{n|V^t|-k}-\delta_{-k,n}\braket{0|V^t|0}  & 0
    \end{pmatrix};
    \begin{pmatrix}
    k,n>0 & k>0,n<0\\
    k<0,n>0 & k,n<0
    \end{pmatrix}
\end{align}
Notice that the solutions for lower-order coefficients are necessary for higher-order equations. 
We can solve these equations, which all have the structure
\begin{equation}
    i\;\text{\text{sgn}}(k)\frac{df(t)}{dt}-\omega_kf(t)=g(t)
\end{equation}
with the general solution
\begin{equation}\label{eq:general_coefficients}
    f(t)=-i\;\text{\text{sgn}}(k)e^{-i\omega_k\text{\text{sgn}}(k)t}\int_{-\infty}^td\tau e^{i\omega_k\text{\text{sgn}}(k)\tau}g(\tau),
\end{equation}
which fulfils the vanishing initial conditions for the coefficients. 
Expressing the change of the time development average value of an observable $A$ from its zero-order mean value in orders of the perturbation we obtain
    \begin{align}\label{resp_esp}
   \delta\braket{A}(t)=A_{V^t}(t)-A_{V^t}(0)=\braket{\tilde{0}|A|\tilde{0}}-\braket{0|A|0}&=  \nonumber\\ 
    & \sum_n e^{-i\omega_n t+ \eta t} \braket{\braket{A;V^{\omega_n}}}_{\omega_n+i\eta}\nonumber\\
    + \frac{1}{2}&\sum_{mn}e^{-i\omega_n t -i\omega_mt+2\eta t}\braket{\braket{A;V^{\omega_n},V^{\omega_m}}}_{\omega_n+i\eta,\omega_m+i\eta}+\nonumber \\
    + \frac{1}{6}&\sum_{lmn}e^{-i\omega_l t -i\omega_mt-i\omega_nt+3\eta t}\braket{\braket{A;V^{\omega_l},V^{\omega_m},V^{\omega_n}}}_{\omega_l+i\eta,\omega_m+i\eta,\omega_n+i\eta}+\dots.
\end{align}
We then define the linear, quadratic and cubic response functions, respectively
\begin{align}\label{exp1}
  & \braket{\braket{A;V^{\omega_n}}}_{\omega_n} \\
 &  \braket{\braket{A;V^{\omega_n},V^{\omega_m}}}_{\omega_n,\omega_m}\\
 &\braket{\braket{A;V^{\omega_l},V^{\omega_m},V^{\omega_n}}}_{\omega_l,\omega_m,\omega_n}.
\end{align}
These functions express the linear, quadratic and cubic time-variation of an observable $A$ as resulting from a perturbation in the frequency domain
\begin{align}
    \delta\braket{A}^{(1)}(\omega_1)&=\braket{\braket{A;V^{\omega_1}}}_{\omega_1}\\
    \delta\braket{A}^{(2)}(\omega_1,\omega_2)&=\frac{1}{2}\braket{\braket{A;V^{\omega_1},V^{\omega_2}}}_{\omega_1,\omega_2}\\
    \delta\braket{A}^{(3)}(\omega_1,\omega_2,\omega_3)&=\frac{1}{6}\braket{\braket{A;V^{\omega_1},V^{\omega_2},V^{\omega_3}}}_{\omega_1,\omega_2,\omega_3},
\end{align}
and can therefore be interpreted as molecular (hyper)polarizabilities.
For instance, for an electric dipole perturbation
\begin{equation}
    V^\omega=-\bm{d}\cdot\bm{E}(\omega),
\end{equation}
the first order variation of the molecular dipole reads
\begin{equation}
\delta\braket{\bm{d}}^{(1)}(\omega)=-\braket{\braket{\bm{d};\bm{d}}}_{\omega}\cdot\bm{E}(\omega)
\end{equation}
and the linear dipole-dipole response function $\braket{\braket{\bm{d};\bm{d}}}_{\omega}$ can then be interpreted as the (negative) time-dependent molecular polarizability at frequency $\omega$.
Using the perturbation expansions of the coefficients and the state wave function Eqs. \eqref{eq:pert_coeff_exp} and \eqref{zero_wave}-\eqref{third_wave}, we obtain
    \begin{align}\label{exp2}
    \braket{A}_{V^t}=& \braket{0|A|0}\nonumber\\
    +&\braket{0^{(1)}|A|0}+\braket{0|A|0^{(1)}}\nonumber\\
    \nonumber
    +&\braket{0^{(1)}|A|0^{(1)}}+\braket{0|A|0^{(2)}}+\braket{0^{(2)}|A|0}\\
    +&\braket{0^{(3)}|A|0}+\braket{0|A|0^{(3)}}+\braket{0^{(1)}|A|0^{(2)}}+\braket{0^{(2)}|A|0^{(1)}}+\dots\nonumber \\
    =&\braket{0|A|0}\nonumber \\
    -&i\sum_nA_{-n}P_n^{(1)}\nonumber\\
    -&i\sum_nA_{-n}P_n^{(2)}+\sum_{n,j>0}P_{-j}^{(1)}A_{j-n}P_{n}^{(1)}\nonumber\\
    -i&\sum_nA_{-n}D^{(3)}_n+\sum_{n,k>0}P^{(2)}_{-k}A_{k-n}P_n^{(1)}+\sum_{n,k>0}P^{(1)}_{-k}A_{k-n}P_n^{(2)}+\dots
\end{align}
By comparing Eqs. \eqref{resp_esp} and \eqref{exp2}, using Eq. \eqref{eq:general_coefficients} for the state transfer parameters, we finally obtain the explicit expressions for the linear, quadratic and cubic response functions in terms of the eigenstates of the Hamiltonian\cite{olsen1985linear}
\begin{align}
    &\braket{\braket{A;V^\omega}}_{\omega} = -\sum_k\frac{\text{\text{sgn}}(k)A_{-k}V_k^\omega}{\omega- \text{\text{sgn}}(k)\omega_k}\label{eq:lin_resp_fun}\\
    &\braket{\braket{A;V^{\omega_1},V^{\omega_2}}}_{\omega_1,\omega_2}=
   \hat{P}(\omega_1,\omega_2)\left[-\sum_{k,n}\frac{A_{-k}V^{\omega_1}_{k-n}V^{\omega_2}_n}{(\omega_1+\omega_2-\text{sgn}(k)\omega_k)(\omega_2-\text{sgn}(n)\omega_n)}+\sum_{k,n>0}\frac{V^{\omega_1}_{-k}A_{k-n}V^{\omega_2}_n}{(\omega_1+\omega_k)(\omega_2-\omega_n)}\right]\label{eq:quad_resp_fun}\\
   &\nonumber\braket{\braket{A;V^{\omega_1},V^{\omega_2},V^{\omega_3}}}_{\omega_1,\omega_2,\omega_3}=
     \nonumber\hat{P}(\omega_1,\omega_2,\omega_3)\left\{\sum_k\frac{A_{-k}\text{sgn}(k)}{\omega_1+\omega_2+\omega_3-\text{sgn}(k)\omega_k} \right.\;\times\left[\sum_n\frac{\theta(kn)V^{\omega_1}_{n}V^{\omega_2}_{-n}V^{\omega_3}_k}{(\omega_2+\text{sgn}(n)\omega_n)(\omega_3-\text{sgn}(k)\omega_k)}\right.\\
     &\left.-\sum_{n,m}\frac{V^{\omega_1}_{k-n}V^{\omega_2}_{n-m}V^{\omega_3}_m}{(\omega_2+\omega_3-\text{sgn}(n)\omega_n)(\omega_3-\text{sgn}(m)\omega_m)}+ \sum_{n>0}\frac{V^{\omega_1}_{k}V^{\omega_2}_nV_{-n}^{\omega_3}}{(\omega_2-\text{sgn}(n)\omega_n)(\omega_3+\text{sgn}(n)\omega_n)}\right]\nonumber\\
    &\left.-\sum_{m,n,k>0}\left( \frac{V^{\omega_1}_{-km}V^{\omega_2}_{-m}V_{n}^{\omega_3}A_{k-n}}{(\omega_1+\omega_2+\omega_k)(\omega_2+\omega_m)(\omega_3-\omega_n)}-\frac{V^{\omega_1}_{-k}V^{\omega_2}_{n-m}V_{m}^{\omega_3}A_{k-n}}{(\omega_2+\omega_3-\omega_n)(\omega_1+\omega_k)(\omega_3-\omega_m)}\right)\right\}\label{cubic:lin_resp_fun}
\end{align}
\noindent where $\hat{P}(\omega_1,\omega_2)$ and $\hat{P}(\omega_1,\omega_2,\omega_3)$ sum all the permutations of the frequencies
\begin{align}
    \hat{P}(\omega_1,\omega_2)f(\omega_1,\omega_2)&=f(\omega_1,\omega_2)+f(\omega_2,\omega_1)\\
    \hat{P}(\omega_1,\omega_2,\omega_3)f(\omega_1,\omega_2,\omega_3)&=f(\omega_1,\omega_2,\omega_3)+f(\omega_1,\omega_3,\omega_2)\nonumber \\
    &+f(\omega_2,\omega_1,\omega_3)+f(\omega_2,\omega_3,\omega_1)\nonumber\\&+f(\omega_3,\omega_1,\omega_2)+f(\omega_3,\omega_2,\omega_1),
\end{align}
and we set the switching parameter $\eta$ to zero.
\end{widetext}

\noindent  These equations are identical to the expressions obtained in molecular response theory,\cite{olsen1985linear, norman2018principles, casida2012progress, helgaker1999, christiansen1998response, sasagane1993higher, langhoff1972aspects, cammi1999linear, helgaker2012recent, lazzeretti2004assessment} since they only rely on the Schr{\"o}dinger equation.
However, the explicit treatment of internal electromagnetic fields as dynamical variables in our system leads to novel perspectives. 
First of all, we must remember that the eigenstates of the system are mixed matter-photon states, so they belong to a larger Hilbert space than the usual molecular one. 
The properties of these field-dressed states will therefore differ from the bare-molecule ones, leading, for instance, to different excitation energies and transition moments. 
Moreover, additional possibilities for probing the properties of the system arise since the external perturbation $V^t$ can now act on both matter and photon degrees of freedom. 
Furthermore, considering the internal electromagnetic fields allows us to explore additional observables connected to the radiation fields and their interactions with matter. \\

\noindent The linear response function Eq. \eqref{eq:lin_resp_fun} describes the first-order variation of an observable $A$ and it is connected to the linear polarizabilities of the system. Its poles occur at the polaritonic excitation energies, and the corresponding residues are connected to the transition moments between the ground and the excited states
\begin{align}
    \lim_{\omega\to\omega_n}(\omega-\omega_n)\braket{\braket{A;B}}_\omega&=\braket{0|A|n}\braket{n|B|0}\\
&\equiv A_{0n}B_{n0}\nonumber\\
\lim_{\omega\to-\omega_n}(\omega+\omega_n)\braket{\braket{A;B}}_\omega&=-\braket{0|B|n}\braket{n|A|0}\\
&\equiv -B_{0n} A_{n0}\nonumber.
\end{align}
 The quadratic response function Eq. \eqref{eq:quad_resp_fun} is connected to the first hyperpolarizabilities of the system since it describes the second-order response to an external perturbation. 
Its residues are connected to the transition moments and matrices between the excited states.
These can be obtained once the transition moments between ground and excited states have been computed
\begin{widetext}
    \begin{align}
    &\lim_{\omega_2\to\omega_m}(\omega_2-\omega_m)\braket{\braket{A;B,C}}_{\omega_1,\omega_2}=   \sum_{k>0}\bigg[\frac{\braket{0|A|k}\big(\braket{k|B|m}-\delta_{mk}\braket{0|B|0}\big)}{(\omega_1+\omega_m-\omega_k)}-\frac{\braket{0|B|k}\big(\braket{k|A|m}-\delta_{mk}\braket{0|A|0}\big)}{\omega_1+\omega_k}\bigg]\braket{m|C|0}\\
    &\lim_{\omega_1\to-\omega_q}(\omega_1+\omega_q)\lim_{\omega_2\to\omega_m}(\omega_2-\omega_m)\braket{\braket{A;B,C}}_{\omega_1,\omega_2}=-
    \braket{0|B|q}\big(\braket{q|A|m}-\delta_{mq}\braket{0|A|0}\big)\braket{m|C|0}.
\end{align} 
\end{widetext}
These expressions also give us access to excited state properties such as the dipole moment.
Finally, the cubic response function describes the third-order variations of an observable as a consequence of an external perturbation, and it is therefore connected to the second hyperpolarizabilities of the system.\\

\noindent The response functions exhibit the following symmetry relations:
\begin{align}\label{lr_symm}
     \braket{\braket{A;B}}_{\omega}=&\braket{\braket{B;A}}_{-\omega}\nonumber\\
     =&\braket{\braket{A;B}}_{-\omega}^*\\
     \nonumber\\
    \braket{\braket{A;B,C}}_{\omega_1,\omega_2}=&\braket{\braket{A;C,B}}_{\omega_2,\omega_1}\nonumber\\
    =&\braket{\braket{C;A,B}}_{-\omega_1-\omega_2,\omega_1}\nonumber\\
    =&\braket{\braket{A;B,C}}_{-\omega_1,-\omega_2}^*\\
         \nonumber\\
    \braket{\braket{A;B,C,D}}_{\omega_1,\omega_2,\omega_3}=&\braket{\braket{A;C,B,D}}_{\omega_2,\omega_1\omega_3}\nonumber\\
    =&\braket{\braket{D;A,B,C}}_{-\omega_1-\omega_2-\omega_3,\omega_1,\omega_2}\nonumber\\
    =&\braket{\braket{A;B,C,D}}_{-\omega_1,-\omega_2, -\omega_3}^*,
\end{align}
for Hermitian operators $A$, $B$, $C$, and $D$.

\subsubsection*{Time-independent limit} \label{QED-TIPT}
The static response theory can be obtained from the previous derivation by setting the external frequencies $\omega$ and the switching parameter $\eta$ to zero. 
Since the perturbation is now time-independent, it is possible to relate the static polaritonic properties to the derivatives of the Hamiltonian eigenvalues.
The perturbation $V$ acting on the system is time-independent and supposed to be smooth and small, and it is usually expressed as an external field $\bm{F}$ acting on a molecular multipole $\bm{x}$
\begin{equation}\label{eq:static_pert_mult}
    V = - \bm{x}\cdot \bm{F}.
\end{equation}
The perturbed Hamiltonian then reads
\begin{gather} \label{Htipt}
    H = H_{0} +  V =H_{0} - \bm{x}\cdot \bm{F} .
\end{gather}
The Hamiltonian in Eq. \eqref{Htipt} is endowed with well-defined eigenstates and eigenvalues, whose small changes compared to the unperturbed eigenvalues and eigenfunctions are commonly studied via the Rayleigh-Schr\"odinger (RS) perturbation theory.\cite{RSpertutheory} 
From the eigenfunctions $\ket{n}$ of the unperturbed Hamiltonian $H_{0}$ (see Eq. \eqref{EigenvalueEquationH}), the ground-state eigenfunction $\ket{\Tilde{0}}$ and the eigenvalue $\Tilde{E}$ of the perturbed operator are expressed as a power series:
\begin{align}
    \ket{\Tilde{0}} &= \ket{0} +  \ket{\Tilde{0}^{(1)}} + \hdots \\
    \Tilde{E} &= E_{0} +  E^{(1)} + E^{(2)} + \hdots
\end{align}
The first- and second-order corrections to the energy read:
\begin{gather}
    E^{(1)} = \bra{0} V \ket{0} \label{E1} \\
    E^{(2)} = \sum_{k=1}^{\infty} \frac{\bra{0} V \ket{k} \bra{k} V \ket{0}}{E_0 - E_{k}} =\frac{1}{2}\braket{\braket{V;V}}_{\omega=0}.\label{E2}
\end{gather}
Eq.~\eqref{E1} is the well-known Helmann-Feynman theorem\cite{hellmann1937,feynman1939} and only requires the knowledge of the unperturbed wave function, while Eq.~\eqref{E2} also includes a sum-over-states contribution from each excited state $\ket{k}$ with energy $E_{k}$.
The derivatives of the energy with respect to the external fields provide us with the ground state molecular properties.
From Eqs. \eqref{eq:static_pert_mult} and \eqref{E1}, we then identify the derivative of the energy with respect to the field $\bm{F}$ as the ground state mean multipole value
\begin{equation}
    -\frac{d\tilde{E}}{d\bm{F}}\bigg|_{\bm{F}=0} = \braket{0|\bm{x}|0}
\end{equation}
From Eq. \eqref{E2}, the second derivative of the energy is then interpreted as the multipole polarizability to the field $\bm{F}$, and from Eq. \eqref{eq:lin_resp_fun} we identify the linear response function for $\omega=0$
\begin{equation}
    \frac{d^2\tilde{E}}{d^2\bm{F}}\bigg|_{\bm{F}=0}=2\sum_{k=1}^{\infty} \frac{\bra{0} \bm{x} \ket{k} \bra{k} \bm{x} \ket{0}}{E_0 - E_{k}} =\braket{\braket{\bm{x};\bm{x}}}_{\omega=0}.
\end{equation}
The second derivatives of the energy are then connected to the molecular polarizabilities.
Higher-order derivatives relate hyperpolarizabilities and nonlinear response functions.
Notice that these expressions are the same as standard molecular static response theory, and they also hold for exact polaritonic states since they are based only on the Schr{\"o}dinger equation. \\

\subsection{On the rotational average of computed molecular properties}\label{sec:rotational_avergage}
In this section, we discuss the problem of the orientational average of the computed polarizabilities, which is straightforwardly accomplished in standard molecular response theory to connect the single-molecule calculation to the macroscopic sample response. 
Since we are now considering states that belong to the extended light-matter Hilbert space in Eq.~\eqref{radiation-matter-hilbert-space}, we have to consider the molecule and the electromagnetic environment, which are defined by the optical device. 
In general, this introduces anisotropy in the system. 
The cavity field can also break the molecular symmetry, possibly allowing for otherwise symmetry-forbidden transitions. 
This anisotropy-symmetry breaking introduces further complications in the computation of molecular properties.\\

\begin{figure}
    \centering
    \includegraphics[width=.45\textwidth]{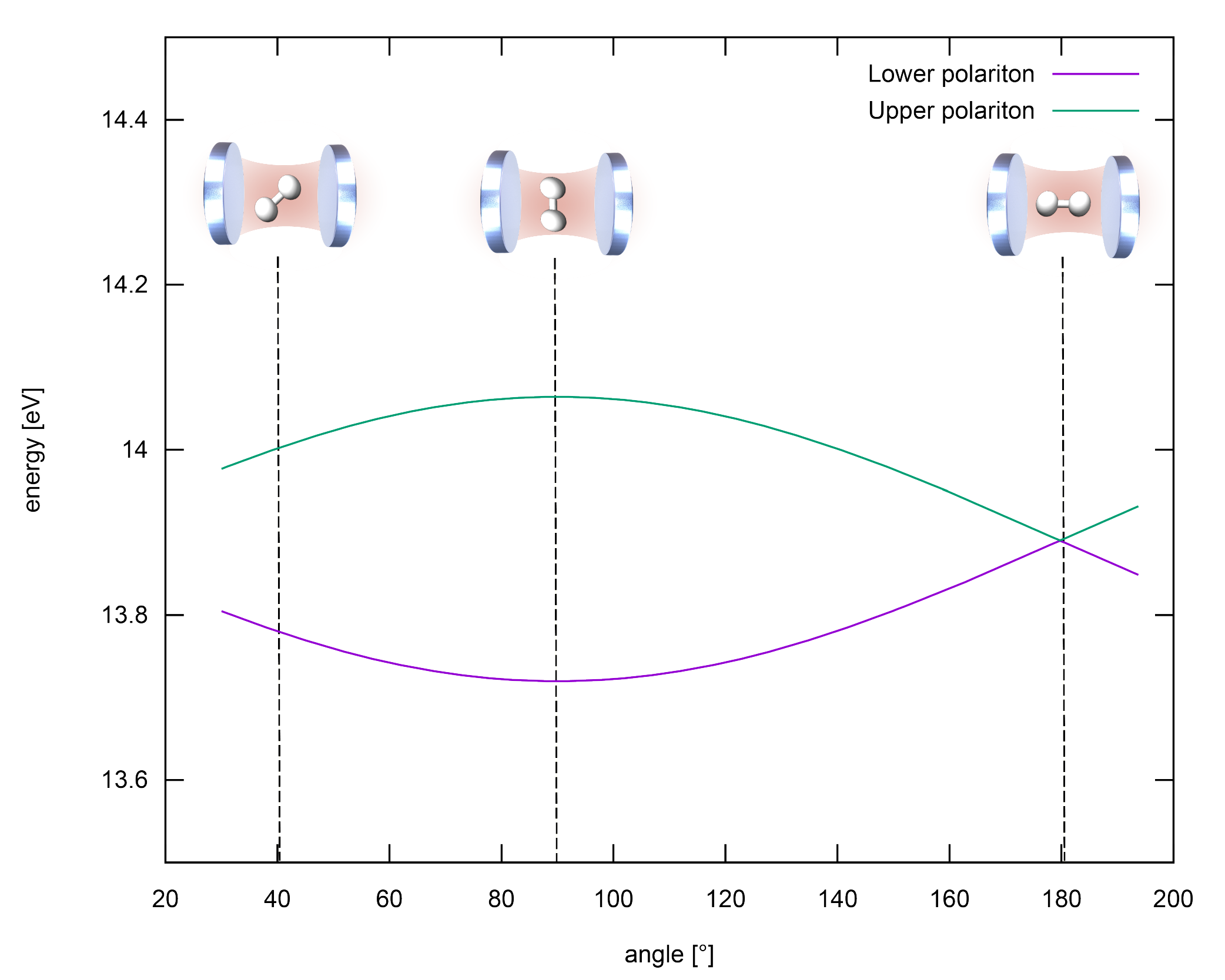}
    \caption{Potential energy surfaces for a hydrogen molecule's upper and lower polaritons resonantly coupled with a quantum cavity, as a function of the molecular orientation. The energies are computed by using the time-dependent quantum electrodynamic Hartree-Fock method with a cc-pVDZ basis set and coupling constant set to $\lambda = 0.01$ a.u. (see section \eqref{Approx_resp_chapt}). When the molecule's transition dipole moment is aligned with the cavity field (90°), the Rabi splitting is maximal, while it decreases to zero when the transition moment is perpendicular to the field's polarization (180°).}
    \label{fig:rotation_energy_H2}
\end{figure}

\noindent The bare molecular states are independent of the spatial orientation of the molecule.
The energy is invariant for rigid-body rotations of the system, while the charge density rotates following the molecular structure, and so do properties such as molecular polarizabilities. 
However, the strong coupling between molecular and cavity states depends on the projection of the molecular dipole operator onto the cavity field, as can be seen in the electron-photon interaction term in Hamiltonian Eq.~\eqref{dipole}. 
This also means {that} there is a non-trivial energy dependence on the relative molecular orientation in the cavity field. 
Excited states will be particularly affected by this orientation dependence.
A bare electronic excitation with a transition moment orthogonal to the cavity field will have a negligible direct coupling with the photonic states. 
On the other hand, if the transition dipole moment is aligned with the field polarization, there will be mixing, giving rise to intense polaritonic excitations of hybrid light-matter character. 
This is illustrated in Fig. \eqref{fig:rotation_energy_H2}, where we plot the upper and lower polaritonic energies of a hydrogen molecule as a function of the angle between the transition dipole moment and the cavity field.
These considerations on the molecular orientation pose some issues for the computation of molecular properties, as the direction of the cavity field is fixed in space by the experimental setup. 
In contrast, inside the cavity, the molecules are usually randomly or quasi-randomly oriented. 
Following a chemical approach, we could compute the sample's properties using the frequency-dependent polarizability Eq.~\eqref{eq:lin_resp_fun}, which must be averaged over the different molecular orientations\cite{barron2009molecular, bernadotte2012origin}. 
The molecular orientation will be defined by a general set of parameters $\Omega$.
For instance, it can be uniquely defined by the three Euler's angles $(\phi,\theta,\chi)$.
Notice that a different set of parameter can be convenient depending on the setup.
For instance, for a Fabry-Pérot resonator with two degenerate modes with wavevector $\mathbf{k}$ and perpendicular polarizations, a rotation of the molecule along $\mathbf{k}$ does not change the Hamiltonian.
Then it is sufficient to use only two parameters instead of the three Euler's angles.
If we consider the Boltzmann weight of each different orientation, we have
\begin{widetext}
    \begin{equation}
    \overline{\braket{\braket{A\;;B}}}_\omega=\int_\Omega  e^{-\frac{E_0(\Omega)}{k_B T}}\;\sum_k\bigg(\frac{A_{0k}(\Omega)B_{k0}(\Omega)}{\omega-\omega_k(\Omega)}-\frac{B_{0k}(\Omega)A_{k0}(\Omega)}{\omega+\omega_k(\Omega)}\bigg)d\Omega\bigg/\int_\Omega  e^{-\frac{E_0(\Omega)}{k_B T}}d\Omega,
\end{equation}
\end{widetext}
where $\overline{\braket{\braket{A\;;B}}}_\omega$ is the averaged response function, $E_0(\Omega)$ is the {ground state} energy of the polaritonic system as a function of the molecular orientation, $k_B$ is the Boltzmann constant, and $T$ is the temperature. 
Even if we disregard the Boltzmann weight and perform an isotropic average, the result does not simply lead to an isotropic polarization tensor.
The excitation energies in the denominator depend on the orientation, as well as the numerator. 
The polarization will be generally anisotropic because of the introduction of the optical device. \\

\noindent As a concrete example, we can consider the computation of absorption spectra.
The linear absorption spectrum is connected to the dipole-dipole polarizability, where $A$ and $B$ in Eq. \eqref{eq:lin_resp_fun} are components of the dipole operator $d_i$\cite{olsen1985linear, norman2018principles, barron2009molecular}
\begin{align}
    &\nonumber\braket{\braket{d_i\;;d_j}}_\omega=\sum_{k>0}\bigg(\frac{\braket{0|d_i|k}\braket{k|d_j|0}}{\omega-\omega_k}-\frac{\braket{0|d_j|k}\braket{k|d_i|0}}{\omega+\omega_k}\bigg).
\end{align}
Neglecting the dependence of the polaritons on the molecular orientation, we obtain the usual expression for absorption by randomly oriented molecules in terms of the oscillator strength
\begin{equation}\label{os_str}
    f_{k0} =\frac{2}{3}\;\omega_k\sum_{i=x,y,z}|\braket{0|d_i|k}|^2.
\end{equation}
This corresponds to fixing the relative molecule-field orientation.
A fictitious peak-broadening is often applied to the computed spectrum $S(\omega)$, e.g. a Lorentzian lineshape with bandwidth broadening parameter $\Delta$
\begin{equation}\label{lineshape}
    \Gamma(\omega,\omega_k) =\frac{1}{\pi} \;\frac{\Delta}{(\omega-\omega_k)^2+\Delta^2} 
\end{equation}
as reported in Refs.\cite{flick2019light, yang2021quantum} so that the computed spectrum is
\begin{equation}\label{spectrum}
    S(\omega)=\sum_k f_{k0}\;\Gamma(\omega,\omega_k).
\end{equation}
However, different orientations could play an important role in determining the properties of the sample, and the choice of a fixed relative orientation does not give the full picture. 
For instance, if we are interested in a specific transition, we could suppose the field polarization to be parallel to the transition dipole moment.
However, this could lead to biased results as certain effects could be enhanced while others suppressed.

\noindent The study of orientational disorder in \textit{ab initio} polaritonic chemistry is still to be addressed. 
A possible approach would be to perform molecular dynamics simulations that sample the different orientations of the molecule inside the cavity and perform an average {of} the spectra obtained for each snapshot. 
This corresponds to an incoherent superposition of the spectra of an ensemble. 
The appearance of collective effects can introduce further complications in the computation of molecular properties. 
If we could perform an \textit{ab initio} simulation including a sufficiently large number of molecules randomly or quasi-randomly oriented, the corresponding computed properties would include both collective and orientational effects.
Alternatively, collective effects can initially be addressed with a simplified model, such as the Tavis-Cumming (TC) model,\cite{tavis1968exact, dicke1954coherence} which considers an ensemble of identical two-level non-interacting systems. 
Moreover, it disregards the dipole self-energy, the molecular dipoles and the so-called counter-rotating terms of the Hamiltonian.
The Rabi splitting is then predicted to scale with the square root of the number of molecules and, therefore, it is sufficient to employ a smaller coupling strength compared to single-molecule calculations.
The generalization of the TC model to an energy-broadened set of two-level molecules\cite{houdre1996vacuum} predicts a Rabi splitting dependent on the \textit{quadratic average} of the coupling strengths and on the energy broadening. 
The predicted spectrum would be very different from the one obtained by an average of single-molecule absorption spectra. \\

\noindent The interplay between single-molecule and collective effects will need a careful comparison between simplified models\cite{jaynes1963comparison, hopfield1958theory, tavis1968exact, dicke1954coherence} and more accurate simulations, and recently some methods to study polaritonic chemistry in an impurity-like approach by an effective embedding have been proposed.\cite{schafer2022shortcut, schafer2022polaritonic, sidler2020polaritonic, pavovsevic2022wavefunction, li2022energy, wang2021defect}
We believe both these effects are important.
However, which one dominates will likely depend on the experimental conditions. 
Both should be carefully considered when accurately modelling polaritonic properties, and further studies in this direction are needed. 
\\

\subsection{Equivalent expressions for polaritonic properties}\label{sec:equivalent_expression}
In this section, we derive equations of motion for polaritonic response functions.
Inserting the expansion Eq.~\eqref{resp_esp} into Ehrenfest theorem Eq.~\eqref{ehrenfest}, we obtain a set of equations as in standard molecular response theory\cite{olsen1985linear}
\begin{widetext}
    \begin{align}
    \omega_1\braket{\braket{A;V^{\omega_1}}}_{\omega_1}&=\braket{\braket{[A,H];V^{\omega_1}}}_{\omega_1}+\braket{0|[A,V^{\omega_1}]|0}\label{fir_ord}\\
    (\omega_1+\omega_2)\braket{\braket{A;V^{\omega_1},V^{\omega_2}}}_{\omega_1,\omega_2}&=\braket{\braket{[A,H];V^{\omega_1},V^{\omega_2}}}_{\omega_1,\omega_2}+\hat{P}(\omega_1,\omega_2)\braket{\braket{[A,V^{\omega_1}];V^{\omega_2}}}_{\omega_2}\label{EOM_2_or_resp}
\end{align}
\end{widetext}
Considering Eq.~\eqref{fir_ord} and the position-momentum operator identity
\begin{equation}
    \bm{p}=-i[\bm{r},H^l]\label{eq1}
\end{equation}
which holds for the dipole Hamiltonian in the length representation Eq.~\eqref{PF_dipole}, we obtain the relation
\begin{equation}
    \omega \braket{\braket{\bm{r};\bm{r}}}_\omega=i\braket{\braket{\bm{p};\bm{r}}}_\omega,
\end{equation}
which states that the frequency-dependent dipole polarizability can be evaluated regarding the position or the conjugate momentum.
This ensures the equivalence of expressions in the dipole and velocity form of transition dipole moments
\begin{equation}\label{dipole-velocity-form}
    \omega_n \braket{0|\bm{r}|n}=i\braket{0|\bm{p}|n},
\end{equation}
where $\ket{n}$ is an excited state and $\omega_n$ is the excitation energy. 
We note that for the velocity representation, in Eq.~\eqref{PF_velocity}, the analogous relation is
\begin{equation}
    \bm{p}+\frac{1}{c}\mathbf{A}(\bm{r})=-i[\bm{r},H^v]\label{eq2},
\end{equation}
which states, exactly as Eq.~\eqref{dipole-velocity-form}, an equivalence between the transition dipole moment and the transition \textit{kinetic} momentum of the electrons.  
Indeed, Eqs.~\eqref{eq1} and~\eqref{eq2} are connected by the transformation in Eq.~\eqref{dipole_transformation}. 
We recall that in second quantization, Eqs.~\eqref{eq1} and \eqref{eq2} are fulfilled only in the limit of a complete basis set.\cite{helgaker2014molecular} \\

An example of a relation introduced by the electromagnetic degrees of freedom is provided by the commutation relation between the Hamiltonian and the annihilation operators in the length form
\begin{equation}
    [b_\alpha,H^l]=\lambda_\alpha\sqrt{\frac{\omega_\alpha}{2}}(\bm{d}\cdot\bm{\epsilon}_\alpha)+\omega_\alpha b_\alpha,
\end{equation}
which through Eq.~\eqref{fir_ord} leads to
\begin{align}\label{photon_eq_ref}
\omega\braket{\braket{b_\alpha;B}}_{\omega}=&\langle{\langle{\lambda_\alpha\sqrt{\frac{\omega_\alpha}{2}}\bm{d}\cdot\bm{\epsilon}_\alpha
+\omega_\alpha b_\alpha;B}}\rangle\rangle_{\omega}\nonumber\\&+\braket{0|[b_\alpha,B]|0}
\end{align}
for any operator $B$.
We obtain from the residues of Eq. \eqref{photon_eq_ref} the relation
\begin{equation}
    (\omega_n-\omega_\alpha)\braket{0|b_\alpha|n}=\lambda_\alpha\sqrt{\frac{\omega_\alpha}{2}}\braket{0|\bm{d}\cdot\bm{\epsilon}_\alpha|n}
\end{equation}
which allows the computation of photonic transition moments in terms of matter transition moments. 
This reflects the entanglement between electronic and photonic degrees of freedom. 
The analogous relation for the creation operators is
\begin{equation}
    (\omega_n+\omega_\alpha)\braket{0|b^\dagger_\alpha|n}=-\lambda_\alpha\sqrt{\frac{\omega_\alpha}{2}}\braket{0|\bm{d}\cdot\bm{\epsilon}_\alpha|n}.
\end{equation}
Moreover, from the relation
\begin{equation}
    [b_\alpha+b_\alpha^\dagger,H^l]=\omega_\alpha(b_\alpha-b^\dagger_\alpha)
\end{equation}
we obtain
\begin{align}
    \omega_n\braket{0|b_\alpha+b^\dagger_\alpha|n}&=\omega_\alpha\braket{0|b_\alpha-b_\alpha^\dagger|n}\label{eq:photo_momenta_conjugate}\\
    &=\lambda_\alpha\sqrt{\frac{\omega_\alpha}{2}}\braket{0|\bm{d}\cdot\bm{\epsilon}_\alpha|n}\frac{2\omega_n\omega_\alpha}{\omega_n^2-\omega_\alpha^2}\label{eq:photo_momenta_conjugate_dipole}
\end{align}
If we introduce the conjugate field displacement $q_\alpha$ and momentum $p_\alpha$ coordinates
\begin{align}
    p_\alpha &= i \sqrt{\frac{\omega_\alpha}{2}}(b_\alpha^\dagger-b_\alpha)\label{eq:momentum_photon}\\
    q_\alpha &=  \frac{1}{\sqrt{2\omega_\alpha}}(b_\alpha^\dagger+b_\alpha)\label{eq:position_photon},
\end{align}
we can write Eq. \eqref{eq:photo_momenta_conjugate} as
\begin{equation}\label{dipole-velocity-photon-form}
        \omega_n \braket{0|q_\alpha|n}=i\braket{0|{p}_\alpha|n}
\end{equation}
which is the equivalent relation of Eq.~\eqref{dipole-velocity-form} for the photonic conjugate momenta. 
Similarly, by employing the Hamiltonian in the velocity form in Eq.~\eqref{PF_velocity} or through the transformation  in Eq.~\eqref{dipole_transformation}, we obtain equivalent relations in the velocity representation.
Notice that relation Eq. \eqref{eq:photo_momenta_conjugate} also holds for the dipole Hamiltonian in the velocity form, but the physical meaning of the operators \eqref{eq:momentum_photon} and \eqref{eq:position_photon} is changed since they refer to a different representation. 
This will be discussed extensively in the following sections. \\

Analogous relations can be obtained from the higher-order equations of motion in Eqs.~\eqref{EOM_2_or_resp}, for instance for the transition dipole moments among excited states we have
\begin{align}\label{eq:transition_dipole_moments_excited_states}
    -i(\omega_m-\omega_n)\big(&\braket{n|r_i|m}-\delta_{mn}\braket{0|r_i|0}\big)=\nonumber\\
    &\braket{n|p_i|m}-\delta_{mn}\braket{0|p_i|0}.
\end{align}
Notice that relations similar to Eq. \eqref{eq:transition_dipole_moments_excited_states} also hold for the photonic conjugate momenta:
\begin{align}
    -i(\omega_m-\omega_n)\big(&\braket{n|q_\alpha|m}-\delta_{mn}\braket{0|q_\alpha|0}\big)=\nonumber\\
    &\braket{n|p_\alpha|m}-\delta_{mn}\braket{0|p_\alpha|0}.
\end{align}

\subsection{Response functions in cavity QED}\label{sec:external_perturbation}

This section provides a discussion of response functions in cavity QED.
Although far from exhaustive, we provide the reader with several examples of matter-photon and photon-photon response functions.
We focus on the peculiarities introduced by the photon dressing, and we explicitly discuss the result of using different mathematical representations of the Hamiltonian.\\

\noindent The perturbation is described by a single frequency component of Eq. \eqref{pert}
\begin{equation}
    V^t=\bigg(V^{\omega}\;e^{-i\omega t}+V^{-\omega}e^{i\omega t}\bigg)\;e^{\eta t},
\end{equation}
where this also includes the case of static perturbations by imposing $\omega=\eta =0$.

\subsubsection{External electromagnetic fields}

Spectroscopic techniques probe the system by means of an external electromagnetic field, whose electric and magnetic components are coupled to the motion of particles.
The external probe is treated as classical (non-quantized) fields, described by its own vector $\mathbf{A}_e(\bm{r},t)$ and scalar $\phi_e(\bm{r},t)$ potentials. 
It is not necessary to describe the internal and external fields through the same gauge.
The Hamiltonian of the system in the Coulomb gauge and Born-Oppenheimer approximation reads\cite{cohen1997photons}
\begin{align} \label{PFwfields}
    H=\sum_i \frac{1}{2}&\bigg(\bm{p}_i-\frac{1}{c}\mathbf{A}(\bm{r}_i)-\frac{1}{c}\mathbf{A}_e(\bm{r}_i)\bigg)^2+V_{coul}\nonumber\\
    & -\sum_i \phi_e(\bm{r}_i) + \sum_{\alpha}\omega_\alpha b^\dagger_\alpha b_\alpha,
\end{align}
where $\alpha$ labels the photon modes and $i$ refers to electrons. 
Expanding the first term, we get the polaritonic Hamiltonian in Eq.~\eqref{eq:Minimal_coupling} and the interaction term $V^t$ with the external fields:
\begin{align}\label{field_pert}
   V^t =& \frac{1}{c^2}\sum_i\mathbf{A}(\bm{r}_i)\cdot \mathbf{A}_e(\bm{r}_i)\nonumber\\
   &+\frac{1}{2c^2}\sum_i\mathbf{A}^2_e(\bm{r}_i)-\sum_i\frac{1}{2c}\big(\mathbf{A}_e(\bm{r}_i)\cdot\bm{p}_i+\bm{p}_i\cdot\mathbf{A}_e(\bm{r}_i)\big)\nonumber\\
   &-\sum_i\phi_e(\bm{r}_i).
\end{align}
While the last three terms of Eq.~\eqref{field_pert} are also found in molecular response theory in the semi-classical approximation,\cite{barron2009molecular, list2020beyond, list2015beyond, lestrange2015consequences} the first term is a purely QED contribution. 
This term cancels the field contribution from the $\bm{p}$ terms and ensures the coupling is only to the matter subsystem. 
When we perform the dipole approximation for the cavity fields and apply the transformation in Eq.~\eqref{dipole_transformation}, we obtain the length Hamiltonian $H^l$. 
The first term in Eq. \eqref{field_pert} is cancelled, leaving only the familiar interaction terms. 
By a suitable expansion of the external potential, the interaction term is obtained in a multipolar fashion as for the standard semiclassical theory\cite{barron2009molecular, list2015beyond, bernadotte2012origin, list2020beyond}
\begin{align}\label{mult}
    H_{mult}=&-\bm{d}\cdot\mathbf{E}_e-\bm{m}\cdot\mathbf{B}_e\nonumber\\
    &-\frac{1}{6}\Theta_{\alpha\beta}\partial_\alpha (E_e)_\beta-\frac{1}{2}(B_e)_\alpha(B_e)_\beta\chi_{\alpha\beta}+\dots
\end{align}
where we defined respectively the electric dipole $\bm{d}$, magnetic dipole $\bm{m}$, electric quadrupole $\Theta_{\alpha\beta}$ and magnetic susceptibility $\chi_{\alpha\beta}$ operators (in atomic units):
\begin{align}
    \bm{d}&=\sum_M Z_M \bm{R}_M - \sum_i \bm{r}_i\\
    \Theta_{\alpha\beta}&=\sum_i(3 r_{i\alpha}  r_{i\beta}-\delta_{\alpha\beta} r_i^2)\\
    \bm{m}&=\sum_i\frac{1}{2c}\bm{l}_i=\sum_i\frac{1}{2c}\bm{r}_i\times\bm{p}_i\\
    \chi_{\alpha\beta}&=\sum_i\frac{1}{4c^2}( r_{i\alpha} r_{i\beta}-\delta_{\alpha\beta} r_i^2).
\end{align}
Note that since we employ the dipole approximation in the length form, these multipolar operators refer exclusively to the matter subsystem and have the same physical meaning as in standard response theory. On the contrary, in the velocity form the operators
\begin{equation}
    \bm{m}=\sum_i\frac{1}{2c}\bm{l}_i=\sum_i\frac{1}{2c}\bm{r}_i\times\bm{p}_i
\end{equation}
has a mixed matter-photon character because of the field component in the conjugate momentum $\bm{p}$ (see Eq. \eqref{velocity_momentum}).\\
 
If we only retain the dipole interaction term in \eqref{mult}, the external electric field $\mathbf{E}_e(0)$ computed at the molecular position is a constant that can be factorized out of the response functions. 
The dipole-dipole response function $\braket{\braket{d_i\;;d_j}}_\omega$ can then be interpreted as the negative molecular polarizability at frequency $\omega$
\begin{align}\label{FDP}
    \braket{\braket{d_i\;;d_j}}_\omega=\sum_{k>0}\bigg(\frac{\braket{0|d_i|k}\braket{k|d_j|0}}{\omega-\omega_k}-\frac{\braket{0|d_j|k}\braket{k|d_i|0}}{\omega+\omega_k}\bigg)
\end{align}
where $i$ and $j$ refer to cartesian components of the dipole operator $\bm{d}$ and $k$ labels excited polaritonic states. 
This function describes the first-order time evolution of the molecular dipole moment subject to an external electric field: 
\begin{equation}
    \delta\braket{\bm{d}}^{(1)}(t)=-\int d\omega\;e^{-i\omega t+ \eta t}\sum_{j=x,y,z}\braket{\braket{\bm{d}\,;d_j}}_{\omega+i\eta}E_j(\omega).
\end{equation}
\begin{equation}\label{uniform_efield}
    V^t=-\bm{d}\cdot\mathbf{E}(t)=-\bm{d}\cdot\bigg(e^{\eta  t}\int d\omega \;e^{-i\omega t}\mathbf{E}(\omega)\bigg)
\end{equation}
The poles occur when the frequency $\omega$ matches the energy difference between the ground and excited states, and the corresponding residues
\begin{align}
    \lim_{\omega\to\omega_k}(\omega-\omega_k)\braket{\braket{d_i\;;d_j}}_\omega&=\braket{0|d_i|k}\braket{k|d_j|0}\\
    \lim_{\omega\to-\omega_k}(\omega+\omega_k)\braket{\braket{d_i\;;d_j}}_\omega&=-\braket{0|d_j|k}\braket{k|d_i|0}
\end{align}
provide information on the transition dipole moments between the ground and the excited states. 
While these expressions are the same as in standard molecular response theory,\cite{olsen1985linear} the states involved are here polaritonic and the predicted properties will consequently be modified.\\

\noindent If we replace one of the electric dipole operators with the magnetic dipole $\bm{m}$ we obtain the electric dipole-magnetic dipole response function
\begin{align}\label{FDPM}
    \braket{\braket{d_i\;;m_j}}_\omega=\sum_{k>0}\bigg(\frac{\braket{0|d_i|k}\braket{k|m_j|0}}{\omega-\omega_k}-\frac{\braket{0|m_j|k}\braket{k|d_i|0}}{\omega+\omega_k}\bigg),
\end{align}
which describes the electric dipole response when a magnetic field is applied 
\begin{equation}
    \delta\braket{\bm{d}}^{(1)}(t)=-\int d\omega\;e^{-i\omega t+ \eta t}\sum_{j=x,y,z}\braket{\braket{\bm{d}\,;m_j}}_{\omega+i\eta}B_j(\omega)
\end{equation}
\begin{equation}
    V^t=-\bm{m}\cdot\mathbf{B}(t)=-\bm{m}\cdot\bigg(e^{\eta  t}\int d\omega \;e^{-i\omega t}\mathbf{B}(\omega)\bigg).
\end{equation}
From the residues of the frequency-dependent electric dipole-magnetic dipole polarizability in Eq.~\eqref{FDPM}, we obtain the rotational strength of optically active molecules
\begin{align}
    \lim_{\omega\to\omega_k}(\omega-\omega_k)\braket{\braket{d_i\;;m_j}}_\omega&=\braket{0|d_i|k}\braket{k|m_j|0}\\
    \lim_{\omega\to-\omega_k}(\omega+\omega_k)\braket{\braket{d_i\;;m_j}}_\omega&=-\braket{0|m_j|k}\braket{k|d_i|0}.
\end{align}
As previously discussed, the operator $\bm{m}$ can only  in the length representation be interpreted as the molecular magnetic dipole operator, while it has a different physical meaning in the velocity Hamiltonian. 
Notice that in polaritonic systems, the optical activity can appear both from molecular and field chirality. 
As discussed in Sec. \eqref{sec:cavities}, several groups have recently developed chiral cavities,\cite{gautier2022planar, voronin2022single, feis2020helicity, beutel2021enhancing, plum2015chiral, scott2020enhanced, yoo2015chiral, liu2020switchable, sofikitis2014evanescent, hodgkinson2000vacuum, graf2019achiral,  hentschel2017chiral, zheng2021discrete, wang2023excitation, govorov2012theory, lan2016self} which can host only one-handedness of light polarization within themselves. 
This opens up the possibility of engineering chiral properties through the chirality of the photon field,\cite{hubener2021engineering, baranov2020circular, guo2021optical, li2023strong, sun2022polariton, gautier2022planar} which is transferred to molecular states via light-matter strong coupling. 
For instance, we expect it to induce chirality in a non-optically active molecule, similar to induced circular dichroism (ICD) when achiral molecules are placed in a chiral solvent.\cite{allenmark2003induced, saeva1971induced, gawronski2003significance, craig1976dynamic}
On the other hand, chiral molecules interacting with a chiral field should give rise to "polaritonic diastereoisomers", possibly leading to novel approaches to chiral discrimination. \\

As the states of the system here belong to the radiation-matter Hilbert space in Eq.~\eqref{radiation-matter-hilbert-space}, we can study how the perturbation of the matter subsystem leads to modifications of the photonic properties. 
If we focus on the perturbation in Eq.~\eqref{uniform_efield}, in the velocity representation, the response function
\begin{align}\label{photon_number_resp}
     \braket{\braket{b^\dagger_\alpha b_\alpha\;; d_j}}_\omega=\sum_{k>0}\bigg(\frac{\braket{0|b^\dagger_\alpha b_\alpha|k}\braket{k|d_j|0}}{\omega-\omega_k}-\frac{\braket{0|d_j|k}\braket{k|b^\dagger_\alpha b_\alpha|0}}{\omega+\omega_k}\bigg)
\end{align}
describes the time evolution of the photon number in the $\alpha$-mode
\begin{equation}
    \delta\braket{b^\dagger_\alpha b_\alpha}^{(1)}=-\int d\omega\;e^{-i\omega t+ \eta t}\sum_{j}\braket{\braket{b^\dagger_\alpha b_\alpha\,;d_j}}_{\omega+i\eta}E_j(\omega).
\end{equation}
The residues computed at the polaritonic excitations
\begin{align}
    \lim_{\omega\to\omega_k}(\omega-\omega_k)\braket{\braket{b^\dagger_\alpha b_\alpha\;;d_j}}_\omega&=\braket{0|b^\dagger_\alpha b_\alpha|k}\braket{k|d_j|0}\\
    \lim_{\omega\to-\omega_k}(\omega+\omega_k)\braket{\braket{b^\dagger_\alpha b_\alpha\;;d_j}}_\omega&=-\braket{0|d_j|k}\braket{k|b^\dagger_\alpha b_\alpha|0}
\end{align}
are now connected to dipole and photon number transition moments between the ground and the excited state. 
Note that to correctly model the coupling to an external field in the velocity representation, we should also include the first interaction term in Eq.~\eqref{field_pert}. 
In the length form, the electric field mode creation (annihilation) operators are described by Eq.~\eqref{length_form_creation}.
The response function equivalent to Eq.~\eqref{photon_number_resp} is then given by
    \begin{align}
    &\langle\langle\big(-ib^\dagger_\alpha -i\sqrt{\frac{\omega_\alpha}{2}}\bm{\lambda}_\alpha\cdot\bm{d}\big)\big(ib_\alpha+i\sqrt{\frac{\omega_\alpha}{2}}\bm{\lambda}_\alpha\cdot\bm{d}\big); d_j\rangle\rangle_\omega.
\end{align}
The response function Eq.~\eqref{photon_number_resp} represents the variation of the number of \textit{displacement field} modes, in agreement with the transformation in Eq.~\eqref{dipole_transformation}.\\

\noindent In an analogous way, we can consider the time evolution of the field displacement coordinates Eqs. \eqref{eq:momentum_photon} and \eqref{eq:position_photon}. Their time development is obtained by means of the response functions
\begin{align}
    \braket{\braket{q_\alpha\;; d_j}}_\omega=\sum_{k>0}\bigg(\frac{\braket{0|q_\alpha|k}\braket{k|d_j|0}}{\omega-\omega_k}-\frac{\braket{0|d_j|k}\braket{k|q_\alpha|0}}{\omega+\omega_k}\bigg)\label{photon_coordinate_resp}\\
    \braket{\braket{p_\alpha\;; d_j}}_\omega=\sum_{k>0}\bigg(\frac{\braket{0|p_\alpha|k}\braket{k|d_j|0}}{\omega-\omega_k}-\frac{\braket{0|d_j|k}\braket{k|p_\alpha|0}}{\omega+\omega_k}\bigg)\label{photon_momentum_resp},
\end{align}
with residues
\begin{align}
    \lim_{\omega\to\omega_k}(\omega-\omega_k)\braket{\braket{q_\alpha\;;d_j}}_\omega&=\braket{0|q_\alpha|k}\braket{k|d_j|0}\\
    \lim_{\omega\to-\omega_k}(\omega+\omega_k)\braket{\braket{q_\alpha\;;d_j}}_\omega&=-\braket{0|d_j|k}\braket{k|q_\alpha|0}\\
    \lim_{\omega\to\omega_k}(\omega-\omega_k)\braket{\braket{p_\alpha\;;d_j}}_\omega&=\braket{0|p_\alpha|k}\braket{k|d_j|0}\\
    \lim_{\omega\to-\omega_k}(\omega+\omega_k)\braket{\braket{p_\alpha\;;d_j}}_\omega&=-\braket{0|d_j|k}\braket{k|p_\alpha|0}
\end{align}
that provide information on the transition dipole and photon moments.\\

\subsubsection{Photonic environment perturbation}
Classical charge currents $\bm{J}_e(\bm{r},t)$ are a source of electromagnetic fields, which means that they can directly perturb the cavity photon field. The interaction term for the Pauli-Fierz Hamiltonian in Eq.~\eqref{eq:Minimal_coupling} coupled to a classical external current reads\cite{cohen1997photons}
\begin{equation}
    H_{\bm{J}_e}=-\frac{1}{c}\int\; d^3r\,\bm{J}_e(\bm{r},t)\cdot\mathbf{A}(\bm{r})\label{H-J-velocity}.
\end{equation}
Following Refs. \cite{tokatly2013time, flick2019light, ruggenthaler2014quantum}, resolving the external currents into modes which act directly on the $\alpha$ mode of the cavity, in the dipole approximation and the length form, the interaction Hamiltonian can be written as
\begin{equation}\label{H-J}
    H_{\bm{J}_e}=\sum_\alpha\frac{j_\alpha(t)}{2\omega_\alpha}\; p_\alpha=\sum_\alpha p_\alpha\;\frac{1}{2\omega_\alpha}\int d\omega\;\tilde{j}_\alpha(\omega)e^{-i\omega t}
\end{equation}
where $p_\alpha$ is the photon conjugate momentum of mode $\alpha$, and ${j}_\alpha$ is connected to the $\alpha$-mode of the time-derivative of the external current $\bm{J}_e$.\cite{tokatly2013time, flick2019light, ruggenthaler2014quantum}
The interaction Hamiltonian Eq.~\eqref{H-J} generates mode excitations into the system. 
Due to the coupling of light and matter, this perturbation provides an indirect way to probe the matter and photon-matter correlation properties. 

\noindent In the length form, the photonic response function
\begin{align}\label{photon_number_resp_ph}
   & \braket{\braket{b^\dagger_\alpha  b_\alpha\;; p_\alpha}}_\omega=\nonumber\\
   &\sum_{k>0}\bigg(\frac{\braket{0|b^\dagger_\alpha b_\alpha|k}\braket{k|p_\alpha|0}}{\omega-\omega_k}-\frac{\braket{0|p_\alpha|k}\braket{k|b^\dagger_\alpha b_\alpha|0}}{\omega+\omega_k}\bigg)
\end{align}
describes the time evolution of the displacement field photon number. Interestingly, the more general response function
\begin{align}\label{photon_number_resp_ph_gen}
     &\braket{\braket{b^\dagger_\beta b_\beta\;;  p_\alpha}}_\omega=\\
     &\sum_{k>0}\bigg(\frac{\braket{0|b^\dagger_\beta b_\beta|k}\braket{k|p_\alpha|0}}{\omega-\omega_k}-\frac{\braket{0|p_\alpha|k}\braket{k|b^\dagger_\beta b_\beta|0}}{\omega+\omega_k}\bigg),
\end{align}
which describes the time evolution of the $\beta$-photon number when the system is coupled to the current $j_\alpha(t)$
\begin{align}\label{ph-ph-r}
    &\braket{b^\dagger_\beta b_\beta}^{(1)}(t)=\int d\omega\;e^{-i\omega t+\eta t} \braket{\braket{b^\dagger_\beta b_\beta, p_\alpha}}_{\omega+i\eta}\frac{\tilde{j}_\alpha(\omega)}{2\omega_\alpha}
\end{align}
is different from zero even when $\alpha\neq\beta$. This happens because of the coupling to matter degrees of freedom, which provide an indirect interaction between different photon modes. This result opens up the possibility of photon transfer between modes with different frequencies or polarization. In the limit of zero coupling, the response function Eq.~\eqref{ph-ph-r} would differ from zero only when $\alpha=\beta$, as photons do not directly interact. Analogously, the response function
\begin{align}\label{photon_coordinate_resp_pg}
    \braket{\braket{p_\alpha\;; p_\beta}}_\omega=\sum_{k>0}\bigg(\frac{\braket{0|p_\alpha|k}\braket{k|p_\beta|0}}{\omega-\omega_k}-\frac{\braket{0|p_\beta|k}\braket{k|p_\alpha|0}}{\omega+\omega_k}\bigg),
\end{align}
describes the time evolution of the $\beta$-cavity mode momentum when perturbing the $\alpha$-cavity mode
\begin{equation}
    \braket{p_\alpha}^{(1)}(t)=\int d\omega\;e^{-i\omega t+\eta t} \braket{\braket{p_\alpha, p_\beta}}_{\omega+i\eta}\frac{\tilde{j}_\beta(\omega)}{2\omega_\beta}.
\end{equation}
As discussed in the other sections, the transformation in Eq.~\eqref{dipole_transformation} links the response functions for the physical observables in the velocity and length form.
For the creation and annihilation operators of the electric photon states, we then need to employ the transformation in Eq.~\eqref{length_form_creation}.\\

\noindent We can also describe the time evolution of matter degrees of freedom when the system is perturbed through Eq.~\eqref{H-J}. The response function
\begin{align}\label{dipole_resp_current}
    \braket{\braket{d_j\;; p_\alpha}}_\omega=\sum_{k>0}\bigg(\frac{\braket{0|d_j|k}\braket{k|p_\alpha|0}}{\omega-\omega_k}-\frac{\braket{0|p_\alpha|k}\braket{k|d_j|0}}{\omega+\omega_k}\bigg)
\end{align}
is analogous to Eq.~\eqref{photon_momentum_resp} as it describes the time evolution of the dipole moment when the photon degrees of freedom are perturbed by ${j}_\alpha(t)$
\begin{equation}
    \braket{\bm{d}}^{(1)}(t)=\int d\omega\;e^{-i\omega t+\eta t} \braket{\braket{\bm{d}, p_\alpha}}_{\omega+i\eta}\frac{\tilde{j}_\alpha(\omega)}{2\omega_\alpha}.
\end{equation}
The response functions in Eqs. ~\eqref{photon_number_resp}, ~\eqref{photon_coordinate_resp}, and ~\eqref{dipole_resp_current} are examples of response functions which describe the cross-talk between photon and matter degrees of freedom. 
In fact, in the limit of $\lambda=0$, the eigenstates of the system would be given by the direct product between bare molecular states and photons states. 
As a consequence, Eqs.~\eqref{photon_number_resp},~\eqref{photon_coordinate_resp}, ~\eqref{photon_momentum_resp}, and~\eqref{dipole_resp_current} would be identically zero.
This means that a perturbation acting on the molecular Hilbert space cannot induce any time evolution in the (decoupled) cavity-radiation space, and vice versa.\\

\subsubsection{Static perturbations and energy derivatives}
This section provides several examples of static perturbations on the polaritonic system and their connection to the energy derivatives.\\

\paragraph{External electric and magnetic fields.}
A static external electric field is described by the scalar potential $\phi_e$ that can be expanded in a Taylor series around the origin
\begin{equation}
    \phi_e(\bm{r})=\phi(0)-r_i E_i^{(0)}-\frac{1}{2}r_i (\partial_i E_j){(0)}r_j+\dots
\end{equation}
where we employed Einstein's summation convention. 
If we retain only the lowest expansion term, the interaction operator in the Hamiltonian is the familiar electric dipole term
\begin{equation}\label{external_static_efield}
    H_{int}=-\bm{d}\cdot \mathbf{E}_e(0).
\end{equation}
Note that the transformation in Eq.~\eqref{dipole_transformation} commutes with this operator, which implies it is the same in both the length and the velocity representation. The static electric dipole polarizability $\bm{\alpha}_0$ then reads
\begin{equation}
    -\bm{\alpha}_0=\frac{d^2{E}}{d^2\mathbf{E}_e}\bigg|_{\mathbf{E}_e=0}=2\sum_{k>0} \frac{\bra{0} \bm{d} \ket{k} \bra{k} \bm{d} \ket{0}}{E_0 - E_{k}} =\braket{\braket{\bm{d};\bm{d}}}_{\omega=0}.
\end{equation}

\noindent An analogous interaction term is introduced when the system is immersed in a static uniform external magnetic field, which can be described by the following vector potential\cite{barron2009molecular}
\begin{equation*}
\mathbf{A}_e=\frac{1}{2} \mathbf{B}_e\times\bm{r}.
\end{equation*}
In the length form, the interaction Hamiltonian reads
\begin{equation}
    H_{int}=-\bm{m}\cdot\mathbf{B}_e-\frac{1}{2}\mathbf{B}_e^T\bm{\chi}\mathbf{B}_e
\end{equation}
where
\begin{align}
    \bm{m}&=\sum_i\frac{1}{2c}\bm{l}_i=\sum_i\frac{1}{2c}\bm{r}_i\times\bm{p}_i\\
    \chi_{\alpha\beta}&=\sum_i\frac{1}{4c^2}( r_{i\alpha} r_{i\beta}-\delta_{\alpha\beta} r_i^2)
\end{align}
and since we are employing the length form, these quantities refer only to the matter subsystem. 
The static molecular magnetizability $\bm{\xi}_0$ is computed as the second derivative of the energy with respect to the magnetic field
\begin{equation}\label{eq:diamagn}
    \bm{\xi}_0=-\frac{d^2{E}}{d^2\mathbf{B}_e}\bigg|_{\mathbf{B}_e=0}=-\braket{\braket{\bm{m};\bm{m}}}_{\omega=0}-\braket{0|\bm{\chi}|0}.
\end{equation}
The first term in Eq. \eqref{eq:diamagn} is called paramagnetic contribution, while the second term is called diamagnetic contribution and usually dominates for closed-shell molecules.
Magnetizabilities have been extensively studied in the framework of molecular response theory,\cite{bak1992first, ruud1993hartree, ruud1994theoretical, ruud1998hartree, aastrand1996magnetizabilities, gauss2007gauge, lutnaes2009benchmarking} but the effect of photon-dressing, to the best of our knowledge, is yet to be explored.
The evaluation of magnetic properties for approximate wave functions is affected by an origin dependence on the choice of the gauge origin of the external field. 
In molecular response theory, the origin independence can be recovered by using gauge invariant atomic orbitals (GIAO).\cite{helgaker1999, london1937}

\paragraph{Spin interactions.} \label{nucspinmagsec} 
Nuclear magnetic resonance (NMR) and electronic paramagnetic resonance (EPR) provide important information on the molecular structure and find wide applications in chemistry. The electron spin is connected to magnetic moment of the electron
\begin{gather}
    \bm{m}^{s}_{i} = - g_e \mu_{B} \bm{s}_{i},
\end{gather}
where $g_e$ is the electronic $g$ factor, $\mu_{B}$ is the Bohr magneton and $\bm{s}_{i}$ is the spin operator of electron $i$. 
In the same way, the nuclear spin is connected to the nuclear magnetic moment
\begin{gather}
    \bm{M}^{s}_{N} = \gamma_{N} \bm{I}_{N}.
\end{gather}
where $\gamma_{N}$ is the magnetogyric ratio and $\bm{I}_N$ is the spin operator of nucleus $N$. 
The spin degrees of freedom modify the Pauli-Fierz Hamiltonian in Eq.~\eqref{PF_velocity}, introducing spin-spin, spin-orbit and spin-external field interaction terms. 
The cavity environment introduces a twofold modification on NMR and EPR spectra: it modifies the ground state wave function, which is now dressed by the photons, and introduces further terms in the Hamiltonian due to the interaction of the spins with the cavity field. 
The effects of these additional QED terms are still to be explored and require going beyond the dipole approximation.\\ 

We now focus on the NMR properties of closed-shell molecules and disregard the electronic spin contribution to the magnetic field.
The total vector potential $\mathbf{A}^{tot}(\bm{r}_i)$ has contributions from the cavity field, the external field, and the nuclear spins
\begin{gather} \label{Atot}
\begin{split}
    \mathbf{A}^{tot}(\bm{r}) &= 
    \mathbf{A}(\bm{r})+ \frac{{\mathbf{B}_{e}} \times \bm{r}}{2} + \frac{1}{c^2} \sum_{N}\left(\frac{\bm{M}_N^{s} \times (\bm{r}-\bm{R}_N)}{{|\bm{r}-\bm{R}_N}|^{3}}\right),
\end{split}
\end{gather}
and the total magnetic induction is
\begin{gather} \label{nmrBA}
    \mathbf{B}^{tot} (\bm{r}) = \nabla \times \mathbf{A}^{tot} (\bm{r}) .
\end{gather}
Therefore, the following interaction terms need to be included in the Hamiltonian in Eq.~\eqref{PF_velocity}:
\begin{gather} \label{nmrH}
    \begin{split}
        H_{int} =  &-  \sum_{i}\bm{m}^{s}_{i} \cdot \mathbf{B}^{tot}(\bm{r}_{i}) - \sum_{N} \bm{M}^{s}_{N} \cdot \mathbf{B}^{tot}(\bm{R}_{N}) \\
        & + \sum_{M > N} \frac{1}{2 c^{2}} \frac{{R}_{MN}^2 \left({\bm{M}^{s}_{M}} \cdot {\bm{M}^{s}_{N}} \right) - 3 \left( {\bm{M}^{s}_{M}} \cdot \bm{R}_{MN} \right) \left( \bm{R}_{MN} \cdot {\bm{M}^{s}_{N}} \right) }{{R}_{MN}^5},
    \end{split}
\end{gather}
where the first term mediates the indirect coupling of nuclear spins, the second term is the nuclear Zeeman interaction, and the last term in Eq.~\eqref{nmrH} is the direct couplings between nuclear dipole magnetic moments. 
Moreover, using Eq.~\eqref{Atot} in the velocity Hamiltonian leads to additional interaction terms. The NMR properties are then connected to the following energy derivatives\cite{helgaker1999}
\begin{gather} \label{magnet}
    \mathbf{E}^{(1,1)}_{N} = \frac{d^2 E(\mathbf{B}_{e},\bm{M}^{s})}{d\mathbf{B}_{e}d\bm{M}^{s}_{N}}\bigg|_{\mathbf{B}_{e}=0,\bm{M}^{s}=0} \\
    \mathbf{E}^{(0,2)}_{M,N} = \frac{d^2 E(\mathbf{B}_{e},\bm{M}^{s})}{d\bm{M}^{s}_{M} d\bm{M}^{s}_{N}} \bigg|_{\mathbf{B}_{e}=0,\bm{M}^{s}=0},
\end{gather}
where $\mathbf{E}^{(1,1)}_{N}$ is related to the nuclear shielding tensor $\bm{\sigma}_{N}$
\begin{gather}\label{shielding}
    \bm{\sigma}_{N} = \bm{1} + \mathbf{E}^{(1,1)}_{N},
\end{gather}
which describes the electron shielding effects on nucleus $N$.
At the same time, $\mathbf{E}^{(0,2)}_{M,N}$ is connected to the direct $\bm{D}_{M,N}$ and indirect (electron-mediated) $\bm{K}_{M,N}$ spin-spin coupling tensors
\begin{gather} \label{jj}
    \mathbf{E}^{(0,2)}_{M,N}=   \bm{K}_{M,N}+ \bm{D}_{M,N}.
\end{gather}
As for magnetizabilities, the computation of NMR properties is affected by origin dependencies, which are usually eliminated using GIAO orbitals. 
The modifications of the ground state density induced by the QED environment change how the nuclei are shielded from the electrons, affecting the NMR properties. Moreover, the novel interaction terms between spins and the cavity magnetic field can possibly affect spin-spin coupling, resulting in modifications to the shape of NMR signals.

\paragraph{QED environment perturbations.}
A modification of the QED environment is reflected in the coupling strength and the frequency of the cavity modes.
If we consider a single-mode and perturb the coupling strength
\begin{equation}
    \bm{\lambda} \to \bm{\lambda}+\Delta\bm{\lambda},
\end{equation}
the Hamiltonian reads
\begin{align}
    H'=&\sum_{pq}h_{pq}E_{pq}+\frac{1}{2}\sum_{pqrs}g_{pqrs}e_{pqrs}+h_{nuc}\nonumber\\
    &+\sqrt{\frac{\omega_\alpha}{2}}(\bm{\lambda}\cdot \bm{d})(b^\dagger+b)+\frac{1}{2}(\bm{\lambda}\cdot \bm{d})^2\nonumber\\
    &+\omega_\alpha\left\{{b}^\dagger{b}+\frac{1}{2}\right\}\nonumber\\
    &+\sqrt{\frac{\omega_\alpha}{2}}(\Delta\bm{\lambda}\cdot \bm{d})(b^\dagger+b)+(\bm{\lambda}\cdot \bm{d})(\Delta\bm{\lambda}\cdot \bm{d})\nonumber\\
    &+\frac{1}{2}(\Delta\bm{\lambda}\cdot \bm{d})^2\\
    =&H^l + V^{(1)} + V^{(2)},
\end{align}
where $H^l$ is the unperturbed dipole Hamiltonian in the length representation Eq. \eqref{dipole}, and we have defined the first and second-order perturbators as
\begin{align}
    V^{(1)} &=\sqrt{\frac{\omega_\alpha}{2}}(\Delta\bm{\lambda}\cdot \bm{d})(b^\dagger+b)+(\bm{\lambda}\cdot \bm{d})(\Delta\bm{\lambda}\cdot \bm{d})\\
    V^{(2)} &=\frac{1}{2}(\Delta\bm{\lambda}\cdot \bm{d})^2.
\end{align}
We therefore obtain expressions for the first and second energy derivatives
\begin{align}
    \frac{dE}{d\Delta\bm{\lambda}}\bigg|_{\bm{\Delta\lambda}=0}&=\braket{0|\bm{d}\bigg(\sqrt{\frac{\omega_\alpha}{2}}(b^\dagger+b)+(\bm{\lambda}\cdot \bm{d})\bigg)|0}\\
    \frac{d^2E}{d\Delta\bm{\lambda}^2}\bigg|_{\bm{\Delta\lambda}=0}&=\braket{0|\bm{d}\bm{d}|0}+\frac{1}{2}\braket{\braket{\frac{dV^{(1)}}{d\Delta\lambda};\frac{dV^{(1)}}{d\Delta\lambda}}}_{\omega=0}\label{eq:polariton_qed_pol}.
\end{align}
These derivatives measure how sensitive the system is to a change in the coupling strength.
In particular, Eq. \eqref{eq:polariton_qed_pol} can be interpreted as the static ground state polarizability with respect to the cavity field fluctuations.
We note that because of the dipole self-energy in $H^l$, at very large coupling strength, these expressions can be approximated as
\begin{align}
    \frac{dE}{d\Delta\bm{\lambda}}\bigg|_{\bm{\Delta\lambda}=0}&\approx\bm{\lambda}\cdot\braket{0|\bm{d} \bm{d}|0}\label{fir_der_en_lambda}\\
    \frac{d^2E}{d\Delta\bm{\lambda}^2}\bigg|_{\bm{\Delta\lambda}=0}&\approx\frac{1}{2}\braket{\braket{(\bm{\lambda}\cdot \bm{d}) \bm{d};(\bm{\lambda}\cdot \bm{d}) \bm{d}}}_{\omega=0}\\
    &=-\sum_{k>0}\frac{|\bm{\lambda}\cdot\bra{0} \bm{d} \bm{d} \ket{k}|^2 }{E_k - E_{0}} \label{second_der_lambda}
\end{align}
As we increase $\bm{\lambda}$, from Eq. \eqref{fir_der_en_lambda} the energy increases since $\braket{0|\bm{d} \bm{d}|0}$ is positive definite, but the Hessian in Eq. \eqref{second_der_lambda} is negative.
Therefore, the system is endowed with a stable ground state.
Riso et al. showed that for infinitely large coupling strength, the polaritonic ground state energy will converge to the SC-QED-HF energy.\cite{riso2022molecular}
On the other hand, if the dipole self-energy is disregarded, the remaining terms are
\begin{align}
    \frac{dE}{d\Delta\bm{\lambda}}\bigg|_{\bm{\Delta\lambda}=0}&\approx\braket{0|\bm{d}\sqrt{\frac{\omega_\alpha}{2}}(b^\dagger+b)|0}\label{eq:firsder_no_dse}\\
    \frac{d^2E}{d\Delta\bm{\lambda}^2}\bigg|_{\bm{\Delta\lambda}=0}&\approx\frac{\omega_\alpha}{4}\braket{\braket{\bm{d}(b^\dagger+b);\bm{d}(b^\dagger+b)}}_{\omega=0}\\
    &=-\frac{\omega_\alpha}{2}\sum_{k>1}\frac{|\bra{0} \bm{d}(b^\dagger+b) \ket{k}|^2 }{E_k - E_{0}} .
\end{align}
While the Hessian is still negative, there is no guarantee from Eq. \eqref{eq:firsder_no_dse} that the energy is increasing, and the ground state energy could decrease indefinitely.
The dipole self-energy has been shown to be fundamental for the stability of the ground state.\cite{schafer2020relevance}\\
\noindent A modification of the cavity environment can effectively change the photon frequency
\begin{equation}
    \omega_\alpha \to \omega_\alpha + \Delta\omega_\alpha.
\end{equation}
From the expansion
\begin{equation}
    \sqrt{\omega_\alpha+\Delta\omega_\alpha}\approx\sqrt{\omega_\alpha}+\frac{1}{2}\sqrt{\omega_\alpha}\frac{\Delta\omega_\alpha}{\omega_\alpha}-\frac{1}{8}\sqrt{\omega_\alpha}\left(\frac{\Delta\omega_\alpha}{\omega_\alpha}\right)^2+\dots
\end{equation}
for $\Delta\omega_\alpha\ll\omega_\alpha$, the perturbed Hamiltonian reads
\begin{align}
    H'&=H^l\nonumber\\
    &+\frac{1}{2}\frac{\Delta\omega_\alpha}{\omega_\alpha}\sqrt{\frac{\omega_\alpha}{2}}(\bm{\lambda}\cdot \bm{d})(b^\dagger+b)+\Delta\omega_\alpha \;b^\dagger b\nonumber\\
    &-\frac{1}{8}\left(\frac{\Delta\omega_\alpha}{\omega_\alpha}\right)^2\sqrt{\frac{\omega_\alpha}{2}}(\bm{\lambda}\cdot \bm{d})(b^\dagger+b)+\dots
\end{align}
The first and second-order perturbation operators are therefore
\begin{align}
    V^{(1)} &=\frac{1}{2}\frac{\Delta\omega_\alpha}{\omega_\alpha}\sqrt{\frac{\omega_\alpha}{2}}(\bm{\lambda}\cdot \bm{d})(b^\dagger+b)+\Delta\omega_\alpha \;b^\dagger b\\
    V^{(2)} &=-\frac{1}{8}\left(\frac{\Delta\omega_\alpha}{\omega_\alpha}\right)^2\sqrt{\frac{\omega_\alpha}{2}}(\bm{\lambda}\cdot \bm{d})(b^\dagger+b).
\end{align}
The energy first and second energy derivatives read
\begin{align}
    \frac{dE}{d\Delta\omega_\alpha}\bigg|_{\Delta\omega_\alpha=0}&=\braket{0|\frac{1}{2\omega_\alpha}\sqrt{\frac{\omega_\alpha}{2}}(\bm{\lambda}\cdot \bm{d})(b^\dagger+b)+\;b^\dagger b|0}\\
    \frac{d^2E}{d\Delta\omega_\alpha^2}\bigg|_{\Delta\omega_\alpha=0}&=-\frac{1}{4}\left(\frac{1}{\omega_\alpha}\right)^2\sqrt{\frac{\omega_\alpha}{2}}\braket{0|(\bm{\lambda}\cdot \bm{d})(b^\dagger+b)|0}\nonumber\\
    &+\frac{1}{2}\braket{\braket{\frac{dV^{(1)}}{d\Delta\omega_\alpha};\frac{dV^{(1)}}{d\Delta\omega_\alpha}}}_{\omega=0}.\label{eq:pol_freq_cavity}
\end{align}
These derivatives measure how sensitive the system is to a change in the photon frequency.\\

\noindent In the following section, we develop response theory for approximate wave functions, focusing on QED-HF and QED-CC response theory.

\section{\textit{Ab initio} approximations}\label{Approx_resp_chapt}
In this section, we derive the linear and quadratic response equations for the \textit{ab initio} QED-HF and QED-CC methods. 

\subsection{QED-HF}

\noindent The QED-HF is the generalization of the HF model to molecules in a QED environment.\cite{haugland2020coupled, haugland2021intermolecular} 
The system is described through the dipole Hamiltonian
\begin{align}
    H=&\sum_{pq}h_{pq}E_{pq}+\frac{1}{2}\sum_{pqrs}g_{pqrs}e_{pqrs}+h_{nuc}\nonumber\\
    &+\sum_\alpha\sqrt{\frac{\omega_\alpha}{2}}(\bm{\lambda}_\alpha\cdot \bm{d})_{pq}E_{pq}(b^\dagger_\alpha+b_\alpha)\\
    &+\frac{1}{2}\sum_{pqrs}\sum_\alpha(\bm{\lambda}_\alpha\cdot \bm{d})_{pq}(\bm{\lambda}_\alpha\cdot \bm{d})_{rs}E_{pq}E_{rs}\nonumber\\
    &+\sum_{\alpha}\omega_{\alpha}\left({b}^\dagger_{\alpha}{b}_{\alpha}+\frac{1}{2}\right)
\end{align}
where the dipole self-energy term can be included in the molecular Hamiltonian by a proper modification of the one and two-electron integrals
\begin{align}
    H_{mol}&=\sum_{pq}h_{pq}E_{pq}+\frac{1}{2}\sum_{pqrs}g_{pqrs}e_{pqrs}\\
    +&\frac{1}{2}\sum_{pqrs}\sum_\alpha(\bm{\lambda}_\alpha\cdot \bm{d})_{pq}(\bm{\lambda}_\alpha\cdot \bm{d})_{rs}E_{pq}E_{rs}\nonumber\\
    &=\sum_{pq}\bar h_{pq}E_{pq}+\frac{1}{2}\sum_{pqrs}\bar g_{pqrs}e_{pqrs} ,
\end{align}
where we have defined
\begin{align}
    \bar g_{pqrs}&=g_{pqrs}+\sum_\alpha(\bm{\lambda}_\alpha\cdot\bm{d})_{pq}(\bm{\lambda}_\alpha\cdot\bm{d})_{rs}\label{two_electron+dse}\\
    \bar h_{pq}&=h_{pq}+\frac{1}{2}\sum_{r}\sum_\alpha(\bm{\lambda}_\alpha\cdot\bm{d})_{pr}(\bm{\lambda}_\alpha\cdot\bm{d})_{rq}\label{one_electron_dse}.
\end{align}
The QED-HF wave function ansatz is a factorized state
\begin{equation}
    \ket{\mathrm{R}}=\ket{\mathrm {HF}}\otimes \sum_{\bm{n}}\prod_\alpha(b^\dagger_\alpha)^{n_\alpha}\ket{0}c_{\bm{n}},
\end{equation}
where $\ket{HF}$ is a single Slater determinant, $\ket{0}$ is the photon vacuum, and $c_{\bm{n}}$ are expansion coefficients for the photon number states.
The QED-HF state is obtained by minimization of the mean value of the Hamiltonian
\begin{equation}\label{hfen}
    E_{\text{QED-HF}}=\braket{\mathrm R|H|\mathrm R}
\end{equation}
with respect to electronic and photonic parameters. 
Haugland et al.\cite{haugland2020coupled, haugland2021intermolecular} showed that the photonic parameters for the ground-state QED-HF wave function define a coherent state through the following transformation
\begin{align}
    \ket{\mathrm R}&=\ket{\mathrm{HF}}\otimes U_{\text{QED-HF}}\ket{0}\equiv U_{\text{QED-HF}}\ket{HF,0}\\
    U_{\text{QED-HF}} &= \prod_\alpha \text{exp}\bigg(-\frac{\bm{\lambda}_\alpha\cdot \braket{\bm{d}}_{\text{QED-HF}}}{\sqrt{2\omega_\alpha}}(b^\dagger_\alpha - b_\alpha)\bigg)\label{coherent-transf}.
\end{align}
By transforming the dipole Hamiltonian using Eq.~\eqref{coherent-transf} we get
\begin{align}\label{QEDHF-transf-Ham}
    U^\dagger_{\text{QED-HF}}H\,U_{\text{QED-HF}}&=\sum_{pq}h_{pq}E_{pq}+\frac{1}{2}\sum_{pqrs}g_{pqrs}e_{pqrs}+h_{nuc}\nonumber\\
    &+\sum_\alpha\sqrt{\frac{\omega_\alpha}{2}}(\bm{\lambda}_\alpha\cdot (\bm{d}-\braket{\bm{d}}_{\text{QED-HF}}))(b^\dagger_\alpha+b_\alpha)\nonumber\\
    &+\frac{1}{2}\sum_\alpha(\bm{\lambda}_\alpha\cdot (\bm{d}-\braket{\bm{d}}_{\text{QED-HF}}))^2\nonumber\\
    &+\sum_{\alpha}\omega_{\alpha}{b}^\dagger_{\alpha}{b}_{\alpha}.
\end{align}
We note that the transformed Hamiltonian Eq. \eqref{QEDHF-transf-Ham} is now origin invariant also for charged systems.
Minimizing the photon-averaged coherent state transformed dipole Hamiltonian in Eq. \eqref{QEDHF-transf-Ham}, with respect to the orbital coefficients, we obtain the QED-Fock matrix
\begin{align}
    &F_{pq}=F_{pq}^e\nonumber\\
    &+\frac{1}{2}\sum_\alpha\bigg(\sum_a(\bm{\lambda}_\alpha\cdot\bm{d}_{pa})(\bm{\lambda}_\alpha\cdot\bm{d}_{aq})-\sum_i (\bm{\lambda}_\alpha\cdot\bm{d}_{pi})(\bm{\lambda}_\alpha\cdot\bm{d}_{iq})\bigg),
\end{align}
where $F^e_{pq}$ is the standard electronic Fock matrix and the indices $a$ and $i$ refer respectively to virtual and occupied orbitals.
The optimization condition that defines the QED-HF molecular orbitals is Brillouin's theorem, $F_{ia}=0$, as in standard HF theory. 
However, the eigenvalues of the Fock matrix $\bm{F}$, usually interpreted as orbital energies, are now origin dependent for charged systems. 
Nevertheless, the total QED-HF energy is origin invariant since the occupied-virtual blocks $F_{ia}$ are origin independent.

\subsubsection{Time-dependent QED-HF}

Once the optimized QED-HF reference state $\ket{\mathrm R}$ has been obtained, the time development due to an external perturbation $V^t$ is parametrized as
\begin{equation}\label{time-dep-qedhf}
    \text{exp}\big(-i\Lambda(t)\big){\ket{\mathrm R}}=\text{exp}\big(-i\Lambda(t)\big)\ket{\mathrm{HF}}\otimes U_{\text{QED-HF}}\ket{0} ,
\end{equation}
where $\Lambda(t)$ is the Hermitian operator
\begin{align}\label{TD-QED-HF}
    \Lambda(t) &= \frac{1}{\sqrt{2}}\sum_{ai}\big(\kappa_{ai}E_{ai}+\kappa^*_{ai}E_{ia}\big)+\sum_\alpha\big(\gamma_\alpha b^\dagger_\alpha +\gamma^*_\alpha b_\alpha\big)\nonumber\\
    &=\bm{\kappa}(t) +\sum_\alpha\big(\gamma_\alpha b^\dagger_\alpha +\gamma^*_\alpha b_\alpha\big) .
\end{align}
The $\kappa_{ai}$ parameters in Eq. \eqref{TD-QED-HF} are orbital rotation parameters, $\gamma_\alpha$ describe the evolution of the QED-HF coherent state, and we compactly defined the orbital rotation operator as
\begin{equation}
    \bm{\kappa}(t)=\frac{1}{\sqrt{2}}\sum_{ai}\big(\kappa_{ai}E_{ai}+\kappa^*_{ai}E_{ia}\big).
\end{equation}
Notice that electronic and photonic operators commute, implying that the exponential in Eq.~\eqref{time-dep-qedhf} can be split into the product of an orbital rotation exponential and a time-dependent coherent state for the field
\begin{widetext}
    \begin{align}
    \text{exp}\big(-i\Lambda(t)\big)\ket{R}&=\text{exp}\big(-i\sum_\alpha\big(\gamma_\alpha b^\dagger_\alpha +\gamma^*_\alpha b_\alpha\big)\big)\text{exp}\big(-i\bm{\kappa}(t)\big)\ket{R}\\
    &=\text{exp}\big(-i\bm{\kappa}(t)\big)\text{exp}\big(-i\sum_\alpha\big(\gamma_\alpha b^\dagger_\alpha +\gamma^*_\alpha b_\alpha\big)\big)\ket{R}\label{tdqedhfpar}
\end{align}
Equivalently, we can parameterize the time dependence as
\begin{equation}\label{tdqedhfalternative}
 \ket{\text{QED-HF}}(t)=\text{exp}\bigg(-\sum_\alpha\frac{\bm{\lambda}_\alpha\cdot \braket{\bm{d}}_{\text{QED-HF}}}{\sqrt{2\omega}}(b^\dagger_\alpha - b_\alpha)\bigg)\text{exp}\big(-i\sum_\alpha\big(\gamma_\alpha b^\dagger_\alpha +\gamma^*_\alpha b_\alpha\big)\big)\text{exp}\big(-i\bm{\kappa}(t)\big)\ket{\mathrm{HF}}\otimes\ket{0},
\end{equation}
where we moved the coherent state transformation to the left. 
Since Eq.~\eqref{tdqedhfalternative} and Eq.~\eqref{tdqedhfpar} differ by an unimportant phase factor, these parameterizations lead to the same time evolution. 
By using the Ehrenfest theorem and developing the equations in orders of the perturbation, we obtain the zero, first and second-order equations
   \begin{align}
    &\braket{[\Omega^{(0)},H]}_{\mathrm{R}}=0\label{zero_ord_qedhf}\\
  & i\braket{\mathrm R|[\frac{\partial}{\partial t}\Lambda^{(1)},\Omega^{(0)}] |\mathrm R} =\braket{\mathrm R|[\Lambda^{(1)},[\Omega^{(0)},H]]|\mathrm R}-i\braket{\mathrm R|[\Omega^{(0)},V^t]|\mathrm R}-i\braket{\mathrm R|[\Omega^{(1)},H]|\mathrm R}\label{first_order_qed_hf}\\
    &\langle \mathrm R|\big[\frac{\partial}{\partial t}\Lambda^{(2)},\Omega^{(0)}\big]+\big[\frac{\partial}{\partial t}\Lambda^{(1)},\Omega^{(1)}\big]+{i}\big[\Lambda^{(1)},\big[\frac{\partial}{\partial t}\Lambda^{(1)},\Omega^{(0)}\big]\big]|\mathrm R\rangle=\braket{\mathrm R|[V^t,\Omega^{(1)}]|\mathrm R}+\braket{\mathrm R|[H,\Omega^{(2)}]|\mathrm R}\nonumber\\
    &+i\langle \mathrm R|\big[\Lambda^{(2)},\big[H,\Omega^{(0)}\big]\big]|R\rangle+i\langle \mathrm R|\big[\Lambda^{(1)},\big[V^t,\Omega^{(0)}\big]\big]|\mathrm R\rangle+i\langle \mathrm R|\big[\Lambda^{(1)},\big[H,\Omega^{(1)}\big]\big]|\mathrm R\rangle-\frac{1}{2}\langle \mathrm R |\big[\Lambda^{(1)},\big[\Lambda^{(1)},[H,\Omega^{(0)}]\big]\big]|\mathrm R\rangle\label{second_order_qed_hf},
\end{align} 
where we also accounted for the possibility that $\Omega$ depends on the external perturbation $\Omega\equiv\Omega(V^t)$.
The apexes $(0)$-$(2)$ refer to the expansion order and
\begin{align}
    \Lambda^{(1)}(t) &=\frac{1}{\sqrt{2}}\sum_{ai}\big(\kappa_{ai}^{(1)}E_{ai}+\kappa^*_{ai}\,^{(1)}\;E_{ia}\big)+\sum_\alpha\big(\gamma_\alpha^{(1)} b^\dagger     +\gamma_\alpha^*\,^{(1)} \;b\big)\label{lambda1}\\
    \Lambda^{(2)}(t) &=\frac{1}{\sqrt{2}}\sum_{ai}\big(\kappa_{ai}^{(2)}E_{ai}+\kappa^*_{ai}\,^{(2)}\;E_{ia}\big)+\sum_\alpha\big(\gamma_\alpha^{(2)} b^\dagger     +\gamma_\alpha^*\,^{(2)} \;b\big)\label{lambda2}.
\end{align}
\end{widetext}
As long as $\Omega$ is a one-electron operator or a purely photonic 1/2 operator, the zero-order condition is satisfied due to the QED-HF optimization condition. 
The first-order equation is non-trivial, and its solution gives us the time evolution of the parameters to first-order in the perturbation. 
Following the derivation of the response equations in molecular response theory,\cite{olsen1985linear, dalgaard1980time} we make use of the operators
\begin{equation}\label{omega_set}
    \Omega \in \bigg\{b_\alpha,\,\frac{1}{\sqrt{2}}\,e^{-i\bm{\kappa}}E_{ia}e^{i\bm{\kappa}},\,b^\dagger_\alpha,\,\frac{1}{\sqrt{2}}\,e^{-i\bm{\kappa}}E_{ai}e^{i\bm{\kappa}}\bigg\}.
\end{equation}
Making this choice, we ensure the response equations are identical to the ones derived from the time-dependent variational principle.\cite{olsen1985linear, mclachlan1964time, di2000hellmann} 
It is convenient to move from the time domain to the frequency domain by performing a Fourier transformation of the parameters
\begin{align}
    \kappa_{ai}(t)&=\int d\omega\;e^{-i\omega t}\kappa_{ai}^\omega+\int \int d\omega\,d\omega'\;e^{-i\omega t}e^{-i\omega' t}\kappa_{ai}^{\omega,\omega'}+\dots\\
    \gamma(t) &=\int d\omega\;e^{-i\omega t}\gamma_\alpha^\omega+\int \int d\omega\,d\omega'\;e^{-i\omega t}e^{-i\omega' t}\gamma_\alpha^{\omega,\omega'}+\dots
\end{align}
such that the operators
\begin{align}
    \Lambda^{\omega} &=\frac{1}{\sqrt{2}}\sum_{ai}\big(\kappa_{ai}^{\omega}E_{ai}+[\kappa_{ai}^{-\omega}]^*\;E_{ia}\big)\nonumber\\
    &+\sum_\alpha\big(\gamma^{\omega}_\alpha b^\dagger_\alpha +[\gamma^{-\omega}_\alpha]^* \;b_\alpha\big)\label{omega-qed-hf-par}\\
    \Lambda^{\omega,\omega'} &=\frac{1}{\sqrt{2}}\sum_{ai}\big(\kappa_{ai}^{\omega,\omega'}E_{ai}+[\kappa_{ai}^{-\omega,-\omega'}]^*\;E_{ia}\big)\nonumber\\
    &+\sum_\alpha\big(\gamma^{\omega,\omega'}_\alpha b^\dagger_\alpha +[\gamma^{-\omega,-\omega'}_\alpha]^* \;b_\alpha\big)
\end{align}
are the Fourier components of Eqs.~\eqref{lambda1} and \eqref{lambda2}
\begin{align}
    &\Lambda^{(1)}(t)=\int d\omega\;e^{-i\omega t}\Lambda^\omega\\
    &\Lambda^{(2)}(t)=\int d\omega\,d\omega'\;e^{-i\omega t}e^{-i\omega' t}\Lambda^{\omega,\omega'}.
\end{align}

\paragraph{Linear response equations.}
From Eq.~\eqref{first_order_qed_hf}, performing a Fourier transform and by using Eq.~\eqref{omega-qed-hf-par} and Eq.~\eqref{omega_set}, after some algebra, we obtain the standard matrix equation
\begin{equation}\label{tdhf_response}
    \left[\begin{pmatrix}
     \mathbf{A}&\mathbf{B}\\
     \mathbf{B}^*&\mathbf{A}^*
    \end{pmatrix}-\omega \begin{pmatrix}
     \bm 1&0\\
     0&-\bm 1
    \end{pmatrix}\right]\begin{pmatrix}
     \bm{X}\\\bm{Y}
    \end{pmatrix}=i\begin{pmatrix}
     \bm{g}_1\\ \bm{g}_2
    \end{pmatrix}
    \equiv i\bm{g},
\end{equation}
where the vectors $\bm{X}$ and $\bm{Y}$ collect the Fourier transformed parameters
\begin{equation}
    \bm{X}=\begin{pmatrix}
    \gamma^\omega_\alpha\\
    \kappa_{ai}^\omega
    \end{pmatrix}\quad \bm{Y}=\begin{pmatrix}
    [\gamma^{-\omega}_\alpha]^*\\
    [\kappa_{ai}^{-\omega}]^*
    \end{pmatrix}.
\end{equation}
The right hand side of Eq.~\eqref{tdhf_response} is the generalized gradient
\begin{equation}\label{eq:generalized_gradient}
    \bm{g}_1=\begin{pmatrix}
    \braket{[b_\alpha,V^\omega]}_\mathrm{R}\\
    \frac{1}{\sqrt{2}}\braket{[E_{ia},V^\omega]}_\mathrm{R}
    \end{pmatrix}\quad \bm{g}_2=\begin{pmatrix}
    \braket{[b^\dagger_\alpha,V^\omega]}_\mathrm{R}\\
    \frac{1}{\sqrt{2}}\braket{[E_{ai},V^\omega]}_\mathrm{R}
    \end{pmatrix}.
\end{equation}
The matrix
\begin{equation}\label{tdhfE}
    \mathbf{E}^{[2]}=\begin{pmatrix}
     \bm{A}&\mathbf{B}\\
     \mathbf{B}^*&\bm{A}^*
    \end{pmatrix}
\end{equation}
is the generalized Hessian matrix while
\begin{equation}\label{tdhfS}
    \bm{S}^{[2]}=\begin{pmatrix}
      \bm 1&0\\
      0&-\bm1
    \end{pmatrix}
\end{equation}
is the generalized metric matrix.
Therefore, Eq.~\eqref{tdhf_response} is a generalization of the Casida equations of TDHF in molecular response theory.\cite{casida2009time, casida1995time} Indeed, the explicit expressions of $\mathbf{A}$ and $\mathbf{B}$ are
\begin{widetext}
    \begin{align}
    \mathbf{A}&=\begin{pmatrix}
     \omega_\alpha\delta_{\alpha\beta}&\sqrt{{\omega_\alpha}}(\bm{\lambda}_\alpha\cdot\bm{d}_{ib})\\
     \sqrt{{\omega_\alpha}}(\bm{\lambda}_\alpha\cdot\bm{d}_{bi})&\mathbf{A}_{el}
    \end{pmatrix}=\begin{pmatrix}
     \braket{[b_\alpha,[H,b^\dagger_\alpha]]}_\mathrm{R}&\frac{1}{\sqrt{2}}\braket{[b_\alpha,[H,E_{ai}]]}_\mathrm{R}\\
     \frac{1}{\sqrt{2}}\braket{[b^\dagger_\alpha,[H,E_{ia}]]}_\mathrm{R}&\frac{1}{2}\braket{[E_{jb},[H,E_{ai}]]}_\mathrm{R}
    \end{pmatrix}\label{Asingle}\\
    \mathbf{B}&=\begin{pmatrix}
    0&-\sqrt{{\omega_\alpha}}(\bm{\lambda}_\alpha\cdot\bm{d}_{ib})\\
    -\sqrt{{\omega_\alpha}}(\bm{\lambda}_\alpha\cdot\bm{d}_{bi})&\mathbf{B}_{el}
    \end{pmatrix}\label{Bsingle}=\begin{pmatrix}
     \braket{[b_\alpha,[H,b_\alpha]]}_\mathrm{R}&\frac{1}{\sqrt{2}}\braket{[b_\alpha,[H,E_{ia}]]}_\mathrm{R}\\
     \frac{1}{\sqrt{2}}\braket{[b^\dagger_\alpha,[H,E_{ai}]]}_\mathrm{R}&\frac{1}{2}\braket{[E_{bj},[H,E_{ai}]]}_\mathrm{R}
    \end{pmatrix}.
\end{align}
The electronic blocks
\begin{align}
    \big(A_{el}\big)_{ia,bj}&=\frac{1}{2}\braket{\mathrm{R}|\left[E_{ia}, \left[H,E_{bj}\right]\right]|\mathrm{R}}=\delta_{ij}F_{ab}-\delta_{ab}F_{ij}+2\bar{g}_{aibj}-\bar{g}_{abji}\label{Aexpl}\\
    \big(B_{el}\big)_{ia,jb}&=\frac{1}{2}\braket{\mathrm{R}|\left[E_{ia}, \left[H,E_{jb}\right]\right]|\mathrm{R}}=\bar{g}_{biaj}-2\bar{g}_{aibj}\label{Bexpl},
\end{align}
\end{widetext}
have the same definition as the standard Casida matrices, $\bm{A}$ and $\mathbf{B}$, in the TDHF theory. 
Nevertheless, we point out that these blocks differ from the bare TDHF matrices for two reasons. 
First, the QED-HF orbitals and orbital energies differ from the bare molecular ones. 
Second, the two-electron integrals now contain the dipole self-energy contributions, as shown in Eq.~\eqref{two_electron+dse}. 
This contribution can be explicitly separated\cite{yang2021quantum} following Eqs. \eqref{two_electron+dse} and \eqref{one_electron_dse}
\begin{align}
    \mathbf{A}_{el}&=\mathbf{A}_{el}^{\text{HF}}+\bm{\Delta}\\
    \mathbf{B}_{el}&=\mathbf{B}_{el}^{\text{HF}}+\bm{\Delta'}
\end{align}
where the $\bm{\Delta}$ and $\bm{\Delta'}$ matrices now include the dipole self-energy contribution to the Hamiltonian, while $\mathbf{A}_{el}^{\text{HF}}$ and $\mathbf{B}_{el}^{\text{HF}}$ only include the standard electronic integrals.
The TD-QED-HF matrices have additional dimensions due to the cavity modes. 
The coupling between the molecular and the photonic parameters is given by the projection of the transition dipole moments onto the coupling strength vector $\bm{\lambda}$ in the off-diagonal terms of Eqs. \eqref{Aexpl} and \eqref{Bexpl}. 
Moreover, the $\mathbf{A}$ matrix has a non-zero photonic block that contains the frequencies $\omega_\alpha$ of the cavity modes. 
As expected, in Eq.~\eqref{Asingle} there is no direct interaction between the photon modes, which are coupled indirectly through matter degrees of freedom. 
We see that these equations are an extension of the familiar TDHF response equations since in the zero coupling limit we recover TDHF solutions and the one photon lines of the empty cavity. 
We can then write the response function from the time evolution of an operator $\Omega$
\begin{widetext}
    \begin{align}
    &\braket{\Omega}(t)-\braket{\Omega}_{\mathrm{R}}=\braket{\Omega}^{(1)}(t)+\dots=\nonumber\\
    &-i\int_{-\infty}^\infty e^{-i\omega t}\left(\sum_{ai}\frac{{\kappa}^\omega_{ai}}{\sqrt{2}}\braket{[\Omega,E_{ai}]}_{\mathrm{R}}+
    \sum_{ai}\frac{[{\kappa}^{-\omega}_{ai}]^*}{\sqrt{2}}\braket{[\Omega,E_{ia}]}_{\mathrm{R}}+{\gamma}^\omega\braket{[\Omega,b_\alpha^\dagger]}_{\mathrm{R}}+[\gamma_\alpha^{-\omega}]^*\braket{[\Omega,b]}_{\mathrm{R}}\right)+\dots
\end{align}
Inverting equation Eq.~\eqref{tdhf_response}, we identify the linear response function
    \begin{align}
    \braket{\braket{\Omega,V^\omega}}_\omega=&-i\bigg(\sum_{ai}\frac{{\kappa}^\omega_{ai}}{\sqrt{2}}\braket{[\Omega,E_{ai}]}_{\mathrm{R}}+
    \sum_{ai}\frac{[{\kappa}^{-\omega}_{ai}]^*}{\sqrt{2}}\braket{[\Omega,E_{ia}]}_{\mathrm{R}}+{\gamma}_\alpha^\omega\braket{[\Omega,b^\dagger]}_{\mathrm{R}}+[\gamma^{-\omega}_\alpha]^*\braket{[\Omega,b]}_{\mathrm{R}}\nonumber\bigg)\\
    =&i\bm{g}_\Omega^\dagger\begin{pmatrix}
     \bm{X}\\\bm{Y}
    \end{pmatrix}=-\bm{g}_\Omega^\dagger\left[\begin{pmatrix}
     \mathbf{A}&\mathbf{B}\\
     \mathbf{B}^*&\mathbf{A}^*
    \end{pmatrix}-\omega \begin{pmatrix}
     1&0\\
     0&-1
    \end{pmatrix}\right]^{-1}\bm{g}\label{qedhf-lin-resp-fun},
\end{align}
where $\bm{g}_\Omega$ has the same structure as $\bm{g}$ with the operator $\Omega$ replacing the perturbation $V^\omega$ in Eq. \eqref{eq:generalized_gradient}.
We note that the response function fulfils the symmetry relation in Eq.~\eqref{lr_symm}. From the generalized eigenvalue equation
\begin{equation}\label{tdhf_response_eig}
    \left[\begin{pmatrix}
     \mathbf{A}&\mathbf{B}\\
     \mathbf{B}^*&\mathbf{A}^*
    \end{pmatrix}-\omega \begin{pmatrix}
     1&0\\
     0&-1
    \end{pmatrix}\right]\begin{pmatrix}
     \bm{X}\\\bm{Y}
    \end{pmatrix}=\begin{pmatrix}
     0\\ 0
    \end{pmatrix},
\end{equation}
we obtain the eigenvectors $(\bm{x}_i,\bm{y}_i)^T$ and the spectral decomposition
    \begin{align}
    \nonumber&\left[\begin{pmatrix}
     \mathbf{A}&\mathbf{B}\\
     \mathbf{B}^*&\mathbf{A}^*
    \end{pmatrix}-\omega \begin{pmatrix}
     1&0\\
     0&-1
    \end{pmatrix}\right]^{-1}=\sum_{i> 0}\left(\frac{1}{\omega_j-\omega}\begin{pmatrix}
     \bm{x}_i\\ \bm{y}_i
    \end{pmatrix}\otimes\begin{pmatrix}
     \bm{x}_i^\dagger&\bm{y}_i^\dagger
    \end{pmatrix}-\frac{1}{\omega_j+\omega}\begin{pmatrix}
     \bm{y}^*_i\\ \bm{x}_i^*
    \end{pmatrix}\otimes\begin{pmatrix}
     \big(\bm{y}_i^*\big)^\dagger&  \big(\bm{x}_i^*\big)^\dagger
    \end{pmatrix}\right).
\end{align}
\end{widetext}
We identify the excitation energies of the system as the eigenvalues of Eq.~\eqref{tdhf_response_eig}. The transition moments are obtained from the residues of the response function, identified as
\begin{equation}\label{QEDHFtr}
    \braket{0|\Omega|k}=
    \begin{pmatrix}
     \bm{x}_k^\dagger&\bm{y}_k^\dagger
    \end{pmatrix}
    \cdot
    \begin{pmatrix}
     \braket{[b_\alpha^\dagger,\Omega]}_{\mathrm{R}}\\
    \frac{1}{\sqrt{2}}\braket{[E_{ai},\Omega]}_{\mathrm{R}}\\
    \braket{[b_\alpha,\Omega]}_{\mathrm{R}}\\
    \frac{1}{\sqrt{2}}\braket{[E_{ia},\Omega]}_{\mathrm{R}}
    \end{pmatrix}.
\end{equation} 
Although the eigenvalue problem in Eq.~\eqref{tdhf_response_eig} is non Hermitian, it is possible to show that if the computed reference state is close to the ground state, Eq. \eqref{tdhf_response_eig} has solutions with real and positive eigenvalues.\cite{olsen1985linear} 
We note that in Eq.~\eqref{tdhf_response}, setting $\omega=0$, we obtain the static coupled-perturbed QED-HF equations.

Several approximations can be proposed for these response equations.
A hierarchy of approximations has been discussed by Yang et al.\cite{yang2021quantum} for their TDDFT-Pauli-Fierz (TDDFT-PF) model, which defined a eigenvalue problem analogous to Eq. \eqref{tdhf_response}. 
They define the Tamm-Dancoff approximation (TDA) by neglecting $\mathbf{B}_{el}$.
However, contrary to standard electronic TDA, this is \textit{not} equivalent to a CI-singles approach. 
For this reason, we suggest it is more natural to define the TDA approximation by neglecting the whole $\mathbf{B}$ matrix, since this is equivalent to a CI problem with singly excited determinants with zero photons and the ground state determinant with one photon. 
Neglecting $\mathbf{B}$ and the dipole self-energy contribution in $\bm{\Delta}$, they define the TDDFT-Jaynes-Cummings approximation, similar to a JC calculation with all the single excited determinants.

\paragraph{Quadratic response equations}
 \noindent The quadratic response equations are obtained in a similar fashion as for the linear response.\cite{olsen1985linear, kjaergaard2008hartree, norman2018principles}
 By employing Eq. \eqref{second_order_qed_hf} and the set of operators \eqref{omega_set}, we obtain the second-order response equation
 \begin{widetext}
     \begin{align}
    &(\omega_1+\omega_2)\braket{\mathrm{R}|[\Lambda^{\omega_1,\omega_2},\bm{\xi}]|\mathrm{R}}-\braket{\mathrm{R}|[\Lambda^{\omega_1,\omega_2},[\bm{\xi},H]]|\mathrm{R}}=\frac{1}{2}\hat{P}(\omega_1,\omega_2)\big(\frac{i}{2}\braket{\mathrm{R}|[\Lambda^{\omega_1},[\Lambda^{\omega_2},[\bm{\xi},H]]]|\mathrm{R}}+\braket{\mathrm{R}|[\Lambda^{\omega_1},[\bm{\xi},V^{\omega_2}]]|\mathrm{R}}\big)\label{eq:qed-hf-2},
\end{align}
where $\hat{P}(\omega_1, \omega_2)$ sums the permutations of $\omega_1$ and $\omega_2$ and we used the compact notation
\begin{equation}
    \bm{\xi}=\begin{pmatrix}
        b_\alpha\\
        \frac{1}{\sqrt{2}}E_{ia}   \\   
        b^\dagger_\alpha\\
        \frac{1}{\sqrt{2}}E_{ai}
    \end{pmatrix}.
\end{equation}
To compute the quadratic response function $\braket{\braket{A;B,C}}_{\omega_1,\omega_2}$, together with the $\mathbf{E}^{[2]}$ and $\bm{S}^{[2]}$ matrices defined in Eqs. \eqref{tdhfE} and \eqref{tdhfS} for the linear response, we need the additional matrices
\begin{align}
    E^{[3]}_{ijk}&=\frac{1}{2}\braket{\mathrm{R}|[\xi_i^\dagger,[\xi_j^\dagger,[H,\xi_k]]]|\mathrm{R}}\\
    X^{[2]}_{ij}&=\braket{\mathrm{R}|[\xi_i^\dagger,[X,{\xi}_j]]|\mathrm{R}},
\end{align}
where $X=A,B$ or $C$. 
The second-order variation of a time-independent observable $A$ is 
    \begin{align}
    \braket{A}^{(2)}(t)&=i\braket{[\Lambda^{(2)},A]}_\mathrm{R}-\frac{1}{2}\braket{[\Lambda^{(1)},[\Lambda^{(1)},A]]}_\mathrm{R}\nonumber\\
    &=\int\int d\omega_1d\omega_2\,\frac{1}{2}\hat{P}(\omega_1, \omega_2)\big(i\braket{[\Lambda^{\omega_1,\omega_2},A]}_\mathrm{R}-\frac{1}{2}\braket{[\Lambda^{\omega_2},[\Lambda^{\omega_1},A]]}_\mathrm{R}\big)e^{-i(\omega_1+\omega_2)t}\label{second_order_var_hf},
\end{align}
where we made the expression explicitly symmetric in the frequencies. 
From Eq.~\eqref{second_order_var_hf}, we can identify the quadratic response function, which has poles where the frequencies or their sum match an excitation energy\cite{dalgaard1982quadratic}
\begin{align}
    \braket{\braket{A;B,C}}_{\omega_1,\omega_2}=\hat{P}(\omega_1, \omega_2)\big(i\braket{[\Lambda^{\omega_1,\omega_2},A]}_\mathrm{R}-\frac{1}{2}\braket{[\Lambda^{\omega_2},[\Lambda^{\omega_1},A]]}_\mathrm{R}\big).
\end{align}
\end{widetext}
For the residues and the response function, we then need the following vectors:
\begin{align*}
   \big[\bm{N}^a(\omega_1+\omega_2)\big]^\dagger &=\bm{g}_A^\dagger[\mathbf{E}^{[2]}
    -(\omega_1+\omega_2)\bm{S}^{[2]}]^{-1}\\
    \bm{N}^b(\omega_1)&=[\mathbf{E}^{[2]}
    -\omega_1\bm{S}^{[2]}]^{-1}\bm{g}_B\\
    \bm{N}^c(\omega_2)&=[\mathbf{E}^{[2]}
    -\omega_2\bm{S}^{[2]}]^{-1}\bm{g}_C.
\end{align*}
As for the linear response equations, these vectors and matrices have additional dimensions due to the photonic parameters.

\subsubsection{On the definition of the photonic character of excited states} \label{sec:photon-character-qedhf}
From the QED-Casida equations, it is possible to define a relative electronic/photonic contribution to the polaritonic excitation.\cite{yang2021quantum, flick2020ab} 
Given a normalized eigenvector of Eq.~\eqref{tdhf_response_eig}, we define the "photonic character" $\chi_n$ as the sum over the cavity modes $\alpha$ of the squares of the photonic response parameters $\gamma_\alpha$\cite{yang2021quantum, flick2020ab}
\begin{equation}\label{photon_contribution}
    \chi_n =\sum_\alpha |\gamma_\alpha|^2.
\end{equation}
The electronic character of the excitation $\rho_n$ is defined as the sum of the orbital rotation $\kappa_{ai}$, such that $\chi_n+\rho_n=1$
\begin{equation}
    \rho_n=\sum_{ai}|k_{ai}|^2=1-\chi.
\end{equation}
However, such a definition of the excitation character is not straightforwardly connected to photon number nor to electronic excitations in a molecular sense. 
In fact, for the QED-HF equations, we used the dipole Hamiltonian in the length form. 
As a consequence of this choice, as pointed out at the end of section \eqref{chapter_dipole_approx}, the photon creation and annihilation operators cannot be simply identified with $b^\dagger$ and $b$.
These operators refer instead to the \textit{displacement field} and therefore include matter contributions from the polarization in Eq. \eqref{eq:polarization_displacement_field}.
Moreover, as the coupling strength increases, the ground state might gain significant contributions from photon states. 
The "photon character" of the excitation becomes, therefore, a slippery concept that requires careful examination also of the ground state. 
A clarification of this photon character ambiguity could be obtained by implementing the quadratic response equations, from which we obtain the expectation values of observables for excited states. 
Accordingly, we may compute the following expectation values
\begin{align}
    &\braket{n|b_\alpha^\dagger b_\alpha|n}\label{eq:displacement_excited_state number}\\
    &\braket{n|\big(b_\alpha^\dagger + \frac{1}{\sqrt{2\omega_\alpha}}\bm{\lambda}_\alpha\cdot \bm{d}\big)\big(b_\alpha + \frac{1}{\sqrt{2\omega_\alpha}}\bm{\lambda}_\alpha\cdot \bm{d}\big)|n}\label{eq:photon_excited_state number}.
\end{align}
In the length representation, Eq. \eqref{eq:displacement_excited_state number} refers to the mean occupation number of the displacement field number states, while Eq. \eqref{eq:photon_excited_state number} to the mean electric field number states. 
We can then compare these calculations to the QED-HF ground state expectation values
\begin{align}
    &\braket{\mathrm R|b_\alpha^\dagger b_\alpha|\mathrm R}=\frac{1}{2\omega_\alpha}\big(\bm{\lambda}_\alpha\cdot \braket{\bm{d}}_{\text{QED-HF}}\big)^2\\
    &\braket{\mathrm R|\big(b_\alpha^\dagger + \frac{1}{\sqrt{2\omega_\alpha}}\bm{\lambda}_\alpha\cdot \bm{d}\big)\big(b_\alpha + \frac{1}{\sqrt{2\omega_\alpha}}\bm{\lambda}_\alpha\cdot \bm{d}\big)|\mathrm R}=0
\end{align}
to obtain an estimate of the cavity-field contribution to the excitations.\\

Employing these less straightforward but more precise definitions might clarify some results presented in the literature. 
When using the definition in Eq. \eqref{photon_contribution}, Yang et al.\cite{yang2021quantum} found that the photon contribution of the lower polariton increases with the coupling strength, while the opposite is found for the upper polariton.\cite{yang2021quantum} 
At the same time, the intensity of the lower (upper) polariton increases (decreases) with the coupling strength, which seems in contradiction with the zero oscillator strength associated with the photon states. 
It is argued that this apparent discrepancy results from the intrusion of higher-energy electronic states, which strengthen the intensity of the lower polariton, overcompensating for the increased photon character of the excitation.\cite{yang2021quantum} 
Nevertheless, in this representation, $b^\dagger$ involves both the photons and the matter polarization, which suggests that this interpretation might be revised.\\

We also note that since QED-HF introduces a coherent state transformation, it would be interesting to investigate the following expectation values
 \begin{widetext}
    \begin{align}
    &\braket{n|b_\alpha^\dagger b_\alpha|n}\label{eq:coherent_state_photon_displacemet_field}\\
    &\braket{n|\big(b_\alpha^\dagger - \frac{1}{\sqrt{2\omega_\alpha}}\bm{\lambda}_\alpha\cdot \braket{\bm{d}}_{\text{QED-HF}}\big)\big(b_\alpha - \frac{1}{\sqrt{2\omega_\alpha}}\bm{\lambda}_\alpha\cdot \braket{\bm{d}}_{\text{QED-HF}}\big)|n}\label{eq:photon_displacemet_field}\\
    &\braket{n|\big(b_\alpha^\dagger + \frac{1}{\sqrt{2\omega_\alpha}}\bm{\lambda}_\alpha\cdot (\bm{d}-\braket{\bm{d}}_{\text{QED-HF}})\big)\big(b_\alpha + \frac{1}{\sqrt{2\omega_\alpha}}\bm{\lambda}_\alpha\cdot (\bm{d}-\braket{\bm{d}}_{\text{QED-HF}})\big)|n}\label{eq:photon_field},
    \end{align}  
 \end{widetext}
where Eq. \eqref{eq:coherent_state_photon_displacemet_field} refers to the occupation of the displacement field coherent states, Eq. \eqref{eq:photon_displacemet_field} refers to the displacement field number states, and Eq. \eqref{eq:photon_field} refers to the occupation of the electric field number states.

\subsubsection{Equivalent transition moments in TD-QED-HF}
Following the same procedure as for standard HF response theory,\cite{dalgaard1980time, olsen1985linear} we can show that Eqs. \eqref{dipole-velocity-form}, \eqref{eq:photo_momenta_conjugate_dipole}, and \eqref{dipole-velocity-photon-form}, derived for exact wave functions, also hold in the QED-HF response framework.
In Tab. \eqref{tab:my_label} we report the TD-QED-HF transition moments for the lower polariton of an ethylene molecule using different basis sets.
We report the oscillator strengths in the length $f_L$ and velocity $f_V$ forms
\begin{align}
    f_L^{0n} = \frac{2}{3}\;\omega_n\sum_{i=x,y,z}|\braket{0|d_i|n}|^2\label{eq:f_L}\\
    f_V^{0n} = \frac{2}{3\omega_n}\sum_{i=x,y,z}|\braket{0|p_i|n}|^2,\label{eq:f_V}
\end{align}
and the transition photon displacement coordinate and momentum
\begin{align}
    \braket{0|p_\alpha|n} &= i \sqrt{\frac{\omega_\alpha}{2}}\braket{0|(b_\alpha^\dagger-b_\alpha)|n}\\
    \braket{0|q_\alpha|n} &=  \frac{1}{\sqrt{2\omega_\alpha}}\braket{0|(b_\alpha^\dagger+b_\alpha)|n}.
\end{align}
We numerically demonstrate the validity of Eqs. \eqref{dipole-velocity-form}, \eqref{eq:photo_momenta_conjugate_dipole}, and \eqref{dipole-velocity-photon-form}.
Notice that since Eq. \eqref{dipole-velocity-form} is fulfilled only in the complete basis set limit, Eqs. \eqref{eq:f_L} and \eqref{eq:f_V} will converge only for large basis sets, as seen from the table. 
At the same time, Eq. \eqref{dipole-velocity-photon-form} holds independently of the basis set size and the photon space truncation.
Notice that although the equivalence expressed by Eqs. \eqref{eq:photo_momenta_conjugate} and \eqref{eq:photo_momenta_conjugate_dipole} do not depend on the quality of the basis set, the computed values of the transition moments change with the basis, as expected from the connection between matter and photon moments expressed by \eqref{eq:photo_momenta_conjugate_dipole}.
In particular, the use of diffuse functions strongly affects the computed results.
Equivalence relations such as Eqs. \eqref{dipole-velocity-form} and \eqref{dipole-velocity-photon-form} are well known in molecular response theory.
However, they have not been explicitly investigated for QED systems. 
Furthermore, these relations do not hold for the Tamm-Dancoff approximation.
The formal equivalence between the transition dipole and velocity momenta can be exploited in the computation of electronic circular dichroism (ECD) and optical rotation.
From standard molecular response theory, the ECD spectrum is known to be proportional to the rotational strength
\begin{equation}\label{eq:rot_str}
    R_{n0}=\Im\{\braket{0|\bm{\mu}|n}\cdot\braket{n|\bm{m}|0}\}.
\end{equation}
The dipole formulation in Eq. \eqref{eq:rot_str} is origin dependent for a finite basis set, which can lead to unphysical results.
If we are interested in the chirality effects promoted by a chiral cavity, such a formulation might be misleading.
However, the equivalent velocity form
\begin{equation}\label{eq:rot_str_vel}
    R_{n0}=\frac{1}{\omega_n}\Re\{\braket{0|\bm{p}|n}\cdot\braket{n|\bm{m}|0}\}
\end{equation}
is always origin independent, and
Eqs. \eqref{eq:rot_str} and \eqref{eq:rot_str_vel} are equivalent in the limit of a complete basis.
\begin{widetext}
\begin{center}
      \begin{table}[ht!]
    \centering
    \begin{tabularx}{\textwidth}{|c| *{7}{Y|}}
       \hline  Basis set& $f_L$ & $f_V$ & $\braket{0|q_\alpha|n}$  [a.u.]& $i\braket{0|p_\alpha|n}$  [a.u.] & $\omega_n \braket{0|q_\alpha|n}$  [a.u.] & $\frac{\braket{0|\bm{d}\cdot\bm{\lambda}_\alpha|n}\omega_n\omega_\alpha}{\omega_n^2-\omega_\alpha^2}$  [a.u.]\\
      \hline  sto-3g &0.00234 & 0.00096&1.36138 & 0.36641 &0.36641 &0.36641\\
      6-31g &0.02540 &0.01902 &1.34118&0.35965  &0.35965& 0.35965\\
    6-311g &0.04362 &0.03158 &1.32054&0.35349 &0.35349 & 0.35349\\
     6-311g* & 0.04978& 0.04891&1.30958&0.35036 &  0.35036&0.35036 \\
        6-311g** & 0.05454& 0.05192&1.30320&0.34852 &  0.34852&0.34852 \\
        6-311+g** & 0.20610& 0.19318&1.03720&0.27413 &  0.27413 &0.27413\\
        6-311++g** & 0.20631& 0.19327&1.03677&0.27401 &  0.27401& 0.27401\\
         cc-pVDZ &0.04267 &0.04121 &1.31895&0.35309 &0.35309 & 0.35309\\
          aug-cc-pVDZ &0.21694 &0.21309 &0.99843&0.26353 &0.26353& 0.26353 \\
          aug-cc-pV5Z &0.22561 &0.22564 &0.96941&0.25559 &0.25559&0.25559\\
          \hline
    \end{tabularx}
    \caption{Oscillator strength in dipole $f_L$ and velocity $f_V$ form and transition photon displacement coordinate $\braket{0|q_\alpha|n}$ and momentum $i\braket{0|p_\alpha|n}$ for the lower polariton of an ethylene molecule, for different basis sets at TD-QED-HF level. The cavity frequency is set to $0.2695$ a.u., and the coupling strength to $0.01$ a.u. with field polarization along the transition dipole moment of the first non-dark excitation of ethylene. Eq. \eqref{dipole-velocity-form} holds only in the complete basis limit, and convergence between the velocity and length form of the oscillator strength strongly depends on the quality of the basis set. On the other hand, Eqs. \eqref{eq:photo_momenta_conjugate} and \eqref{eq:photo_momenta_conjugate_dipole} hold independently of the truncation of the electronic or photonic space.}
    \label{tab:my_label}
\end{table}  
\end{center}
\end{widetext}

\subsubsection{QED-HF static response equations}

For static perturbations, the response equations for the optimized QED-HF state $\ket{\mathrm R}$ can be derived using the same parametrization shown in Eq.~\eqref{time-dep-qedhf}, which now is time-independent 
\begin{equation}\label{QEDHF_}
    \ket{\text{QED-HF}}=\text{exp}\big(i\Lambda\big)\ket{\mathrm R}.
\end{equation}
Here $\Lambda$ has the same form showed in Eq.~\eqref{TD-QED-HF} and may be written compactly as
\begin{gather}
    \Lambda =  \bm{\xi}^\dagger \cdot\bm{\Theta},
\end{gather}
where $\bm{\Theta}$ collects both photon and electron parameters and $\bm{\xi}$ the respective operators:
\begin{equation*}
    \bm{\Theta}=\begin{pmatrix}
        \gamma_\alpha\\
        \kappa_{ai}\\
        \gamma_\alpha^*\\
        \kappa_{ai}^*
        \end{pmatrix};\quad\bm{\xi}=\begin{pmatrix}
        b_\alpha\\
        \frac{1}{\sqrt{2}}E_{ia}   \\   
        b^\dagger_\alpha\\
        \frac{1}{\sqrt{2}}E_{ai}
    \end{pmatrix}.
\end{equation*}
Following the perturbation expansion of the energy in Sec.~\eqref{QED-TIPT}, the first-order energy derivative is
\begin{gather}  \label{hfthm-qed-hf}
    E^{(1)}_{\text{QED-HF}} = \braket{\mathrm R|H^{(1)}|\mathrm R},
\end{gather}
which expresses the Hellmann-Feynman theorem for the QED-HF state.
In Eq. \eqref{hfthm-qed-hf}, $H^{(1)}$ refers to the first-order interactions.
The second-order energy derivative is
\begin{gather}
    E_{\text{QED-HF}}^{(2)} = \braket{\mathrm R|H^{(2)}|\mathrm R} + \braket{\mathrm R|[\Lambda^{(1)}, H^{(1)}]|\mathrm R},
\end{gather}
where $H^{(2)}$ is the second-order interaction and $\Lambda^{(1)}$ is the first-order correction to the parameterization. 
Since QED-HF state is a variational theory,\cite{haugland2020coupled} the optimization conditions are equivalent to those in standard HF\cite{helgaker1999}
\begin{align} 
    \braket{\mathrm R|[\bm{\xi}, H^{(0)}]|\mathrm R} &= 0  \label{brillouin-qed-hf} \\
        \sum_{j} \braket{\mathrm R|[ {\xi}^\dagger_{i} , [ { H^{(0)}, {\xi}}_{j} ]]|\mathrm R} {\Theta}_{j}^{(1)} &= i\braket{\mathrm R|[{\xi}_{i}, H^{(1)}]|\mathrm R} \label{cp-qedhf},
\end{align}
where Eq.~\eqref{brillouin-qed-hf} corresponds to the QED-HF Brillouin's theorem ,\cite{haugland2020coupled} and Eq.~\eqref{cp-qedhf} to the first-order static response equations. 
The left-hand side of Eq.~\eqref{cp-qedhf} includes the generalized Hessian matrix, whose blocks are given in Eq.~\eqref{Asingle}-Eq.~\eqref{Bsingle}, and the first-order parameters $\bm{\Theta}^{(1)}$. The right-hand side contains the first-order Hamiltonian, which includes the first-order interaction terms. We note that Eq. \eqref{cp-qedhf} is equivalent to the time-dependent response equation \eqref{tdhf_response} with $\omega=0$.

\subsection{QED-CC}

The reference wave function used in QED-CC is QED-HF, where we employ the coherent-state transformed Hamiltonian in Eq.~\eqref{QEDHF-transf-Ham}.\cite{haugland2020coupled}
The QED-CC state is defined as
\begin{align}
    \ket{\text{QED-CC}}&=\exp(T)\ket{\text{HF},0},
\end{align}
where $T$ is the cluster operator
\begin{align}
    T&=T_e+T_p+T_{int}.
\end{align}
The electronic cluster $T_e$ is the standard electronic excitation operator\cite{helgaker2014molecular}
\begin{gather}
    T_e=\sum_\mu t_\mu\tau_\mu\label{T_e_CC}\\
    \tau_\mu\ket{\text{HF}}=\ket{\mu},
\end{gather}
where $\ket{\mu}$ is an excited HF determinant.
The cluster operator $T_p$ includes pure photonic excitations
\begin{gather}\label{Tp}
    T_p=\sum_{\bm{n}}\Gamma_{\bm{n}}=\sum_{\bm{n}}\gamma_{\bm{n}}\prod_\alpha(b^\dagger_\alpha)^{n_\alpha},
\end{gather}
where ${\bm{n}}$ is a vector of integers $n_\alpha$ referring to the $\alpha$-mode.
Finally, the interaction operator $T_{int}$ includes simultaneous excitations of matter and cavity modes
\begin{gather}\label{T_ep_QEDCC}
    T_{int}=\sum_{\bm{n}}(S_1^{\bm{n}}+S_2^{\bm{n}}+\dots+S^{\bm{n}}_{N_e}),
\end{gather}
where fir instance
\begin{align}
    S_1^{\bm{n}}&=\sum_{ai}s_{ai}^{\bm{n}}E_{ai}\prod_\alpha(b^\dagger)^{n_\alpha}\\
    S_2^{\bm{n}}&=\frac{1}{2}\sum_{aibj}s_{aibj}^{\bm{n}}E_{ai}E_{bj}\prod_\alpha(b^\dagger)^{n_\alpha}.
\end{align}
The Schr{\"o}dinger equation is solved by projection onto the space ${S}$ spanned by the electronic determinants and the number states\cite{haugland2020coupled}
\begin{equation}\label{CC_space}
    {S}=\text{Span}\big\{ \ket{\text{HF},0}, \ket{\mu,0}, \ket{\text{HF},{\bm{n}}}, \ket{\mu,{\bm{n}}} \big\}.
\end{equation}
In Eq. \eqref{CC_space}, the state $\ket{\mu,{\bm{n}}}$ is a simultaneous excitation to the $\mu$ electronic excited determinant and the ${\bm{n}}$ photon state
\begin{equation}
    \ket{\mu,\bm n}=\ket{\mu}\otimes\ket{\bm n}=\tau_\mu\prod_\alpha\frac{(b_\alpha^\dagger)^{n_\alpha}}{\sqrt{n_\alpha!}}\ket{\text{HF},0}.
\end{equation}
The ground state QED-CC equations are therefore
\begin{align}
    &\braket{\mu,\bm{n}|e^{-T}He^{T}|\text{HF},0}=0\\
    &\braket{\text{HF},0|H|\text{QED-CC}}=E_{\text{QED-CC}}.
\end{align}
The QED-CC dual state is defined as in standard CC theory
\begin{equation}
    \bra{\Lambda}=\bra{\text{HF},0}+\sum\bar{t}_{\mu {\bm{n}}}\bra{\mu,{\bm{n}}}e^{-T},
\end{equation}
where $\bar{t}_{\mu \bm{n}}$ are the Lagrangian multipliers.\cite{haugland2020coupled, helgaker2014molecular}
A collective index ${t}_{\mu \bm{n}}$ can also be defined for the cluster operator such that we can write $T$ more compactly as
\begin{equation}
    T=\sum_{\mu, \bm{n}}{t}_{\mu \bm{n}}\tau_\mu\prod_\alpha(b^\dagger_\alpha)^{n_\alpha},
\end{equation}
where
\begin{align}
    t_{\mu0} &=t_\mu \\
  t_{\mu \bm{n}}   &= s^{\bm{n}}_\mu\\
 t_{\text{HF}\bm{n}}  &=  \gamma_{\bm{n}}.
\end{align}
As in standard CC theory, QED-CC is based on a hierarchy of approximations where the cluster operators and the projection space are truncated.\cite{helgaker2014molecular, haugland2020coupled} 
Since the space ${S}$ is larger than in standard CC theory, as it also includes the photonic excitations, the QED-CC Jacobian $\mathbf{A}$ has additional dimensions
\begin{equation}
    A_{\mu {\bm{n}}, \nu \bm{m}}= \braket{\mu,{\bm{n}}|[e^{-T}{H}e^{T},\tau_\nu\prod_\alpha(b^\dagger_\alpha)^{m_\alpha}]|\text{HF},0}.
\end{equation}
In addition to the electronic block $\mathbf{A}_{e,e}$ similar to electronic CC, there is also a photonic block $\mathbf{A}_{p,p}$ and blocks involving the electronic-photonic parameters
\begin{equation}
\mathbf{A}=\begin{pmatrix}
\mathbf{A}_{e,e}&\mathbf{A}_{e,ep}&\mathbf{A}_{e,p}\\
\mathbf{A}_{ep,e}&\mathbf{A}_{ep,ep}&\mathbf{A}_{ep,p}\\
\mathbf{A}_{p,e}&\mathbf{A}_{p,ep}&\mathbf{A}_{p,p}
\end{pmatrix}.
\end{equation}\\
From the Jacobian $\mathbf{A}$ and the $\bm{\eta}$ vector
\begin{equation}
    \eta_{\mu{\bm{n}}}=\braket{\text{HF},0|e^{-T}{H}e^{T}|\mu,{\bm{n}}},
\end{equation}
we obtain the Lagrangian multipliers and the equation of motion (EOM)-QED-CC formalism for properties and excited states.\cite{haugland2020coupled}

\subsubsection{Time-dependent QED-CC}

\noindent Following the response theory for electronic CC,\cite{pedersen1997coupled, koch1990coupled, stanton1993equation} we parametrize the time-evolution of the QED-CC state by a time-dependent cluster operator $T(t)$
\begin{equation}
    \ket{\text{QED-CC}}(t)=e^{T(t)}\ket{\text{HF},0}e^{i\epsilon(t)}.
\end{equation}
The time-dependent parameters are obtained by projection of the time dependent Schr{\"o}dinger equation
\begin{equation}
    i\frac{d}{dt}\big(e^{T(t)}\ket{\text{HF},0}e^{i\epsilon(t)}\big)=(H+V^t)e^{T(t)}\ket{\text{HF},0}e^{i\epsilon(t)}
\end{equation}
onto the space ${S}$. 
We obtain the following response equations for the cluster parameters
\begin{align}
    \frac{d\epsilon}{dt}&=-\braket{\text{HF},0|(H+V^t)e^{T(t)}|\text{HF},0}\label{eq:epsilon_CC_resp}\\
    \frac{dt_\mu}{dt}&=-i\braket{\mu,0|e^{-T(t)}(H+V^t)e^{T(t)}|\text{HF},0}\label{eq:electronic_CC_resp}\\
    \frac{d\gamma_{\bm{n}}}{dt}&=-i\braket{\text{HF},{\bm{n}}|e^{-T(t)}(H+V^t)e^{T(t)}|\text{HF},0}\label{eq:photon_CC_resp}\\
    \frac{ds_{\mu}^{\bm{n}}}{dt}&=-i\braket{\mu,{\bm{n}}|e^{-T(t)}(H+V^t)e^{T(t)}|\text{HF},0}\label{eq:interaction_CC_resp},
\end{align}
where Eqs. \eqref{eq:epsilon_CC_resp} and \eqref{eq:electronic_CC_resp} are analogous to standard CC response theory,\cite{koch1990coupled, pedersen1997coupled} while Eqs. \eqref{eq:photon_CC_resp} and \eqref{eq:interaction_CC_resp} are additional equations for the photon and electron-photon parameters.
We then perform a perturbative expansion of the amplitudes and a Fourier decomposition
\begin{widetext}
  \begin{align}\label{pert-qedcc}
    t_{\mu n}=&t_{\mu n}^{(0)}+t_{\mu n}^{(1)}+t_{\mu n}^{(2)}+\dots  \nonumber\\
    =&t_{\mu n}^{(0)}+\int d\omega_1 X_{\mu n}^{(1)}(\omega_1)e^{-i\omega_1t}+\int d\omega_1\int d\omega_2 X_{\mu n}^{(2)}(\omega_1,\omega_2)e^{-i\omega_1t-i\omega_2 t}+\dots
\end{align}
Focusing now on a single cavity mode, although the generalization to a multimode system is straightforward, we obtain the following set of equations
\begin{align}
    &\braket{{\mu,n}|e^{-T^{(0)}}He^{T^{(0)}}|\text{HF},0}=0 \label{usual_CC}\\
    &\sum_{\nu m}\left(\omega_1\bm{I}-\mathbf{A}\right)_{\mu n, \nu m}X^{(1)}_{\nu m}=\xi^{(1)}_{\mu n}\\
    &\sum_{\nu m}\left((\omega_1+\omega_2)\bm{I}-\mathbf{A}\right)_{\mu n, \nu m}X^{(2)}_{\nu m}=\xi_{\mu n}^{(2)},
\end{align}
where
\begin{align}
    \xi_{\mu n}^{(1)}(\omega_1)&=\braket{\overline{\mu,n}|V^{\omega_1}|\text{QED-CC}}\\
    \xi_{\mu n}^{(2)}(\omega_1,\omega_2)&=\frac{1}{2}\hat{P}(\omega_1\, \omega_2)\left(\braket{\overline{\mu,n}|[V^{\omega_1},T^{(1)}(\omega_2)]|\text{QED-CC}}+\frac{1}{2}\braket{\overline{\mu,n}|[[H,T^{(1)}(\omega_1)],T^{(1)}(\omega_2)]|\text{QED-CC}}\right),
\end{align}   
and we defined the QED-CC-transformed states
\begin{align}
    &\bra{\overline{\mu,n}}=\bra{\mu,n}e^{-T^{(0)}}\\
    &\ket{\text{QED-CC}}=e^{T^{(0)}}\ket{\text{HF},0}.
\end{align}
The zero, first and second-order cluster operators read
\begin{align}
    T^{(0)}&=\sum_{\mu n}t_{\mu n}\tau_{\mu n}\\
    T^{(1)}(\omega_1)&=\sum_{\mu n} X^{(1)}_{\mu n}(\omega_1)\tau_{\mu n}\\
    T^{(2)}(\omega_1,\omega_2)&=\sum_{\mu n} X^{(2)}_{\mu n}(\omega_1,\omega_2)\tau_{\mu n},
\end{align}
and we note that the zero-order equation in Eq \eqref{usual_CC} is the ground state QED-CC optimization condition.
In an analogous way, we can parametrize the time dependence of the dual QED-CC state
\begin{equation}
    \bra{\Lambda(t)}=\left(\bra{\text{HF}}+\sum_{\mu n}\bar{t}_{\mu n}(t)\bra{\mu,n}e^{-T(t)}\right)e^{-i\epsilon(t)},
\end{equation}
where we have the standard normalization condition of CC at all times\cite{helgaker2014molecular}
\begin{equation}
    \braket{\Lambda(t)|\text{QED-CC}(t)}=1.
\end{equation}
The equations for the Lagrange multipliers are obtained by projection onto ${S}$
\begin{align}
    \frac{d\bar{t}_{\mu n}}{dt}&=i\braket{\Lambda(t)|[H+V_t,\tau_\mu(b^\dagger)^n]|\text{QED-CC}(t)}
\end{align}
and using the same expansion as Eq.~\eqref{pert-qedcc} we can write
\begin{align}\label{pert_lagr}
    \bar{t}_{\mu n}=&\bar{t}_{\mu n}^{(0)}+\bar{t}_{\mu n}^{(1)}+\bar{t}_{\mu n}^{(2)}+\dots  \nonumber\\
    =&\bar{t}_{\mu n}^{(0)}+\int d\omega_1 Y_{\mu n}^{(1)}(\omega_1)e^{-i\omega_1t}+\int d\omega_1\int d\omega_2 Y_{\mu n}^{(2)}(\omega_1,\omega_2)e^{-i\omega_1t-i\omega_2 t}+\dots
\end{align}
We then obtain the equations
\begin{align}
   & \sum_{\nu n}\bar{t}_{\nu n}^{(0)}A_{\nu n,\mu m}=-\braket{R|[H,\tau_\mu(b^\dagger)^n]|\text{QED-CC}}\label{usual_mult_CC}\\
   &\sum_{\nu n}Y^{(1)}_{\nu n}(\omega_1\bm{I}+\mathbf{A})_{\nu n,\mu m}=-\eta_{\mu m}^{(1)}-\sum_{\gamma k}F_{\mu m ,\gamma k}X^{(1)}_{\gamma k}\\
   &\sum_{\nu n}Y^{(2)}_{\nu n}((\omega_1+\omega_2)\bm{I}+\mathbf{A})_{\nu n,\mu m}=-\eta_{\mu m}^{(2)}-\sum_{\gamma k}F_{\mu m,\gamma k}X^{(2)}_{\gamma k}.
\end{align}
The zero-order equation in Eq. \eqref{usual_mult_CC} is the ground state multiplier equation, and the vectors $\eta$ are defined as
    \begin{align}
    \eta_{\mu n}^{(1)}(\omega_1)=&\braket{\Lambda|[V^{\omega_1},\tau_\mu(b^\dagger)^n]|\text{QED-CC}}\\
    \eta_{\mu n}^{(2)}(\omega_1,\omega_2)=&\frac{1}{2}\hat{P}(\omega_1, \omega_2)\bigg(\braket{\Lambda|[[V^{\omega_1},\tau_\mu(b^\dagger)^n],T^{(1)}(\omega_2)]|\text{QED-CC}}+\frac{1}{2}\braket{\Lambda|[[[H,\tau_\mu(b^\dagger)^n],T^{(1)}(\omega_1)],T^{(1)}(\omega_2)]|\text{QED-CC}} \nonumber\\
    +  &\sum_{\nu m}Y^{(1)}_{\nu,n}(\omega_1)\braket{\overline{\nu,m}|[[H,\tau_\mu(b^\dagger)^n],T^{(1)}(\omega_2)]+[V^{\omega_2},\tau_\mu(b^\dagger)^n]|\text{QED-CC}}\bigg),\\
\end{align}  
where the matrix $\bm{F}$ as
\begin{equation}
    F_{\mu m, \nu k} = \braket{\Lambda|[[H,\tau_\mu(b^\dagger)^m],\tau_\nu(b^\dagger)^k]|\text{QED-CC}}.
\end{equation}
As for QED-HF, the response matrices and vectors have additional dimensions connected to the electromagnetic degrees of freedom. 
The $X_{\mu n}$ and $Y_{\nu m}$ vectors include, together with the standard CC electronic excitation response parameters, the purely photonic and the simultaneous electronic-photonic response parameters of Eqs. \eqref{Tp} and \eqref{T_ep_QEDCC}. 
As for the Jacobian matrix $\mathbf{A}$, the $\bm{F}$ matrix has additional photon and electron-photon blocks
    \begin{equation}
\bm{F}=\begin{pmatrix}
\bm{F}_{e,e}&\bm{F}_{e,ep}&\bm{F}_{e,p}\\
\bm{F}_{ep,e}&\bm{F}_{ep,ep}&\bm{F}_{ep,p}\\
\bm{F}_{p,e}&\bm{F}_{p,ep}&\bm{F}_{p,p}
\end{pmatrix}.
\end{equation}
Moreover, the perturbation $V^\omega$ can act on both the electronic and the photon degrees of freedom, as in Eq. \eqref{H-J}.\\

\noindent The CC expectation value is defined as\cite{pedersen1997coupled}
\begin{equation}\label{mean}
    \braket{A}_{\text{QED-CC}}(t)=\frac{1}{2}\big(\braket{\Lambda(t)|A|\text{QED-CC}(t)}+\braket{\Lambda(t)|A|\text{QED-CC}(t)}^*\big).
\end{equation}
\noindent From the perturbative expansion
\begin{align}
    \braket{A}_{\text{QED-CC}}(t)&=\braket{A}_{\text{QED-CC}}+\int d\omega_1\braket{\braket{A,V^{\omega_1}}}_{\omega_1+i\eta}e^{-i\omega_1t+i\eta t}+\dots\nonumber
\end{align}
we can identify the QED-CC response functions
    \begin{align}
    \braket{\braket{A,V^{\omega_1}}}_{\omega_1}&=\frac{1}{2}\left(\mathcal{F}^{A,V^{\omega_1}}_{\omega_1}+\left(\mathcal{F}^{A,V^{-\omega_1}}_{-\omega_1}\right)^*\right)\\
    \braket{\braket{A,V^{\omega_1},V^{\omega_2}}}_{\omega_1,\omega_2}&=\frac{1}{2}\left(\mathcal{F}^{A,V^{\omega_1},V^{\omega_2}}_{\omega_1,\omega_2}+\left(\mathcal{F}^{A,V^{-\omega_1},V^{\omega_2}}_{-\omega_1,-\omega_2}\right)^*\right)
\end{align}
where
    \begin{align}
    \mathcal{F}^{A,V^{\omega_1}}_{\omega_1} =& \sum_{\mu n}\left(Y^{(1)}_{\mu n}(\omega_1)\braket{\overline{\mu, n}|A|\text{QED-CC}}+X_{\mu n}^{(1)}(\omega_1)\braket{\Lambda|[A,\tau_{\mu n}]|\text{QED-CC}}\right)\\
    \mathcal{F}^{A,V^{\omega_1},V^{\omega_2}}_{\omega_1,\omega_2}=&2\sum_{\mu n} \left(Y_{\mu n}^{(2)}(\omega_1,\omega_2)\braket{\overline{\mu, n}|A|\text{QED-CC}}+X_{\mu n}^{(2)}(\omega_1,\omega_2)\braket{\Lambda|[A,\tau_\mu(b^\dagger)^n]|\text{QED-CC}}\right)\nonumber\\
    +&\sum_{\mu n, \nu m}\left(X^{(1)}_{\mu n}(\omega_1)F^A_{\mu n, \nu m}X^{(1)}_{\nu m}(\omega_2)+\hat{P}(\omega_1,\omega_2)Y^{(1)}_{\mu n}(\omega_1)\braket{\overline{\mu, m}|[A,\tau_\nu(b^\dagger)^n]|\text{QED-CC}}X^{(1)}_{\nu n}(\omega_2)\right).
\end{align}
Here we have used the notation
\begin{equation}
    F^A_{\mu m, \nu k}=\braket{\Lambda|[[A,\tau_\mu(b^\dagger)^m],\tau_\nu(b^\dagger)^k]|\text{QED-CC}}.
\end{equation}
\end{widetext}
As for QED-HF, the static response equations can be obtained by setting the frequency of the external to zero.
The poles and the residues of the response functions can be obtained assuming that $\mathbf{A}$ can be diagonalized
\begin{equation}
    (\bm{S}^{-1}\mathbf{A}\bm{S})_{mn}=\delta_{mn}\omega_n,\label{diagonalCC}
\end{equation}
where $\omega_n$ is the n-th (real) excitation energy.
We further introduce the notation
\begin{align}\label{cc-resp-op}
    &\tilde\tau_n=\sum_{\mu,m}S_{\mu m, n}\tau_\mu(b^\dagger)^m\\
    &\tilde\tau_n^\dagger=\sum_{\mu,m}S^{-1}_{n, \mu m}\tau_\mu^\dagger b^m
\end{align}
to describe the diagonal representation in Eq. \eqref{diagonalCC}.
From the poles of the linear response function, we identify the residues
    \begin{align}
    &\lim_{\omega\to\omega_k}(\omega-\omega_k)\braket{\braket{A;B}}_\omega=\nonumber\\
    &\frac{1}{2}\bigg(\Gamma^A_k\Theta^B_k+\big(\Gamma^B_k\Theta^A_k\big)^*\bigg)\equiv\braket{0|A|k}\braket{k|B|0}\\
    &\lim_{\omega\to-\omega_k}(\omega+\omega_k)\braket{\braket{A;B}}_\omega=\nonumber\\
    &-\frac{1}{2}\bigg(\big(\Gamma^A_k\Theta^B_k\big)^*+\Gamma^B_k\Theta^A_k\bigg)\equiv-\big(\braket{0|A|k}\braket{k|B|0}\big)^*
    \end{align}
where
\begin{align}
    \Theta^A_k&=\braket{{\text{QED-HF}}|\tilde\tau_k^\dagger e^{-T^{(0)}}A|\text{QED-CC}}\\
    \Gamma^A_k&=\braket{\Lambda|[A,\tilde\tau_{k}]|\text{QED-CC}}\nonumber\\
    &-\sum_n\frac{\braket{{\text{QED-HF}}|\tilde\tau_k^\dagger e^{-T^{(0)}}A|\text{QED-CC}}F_{nk}}{\omega_n+\omega_k}.
\end{align}
From these equations, we can compute the polaritonic properties.\cite{pedersen1997coupled}
Since we employ the QED-HF coherent-state transformed Hamiltonian in Eq. \eqref{QEDHF-transf-Ham}, the operators and the perturbations in the response functions must be described in the same representation.

\subsubsection{QED-CC electronic and photonic excitation character}

Since QED-CC is a highly correlated method, defined for the QED-HF coherent-state transformed Hamiltonian in Eq. \eqref{QEDHF-transf-Ham}, the definition of the electronic or photonic character of the excitation is not trivial. 
In the EOM framework, Haugland et. al\cite{haugland2020coupled} defined the electronic weight $w_{el}$ of the state by means of projection operators
\begin{equation}
    w^k_{el}=\sqrt{\frac{\braket{\Lambda_k|P_{el}|R_k}}{\braket{\Lambda_k|P|R_k}}}
\end{equation}
where $\Lambda_k$ and $R_k$ are the $k$-th left and right state, $P$ is the projection operator onto the space $ S$ of Eq. \eqref{CC_space} and $P_{el}$ is the projector onto the states of $ S$ with zero photons.
In the response framework, a similar definition can be obtained by considering the electronic components of the $X_{\mu n}$ and $Y_{\mu n}$ vectors, for instance
\begin{align}
    w_{el}&=\sqrt{\sqrt{\frac{\sum_\mu |X_{\mu 0}|^2}{\sum_{\mu n} |X_{\mu n}|^2}}\times \sqrt{\frac{\sum_\mu |Y_{\mu 0}|^2}{\sum_{\mu n} |Y_{\mu n}|^2}}}.
\end{align}
Notice, however, that such a definition does not consider the contribution of the simultaneous electron-photon excitations.
A more consistent definition of the photonic character could be obtained by considering the mean values of the field number operators from the quadratic response function, and comparing them to the ground state values, as discussed for QED-HF
 \begin{widetext}
    \begin{align}
    &\braket{\Lambda_k|b^\dagger b|R_k}\label{eq:CCcoherent_state_photon_displacemet_field}\\
    &\braket{\Lambda_k|\big(b_\alpha^\dagger - \frac{1}{\sqrt{2\omega_\alpha}}\bm{\lambda}_\alpha\cdot \braket{\bm{d}}_{HF}\big)\big(b_\alpha - \frac{1}{\sqrt{2\omega_\alpha}}\bm{\lambda}_\alpha\cdot \braket{\bm{d}}_{HF}\big)|R_k}\label{eq:CCphoton_displacemet_field}\\
    &\braket{\Lambda_k|\big(b_\alpha^\dagger + \frac{1}{\sqrt{2\omega_\alpha}}\bm{\lambda}_\alpha\cdot (\bm{d}-\braket{\bm{d}}_{HF})\big)\big(b_\alpha + \frac{1}{\sqrt{2\omega_\alpha}}\bm{\lambda}_\alpha\cdot (\bm{d}-\braket{\bm{d}}_{HF})\big)|R_k}\label{eq:CCphoton_field}.
    \end{align}  
 \end{widetext}

\section{Concluding remarks} \label{conclusion_chapt}

In this paper, we proposed a systematic discussion of polaritonic response theory based on the well-established molecular response theory routinely employed in quantum chemistry. 
The fundamental definitions and features of the response functions are still valid, but the explicit treatment of the electromagnetic degrees of freedom allows for novel perspectives. 
Additional equivalence relations between matter and photonic observables are introduced, and novel ways to probe the system are discussed.
Particular care is needed when using different mathematical representations of the operators, as this can lead to misinterpretations of the computed results.
We also provided QED-HF and QED-CC response equations that resemble the standard electronic response theory, providing the reader with a general framework for \textit{ab inito} QED response theory.
While significant progress in the theoretical description of polaritonic systems has been made, we emphasize several challenges that future research will have to address.
These include a description of light-matter interaction beyond the electric dipole approximation, the issue of disorder in optical devices, the role of collective effects on polaritonic properties and chemical reactions, and chiral polaritonics.

\section{Acknowledgments}
We acknowledge Tor S. Haugland for insightful discussions.

\section{Funding information}
M.C., A.B., and H.K. acknowledge funding from the European Research Council (ERC) under the European Union’s Horizon 2020 Research and Innovation Programme (grant agreement No. 101020016).
R.R.R and H.K. acknowledge funding from the Research Council of Norway through FRINATEK Project No. 275506. 
E.R acknowledges funding from the European Research Council (ERC) under the European Union’s Horizon Europe Research and Innovation Programme (Grant n. ERC-StG-2021-101040197 - QED-SPIN).
We acknowledge computing resources through UNINETT Sigma2—the National Infrastructure for High Performance Computing and Data Storage in Norway, through Project No. NN2962k.

\section{Conflict of interest}
The authors declare no conflict of interest for this paper.


\begin{thebibliography}{100}

\bibitem{huang1951lattice}
Huang K.
\newblock Lattice vibrations and optical waves in ionic crystals.
\newblock Nature. 1951;167(4254):779-80.

\bibitem{hopfield1958theory}
Hopfield J.
\newblock Theory of the contribution of excitons to the complex dielectric
  constant of crystals.
\newblock Physical Review. 1958;112(5):1555.

\bibitem{tolpygo1950physical}
Tolpygo K.
\newblock Physical properties of a rock salt lattice made up of deformable
  ions.
\newblock Zh eksp teor fiz. 1950;20(6):497.

\bibitem{jaynes1963comparison}
Jaynes ET, Cummings FW.
\newblock Comparison of quantum and semiclassical radiation theories with
  application to the beam maser.
\newblock Proceedings of the IEEE. 1963;51(1):89-109.

\bibitem{rempe1987observation}
Rempe G, Walther H, Klein N.
\newblock Observation of quantum collapse and revival in a one-atom maser.
\newblock Physical review letters. 1987;58(4):353.

\bibitem{brune1996quantum}
Brune M, Schmidt-Kaler F, Maali A, Dreyer J, Hagley E, Raimond J, et~al.
\newblock Quantum Rabi oscillation: A direct test of field quantization in a
  cavity.
\newblock Physical review letters. 1996;76(11):1800.

\bibitem{long2015coherent}
Long JP, Simpkins B.
\newblock Coherent coupling between a molecular vibration and Fabry--Perot
  optical cavity to give hybridized states in the strong coupling limit.
\newblock ACS photonics. 2015;2(1):130-6.

\bibitem{gordon1964equivalence}
Gordon J, Kogelnik H.
\newblock Equivalence relations among spherical mirror optical resonators.
\newblock Bell System Technical Journal. 1964;43(6):2873-86.

\bibitem{fox1961resonant}
Fox AG, Li T.
\newblock Resonant modes in a maser interferometer.
\newblock Bell System Technical Journal. 1961;40(2):453-88.

\bibitem{kojima2002laser}
Kojima J, Nguyen QV.
\newblock Laser pulse-stretching with multiple optical ring cavities.
\newblock Applied optics. 2002;41(30):6360-70.

\bibitem{schmidt2016quantum}
Schmidt MK, Esteban R, Gonz{\'a}lez-Tudela A, Giedke G, Aizpurua J.
\newblock Quantum mechanical description of Raman scattering from molecules in
  plasmonic cavities.
\newblock ACS nano. 2016;10(6):6291-8.

\bibitem{santhosh2016vacuum}
Santhosh K, Bitton O, Chuntonov L, Haran G.
\newblock Vacuum Rabi splitting in a plasmonic cavity at the single quantum
  emitter limit.
\newblock Nature communications. 2016;7(1):1-5.

\bibitem{weisbuch1992observation}
Weisbuch C, Nishioka M, Ishikawa A, Arakawa Y.
\newblock Observation of the coupled exciton-photon mode splitting in a
  semiconductor quantum microcavity.
\newblock Physical review letters. 1992;69(23):3314.

\bibitem{yakovlev1975surface}
Yakovlev V, Nazin V, Zhizhin G.
\newblock The surface polariton splitting due to thin surface film LO
  vibrations.
\newblock Optics Communications. 1975;15(2):293-5.

\bibitem{lidzey1998strong}
Lidzey DG, Bradley D, Skolnick M, Virgili T, Walker S, Whittaker D.
\newblock Strong exciton--photon coupling in an organic semiconductor
  microcavity.
\newblock Nature. 1998;395(6697):53-5.

\bibitem{fujita1998tunable}
Fujita T, Sato Y, Kuitani T, Ishihara T.
\newblock Tunable polariton absorption of distributed feedback microcavities at
  room temperature.
\newblock Physical Review B. 1998;57(19):12428.

\bibitem{hutchison2012modifying}
Hutchison JA, Schwartz T, Genet C, Devaux E, Ebbesen TW.
\newblock Modifying chemical landscapes by coupling to vacuum fields.
\newblock Angewandte Chemie International Edition. 2012;51(7):1592-6.

\bibitem{thomas2016ground}
Thomas A, George J, Shalabney A, Dryzhakov M, Varma SJ, Moran J, et~al.
\newblock Ground-state chemical reactivity under vibrational coupling to the
  vacuum electromagnetic field.
\newblock Angewandte Chemie. 2016;128(38):11634-8.

\bibitem{lather2019cavity}
Lather J, Bhatt P, Thomas A, Ebbesen TW, George J.
\newblock Cavity catalysis by cooperative vibrational strong coupling of
  reactant and solvent molecules.
\newblock Angewandte Chemie. 2019;131(31):10745-8.

\bibitem{thomas2019tilting}
Thomas A, Lethuillier-Karl L, Nagarajan K, Vergauwe RM, George J, Chervy T,
  et~al.
\newblock Tilting a ground-state reactivity landscape by vibrational strong
  coupling.
\newblock Science. 2019;363(6427):615-9.

\bibitem{canaguier2013thermodynamics}
Canaguier-Durand A, Devaux E, George J, Pang Y, Hutchison JA, Schwartz T,
  et~al.
\newblock Thermodynamics of molecules strongly coupled to the vacuum field.
\newblock Angewandte Chemie International Edition. 2013;52(40):10533-6.

\bibitem{sau2021modifying}
Sau A, Nagarajan K, Patrahau B, Lethuillier-Karl L, Vergauwe RM, Thomas A,
  et~al.
\newblock Modifying Woodward--Hoffmann stereoselectivity under vibrational
  strong coupling.
\newblock Angewandte Chemie International Edition. 2021;60(11):5712-7.

\bibitem{eizner2019inverting}
Eizner E, Martínez-Martínez LA, Yuen-Zhou J, K{\'e}na-Cohen S.
\newblock Inverting singlet and triplet excited states using strong
  light-matter coupling.
\newblock Science advances. 2019;5(12):eaax4482.

\bibitem{takahashi2019singlet}
Takahashi S, Watanabe K, Matsumoto Y.
\newblock Singlet fission of amorphous rubrene modulated by polariton
  formation.
\newblock The Journal of Chemical Physics. 2019;151(7):074703.

\bibitem{martinez2018polariton}
Mart{\'\i}nez-Mart{\'\i}nez LA, Du M, Ribeiro RF, K{\'e}na-Cohen S, Yuen-Zhou
  J.
\newblock Polariton-assisted singlet fission in acene aggregates.
\newblock The Journal of Physical Chemistry Letters. 2018;9(8):1951-7.

\bibitem{stranius2018selective}
Stranius K, Hertzog M, B{\"o}rjesson K.
\newblock Selective manipulation of electronically excited states through
  strong light--matter interactions.
\newblock Nature Communications. 2018;9(1):1-7.

\bibitem{yu2021barrier}
Yu Y, Mallick S, Wang M, B{\"o}rjesson K.
\newblock Barrier-free reverse-intersystem crossing in organic molecules by
  strong light-matter coupling.
\newblock Nature communications. 2021;12(1):1-8.

\bibitem{ulusoy2019modifying}
Ulusoy IS, Gomez JA, Vendrell O.
\newblock Modifying the nonradiative decay dynamics through conical
  intersections via collective coupling to a cavity mode.
\newblock The Journal of Physical Chemistry A. 2019;123(41):8832-44.

\bibitem{joseph2021supramolecular}
Joseph K, Kushida S, Smarsly E, Ihiawakrim D, Thomas A, Paravicini-Bagliani GL,
  et~al.
\newblock Supramolecular assembly of conjugated polymers under vibrational
  strong coupling.
\newblock Angewandte Chemie International Edition. 2021;60(36):19665-70.

\bibitem{hirai2021selective}
Hirai K, Ishikawa H, Chervy T, Hutchison JA, Uji-i H.
\newblock Selective crystallization via vibrational strong coupling.
\newblock Chemical science. 2021;12(36):11986-94.

\bibitem{garcia2021manipulating}
Garcia-Vidal FJ, Ciuti C, Ebbesen TW.
\newblock Manipulating matter by strong coupling to vacuum fields.
\newblock Science. 2021;373(6551):eabd0336.

\bibitem{chervy2018vibro}
Chervy T, Thomas A, Akiki E, Vergauwe RM, Shalabney A, George J, et~al.
\newblock Vibro-polaritonic IR emission in the strong coupling regime.
\newblock ACS Photonics. 2018;5(1):217-24.

\bibitem{george2015ultra}
George J, Wang S, Chervy T, Canaguier-Durand A, Schaeffer G, Lehn JM, et~al.
\newblock Ultra-strong coupling of molecular materials: spectroscopy and
  dynamics.
\newblock Faraday discussions. 2015;178:281-94.

\bibitem{xue2018ultrastrong}
Xue B, Wang D, Tu L, Sun D, Jing P, Chang Y, et~al.
\newblock Ultrastrong absorption meets ultraweak absorption: unraveling the
  energy-dissipative routes for dye-sensitized upconversion luminescence.
\newblock The Journal of Physical Chemistry Letters. 2018;9(16):4625-31.

\bibitem{del2015signatures}
del Pino J, Feist J, Garcia-Vidal F.
\newblock Signatures of vibrational strong coupling in Raman scattering.
\newblock The Journal of Physical Chemistry C. 2015;119(52):29132-7.

\bibitem{baranov2020circular}
Baranov DG, Munkhbat B, L{\"a}nk NO, Verre R, K{\"a}ll M, Shegai T.
\newblock Circular dichroism mode splitting and bounds to its enhancement with
  cavity-plasmon-polaritons.
\newblock Nanophotonics. 2020;9(2):283-93.

\bibitem{guo2021optical}
Guo J, Song G, Huang Y, Liang K, Wu F, Jiao R, et~al.
\newblock Optical Chirality in a Strong Coupling System with Surface Plasmons
  Polaritons and Chiral Emitters.
\newblock ACS Photonics. 2021;8(3):901-6.

\bibitem{itoh2018reproduction}
Itoh T, Yamamoto YS.
\newblock Reproduction of surface-enhanced resonant Raman scattering and
  fluorescence spectra of a strong coupling system composed of a single silver
  nanoparticle dimer and a few dye molecules.
\newblock The Journal of Chemical Physics. 2018;149(24):244701.

\bibitem{herrera2017absorption}
Herrera F, Spano FC.
\newblock Absorption and photoluminescence in organic cavity QED.
\newblock Physical Review A. 2017;95(5):053867.

\bibitem{wang2020coherent}
Wang S, Scholes GD, Hsu LY.
\newblock Coherent-to-incoherent transition of molecular fluorescence
  controlled by surface plasmon polaritons.
\newblock The Journal of Physical Chemistry Letters. 2020;11(15):5948-55.

\bibitem{takele2021scouting}
Takele WM, Piatkowski L, Wackenhut F, Gawinkowski S, Meixner AJ, Waluk J.
\newblock Scouting for strong light--matter coupling signatures in Raman
  spectra.
\newblock Physical Chemistry Chemical Physics. 2021;23(31):16837-46.

\bibitem{barachati2018tunable}
Barachati F, Simon J, Getmanenko YA, Barlow S, Marder SR, K{\'e}na-Cohen S.
\newblock Tunable third-harmonic generation from polaritons in the ultrastrong
  coupling regime.
\newblock Acs Photonics. 2018;5(1):119-25.

\bibitem{mund2020optical}
Mund J, Yakovlev DR, Semina MA, Bayer M.
\newblock Optical harmonic generation on the exciton-polariton in ZnSe.
\newblock Physical Review B. 2020;102(4):045203.

\bibitem{ebadian2017extending}
Ebadian H, Mohebbi M.
\newblock Extending the high-order-harmonic spectrum using surface plasmon
  polaritons.
\newblock Physical Review A. 2017;96(2):023415.

\bibitem{wang2021large}
Wang K, Seidel M, Nagarajan K, Chervy T, Genet C, Ebbesen T.
\newblock Large optical nonlinearity enhancement under electronic strong
  coupling.
\newblock Nature Communications. 2021;12(1):1-9.

\bibitem{wang2014quantum}
Wang S, Chervy T, George J, Hutchison JA, Genet C, Ebbesen TW.
\newblock Quantum yield of polariton emission from hybrid light-matter states.
\newblock The journal of physical chemistry letters. 2014;5(8):1433-9.

\bibitem{imperatore2021reproducibility}
Imperatore MV, Asbury JB, Giebink NC.
\newblock Reproducibility of cavity-enhanced chemical reaction rates in the
  vibrational strong coupling regime.
\newblock The Journal of Chemical Physics. 2021;154(19):191103.

\bibitem{ruggenthaler2022understanding}
Ruggenthaler M, Sidler D, Rubio A.
\newblock Understanding polaritonic chemistry from ab initio quantum
  electrodynamics.
\newblock arXiv preprint arXiv:221104241. 2022.

\bibitem{fregoni2022theoretical}
Fregoni J, Garcia-Vidal FJ, Feist J.
\newblock Theoretical challenges in polaritonic chemistry.
\newblock ACS photonics. 2022;9(4):1096-107.

\bibitem{hirai2020recent}
Hirai K, Hutchison JA, Uji-i H.
\newblock Recent progress in vibropolaritonic chemistry.
\newblock ChemPlusChem. 2020;85(9):1981-8.

\bibitem{feist2018polaritonic}
Feist J, Galego J, Garcia-Vidal FJ.
\newblock Polaritonic chemistry with organic molecules.
\newblock ACS Photonics. 2018;5(1):205-16.

\bibitem{sidler2022perspective}
Sidler D, Ruggenthaler M, Sch{\"a}fer C, Ronca E, Rubio A.
\newblock A perspective on ab initio modeling of polaritonic chemistry: The
  role of non-equilibrium effects and quantum collectivity.
\newblock The Journal of Chemical Physics. 2022;156(23):230901.

\bibitem{hertzog2019strong}
Hertzog M, Wang M, Mony J, B{\"o}rjesson K.
\newblock Strong light--matter interactions: a new direction within chemistry.
\newblock Chemical Society Reviews. 2019;48(3):937-61.

\bibitem{nagarajan2021chemistry}
Nagarajan K, Thomas A, Ebbesen TW.
\newblock Chemistry under vibrational strong coupling.
\newblock Journal of the American Chemical Society. 2021;143(41):16877-89.

\bibitem{schafer2022shining}
Sch{\"a}fer C, Flick J, Ronca E, Narang P, Rubio A.
\newblock Shining light on the microscopic resonant mechanism responsible for
  cavity-mediated chemical reactivity.
\newblock Nature Communications. 2022;13(1):7817.

\bibitem{ruggenthaler2014quantum}
Ruggenthaler M, Flick J, Pellegrini C, Appel H, Tokatly IV, Rubio A.
\newblock Quantum-electrodynamical density-functional theory: Bridging quantum
  optics and electronic-structure theory.
\newblock Physical Review A. 2014;90(1):012508.

\bibitem{buchholz2019reduced}
Buchholz F, Theophilou I, Nielsen SE, Ruggenthaler M, Rubio A.
\newblock Reduced density-matrix approach to strong matter-photon interaction.
\newblock ACS photonics. 2019;6(11):2694-711.

\bibitem{mallory2022reduced}
Mallory JD, DePrince~III AE.
\newblock Reduced-density-matrix-based ab initio cavity quantum
  electrodynamics.
\newblock Physical Review A. 2022;106(5):053710.

\bibitem{mordovina2020polaritonic}
Mordovina U, Bungey C, Appel H, Knowles PJ, Rubio A, Manby FR.
\newblock Polaritonic coupled-cluster theory.
\newblock Physical Review Research. 2020;2(2):023262.

\bibitem{haugland2020coupled}
Haugland TS, Ronca E, Kj{\o}nstad EF, Rubio A, Koch H.
\newblock Coupled cluster theory for molecular polaritons: Changing ground and
  excited states.
\newblock Physical Review X. 2020;10(4):041043.

\bibitem{haugland2021intermolecular}
Haugland TS, Sch{\"a}fer C, Ronca E, Rubio A, Koch H.
\newblock Intermolecular interactions in optical cavities: An ab initio QED
  study.
\newblock The Journal of Chemical Physics. 2021;154(9):094113.

\bibitem{mandal2020polarized}
Mandal A, Montillo~Vega S, Huo P.
\newblock Polarized Fock states and the dynamical Casimir effect in molecular
  cavity quantum electrodynamics.
\newblock The Journal of Physical Chemistry Letters. 2020;11(21):9215-23.

\bibitem{pavosevic2021polaritonic}
Pavosevic F, Flick J.
\newblock Polaritonic unitary coupled cluster for quantum computations.
\newblock The Journal of Physical Chemistry Letters. 2021;12(37):9100-7.

\bibitem{riso2022molecular}
Riso RR, Haugland TS, Ronca E, Koch H.
\newblock Molecular orbital theory in cavity QED environments.
\newblock Nature communications. 2022;13(1):1-8.

\bibitem{bauer2023perturbation}
Bauer MM, Dreuw A.
\newblock Perturbation theoretical approaches to strong light-matter coupling
  in ground and excited electronic states for the description of molecular
  polaritons.
\newblock The Journal of Chemical Physics. 2023.

\bibitem{salmon2022gauge}
Salmon W, Gustin C, Settineri A, Di~Stefano O, Zueco D, Savasta S, et~al.
\newblock Gauge-independent emission spectra and quantum correlations in the
  ultrastrong coupling regime of open system cavity-QED.
\newblock Nanophotonics. 2022;11(8):1573-90.

\bibitem{dicke1954coherence}
Dicke RH.
\newblock Coherence in spontaneous radiation processes.
\newblock Physical review. 1954;93(1):99.

\bibitem{tavis1968exact}
Tavis M, Cummings FW.
\newblock Exact solution for an N-molecule—radiation-field Hamiltonian.
\newblock Physical Review. 1968;170(2):379.

\bibitem{frisk2019ultrastrong}
Frisk~Kockum A, Miranowicz A, De~Liberato S, Savasta S, Nori F.
\newblock Ultrastrong coupling between light and matter.
\newblock Nature Reviews Physics. 2019;1(1):19-40.

\bibitem{imamouglu2009cavity}
Imamo{\u{g}}lu A.
\newblock Cavity QED based on collective magnetic dipole coupling: spin
  ensembles as hybrid two-level systems.
\newblock Physical review letters. 2009;102(8):083602.

\bibitem{knight1978super}
Knight J, Aharonov Y, Hsieh G.
\newblock Are super-radiant phase transitions possible?
\newblock Physical Review A. 1978;17(4):1454.

\bibitem{vukics2012adequacy}
Vukics A, Domokos P.
\newblock Adequacy of the Dicke model in cavity QED: A counter-no-go statement.
\newblock Physical Review A. 2012;86(5):053807.

\bibitem{grynberg2010introduction}
Grynberg G, Aspect A, Fabre C.
\newblock Introduction to quantum optics: from the semi-classical approach to
  quantized light.
\newblock Cambridge university press; 2010.

\bibitem{f2018theory}
F~Ribeiro R, Dunkelberger AD, Xiang B, Xiong W, Simpkins BS, Owrutsky JC,
  et~al.
\newblock Theory for nonlinear spectroscopy of vibrational polaritons.
\newblock The journal of physical chemistry letters. 2018;9(13):3766-71.

\bibitem{gonzalez2016uncoupled}
Gonzalez-Ballestero C, Feist J, Bad{\'\i}a EG, Moreno E, Garcia-Vidal FJ.
\newblock Uncoupled dark states can inherit polaritonic properties.
\newblock Physical review letters. 2016;117(15):156402.

\bibitem{galego2015cavity}
Galego J, Garcia-Vidal FJ, Feist J.
\newblock Cavity-induced modifications of molecular structure in the
  strong-coupling regime.
\newblock Physical Review X. 2015;5(4):041022.

\bibitem{luk2017multiscale}
Luk HL, Feist J, Toppari JJ, Groenhof G.
\newblock Multiscale molecular dynamics simulations of polaritonic chemistry.
\newblock Journal of chemical theory and computation. 2017;13(9):4324-35.

\bibitem{galego2019cavity}
Galego J, Climent C, Garcia-Vidal FJ, Feist J.
\newblock Cavity Casimir-Polder forces and their effects in ground-state
  chemical reactivity.
\newblock Physical Review X. 2019;9(2):021057.

\bibitem{li2021cavity}
Li TE, Nitzan A, Subotnik JE.
\newblock Cavity molecular dynamics simulations of vibrational
  polariton-enhanced molecular nonlinear absorption.
\newblock The Journal of Chemical Physics. 2021;154(9):094124.

\bibitem{li2020cavity}
Li TE, Subotnik JE, Nitzan A.
\newblock Cavity molecular dynamics simulations of liquid water under
  vibrational ultrastrong coupling.
\newblock Proceedings of the National Academy of Sciences.
  2020;117(31):18324-31.

\bibitem{fregoni2020photochemistry}
Fregoni J, Corni S, Persico M, Granucci G.
\newblock Photochemistry in the strong coupling regime: A trajectory surface
  hopping scheme.
\newblock Journal of Computational Chemistry. 2020;41(23):2033-44.

\bibitem{fregoni2018manipulating}
Fregoni J, Granucci G, Coccia E, Persico M, Corni S.
\newblock Manipulating azobenzene photoisomerization through strong
  light--molecule coupling.
\newblock Nature communications. 2018;9(1):4688.

\bibitem{todorov2012intersubband}
Todorov Y, Sirtori C.
\newblock Intersubband polaritons in the electrical dipole gauge.
\newblock Physical Review B. 2012;85(4):045304.

\bibitem{olsen1985linear}
Olsen J, J{\o}rgensen P.
\newblock Linear and nonlinear response functions for an exact state and for an
  MCSCF state.
\newblock The Journal of chemical physics. 1985;82(7):3235-64.

\bibitem{norman2018principles}
Norman P, Ruud K, Saue T.
\newblock Principles and practices of molecular properties: Theory, modeling,
  and simulations.
\newblock John Wiley \& Sons; 2018.

\bibitem{casida2012progress}
Casida ME, Huix-Rotllant M.
\newblock Progress in time-dependent density-functional theory.
\newblock Annual review of physical chemistry. 2012;63:287-323.

\bibitem{helgaker1999}
Helgaker T, Jaszuński M, Ruud K.
\newblock Ab Initio Methods for the Calculation of NMR Shielding and Indirect
  Spin-Spin Coupling Constants.
\newblock Chemical Reviews. 1999;99:293-352.

\bibitem{christiansen1998response}
Christiansen O, J{\o}rgensen P, H{\"a}ttig C.
\newblock Response functions from Fourier component variational perturbation
  theory applied to a time-averaged quasienergy.
\newblock International Journal of Quantum Chemistry. 1998;68(1):1-52.

\bibitem{sasagane1993higher}
Sasagane K, Aiga F, Itoh R.
\newblock Higher-order response theory based on the quasienergy derivatives:
  The derivation of the frequency-dependent polarizabilities and
  hyperpolarizabilities.
\newblock The Journal of chemical physics. 1993;99(5):3738-78.

\bibitem{langhoff1972aspects}
Langhoff P, Epstein S, Karplus M.
\newblock Aspects of time-dependent perturbation theory.
\newblock Reviews of Modern Physics. 1972;44(3):602.

\bibitem{cammi1999linear}
Cammi R, Mennucci B.
\newblock Linear response theory for the polarizable continuum model.
\newblock The Journal of chemical physics. 1999;110(20):9877-86.

\bibitem{helgaker2012recent}
Helgaker T, Coriani S, J{\o}rgensen P, Kristensen K, Olsen J, Ruud K.
\newblock Recent advances in wave function-based methods of molecular-property
  calculations.
\newblock Chemical reviews. 2012;112(1):543-631.

\bibitem{lazzeretti2004assessment}
Lazzeretti P.
\newblock Assessment of aromaticity via molecular response properties.
\newblock Physical chemistry chemical physics. 2004;6(2):217-23.

\bibitem{flick2019light}
Flick J, Welakuh DM, Ruggenthaler M, Appel H, Rubio A.
\newblock Light--matter response in nonrelativistic quantum electrodynamics.
\newblock ACS photonics. 2019;6(11):2757-78.

\bibitem{yang2021quantum}
Yang J, Ou Q, Pei Z, Wang H, Weng B, Shuai Z, et~al.
\newblock Quantum-electrodynamical time-dependent density functional theory
  within Gaussian atomic basis.
\newblock The Journal of Chemical Physics. 2021;155(6):064107.

\bibitem{welakuh2023tunable}
Welakuh DM, Narang P.
\newblock Tunable Nonlinearity and Efficient Harmonic Generation from a
  Strongly Coupled Light--Matter System.
\newblock ACS Photonics. 2023;10(2):383-93.

\bibitem{welakuh2022frequency}
Welakuh DM, Flick J, Ruggenthaler M, Appel H, Rubio A.
\newblock Frequency-Dependent Sternheimer Linear-Response Formalism for
  Strongly Coupled Light--Matter Systems.
\newblock Journal of Chemical Theory and Computation. 2022.

\bibitem{bonini2022ab}
Bonini J, Flick J.
\newblock Ab initio linear-response approach to vibro-polaritons in the cavity
  Born--Oppenheimer approximation.
\newblock Journal of Chemical Theory and Computation. 2022;18(5):2764-73.

\bibitem{flick2020ab}
Flick J, Narang P.
\newblock Ab initio polaritonic potential-energy surfaces for excited-state
  nanophotonics and polaritonic chemistry.
\newblock The Journal of Chemical Physics. 2020;153(9):094116.

\bibitem{welakuh2022transition}
Welakuh DM, Narang P.
\newblock Transition from Lorentz to Fano Spectral Line Shapes in
  Nonrelativistic Quantum Electrodynamics.
\newblock ACS Photonics. 2022;9(9):2946-55.

\bibitem{fregoni2021strong}
Fregoni J, Haugland TS, Pipolo S, Giovannini T, Koch H, Corni S.
\newblock Strong coupling between localized surface plasmons and molecules by
  coupled cluster theory.
\newblock Nano Letters. 2021;21(15):6664-70.

\bibitem{jackson1977classical}
Jackson JD.
\newblock Classical electrodynamics.
\newblock Wiley New York; 1977.

\bibitem{fabry1899theorie}
Fabry C.
\newblock Theorie et applications d'une nouvelle methods de spectroscopie
  intereferentielle.
\newblock Ann Chim Ser 7. 1899;16:115-44.

\bibitem{pfeifer2022achievements}
Pfeifer H, Ratschbacher L, Gallego J, Saavedra C, Fa{\ss}bender A, von Haaren
  A, et~al.
\newblock Achievements and perspectives of optical fiber Fabry--Perot cavities.
\newblock Applied Physics B. 2022;128(2):29.

\bibitem{muller2010ultrahigh}
Muller A, Flagg EB, Lawall JR, Solomon GS.
\newblock Ultrahigh-finesse, low-mode-volume Fabry--Perot microcavity.
\newblock Optics letters. 2010;35(13):2293-5.

\bibitem{steinmetz2006stable}
Steinmetz T, Colombe Y, Hunger D, H{\"a}nsch T, Balocchi A, Warburton R, et~al.
\newblock Stable fiber-based Fabry-P{\'e}rot cavity.
\newblock Applied Physics Letters. 2006;89(11):111110.

\bibitem{rakhmanov2002dynamic}
Rakhmanov M, Savage~Jr R, Reitze D, Tanner D.
\newblock Dynamic resonance of light in Fabry--Perot cavities.
\newblock Physics Letters A. 2002;305(5):239-44.

\bibitem{schlawin2022cavity}
Schlawin F, Kennes DM, Sentef MA.
\newblock Cavity quantum materials.
\newblock Applied Physics Reviews. 2022;9(1):011312.

\bibitem{schouwink2002dependence}
Schouwink P, Berlepsch H, D{\"a}hne L, Mahrt R.
\newblock Dependence of Rabi-splitting on the spatial position of the optically
  active layer in organic microcavities in the strong coupling regime.
\newblock Chemical physics. 2002;285(1):113-20.

\bibitem{wang2022manipulating}
Wang Y, Ren Y, Luo X, Li B, Chen Z, Liu Z, et~al.
\newblock Manipulating cavity photon dynamics by topologically curved space.
\newblock Light: Science \& Applications. 2022;11(1):308.

\bibitem{mckeever2003experimental}
McKeever J, Boca A, Boozer AD, Buck JR, Kimble HJ.
\newblock Experimental realization of a one-atom laser in the regime of strong
  coupling.
\newblock Nature. 2003;425(6955):268-71.

\bibitem{culver2016collective}
Culver R, Lampis A, Megyeri B, Pahwa K, Mudarikwa L, Holynski M, et~al.
\newblock Collective strong coupling of cold potassium atoms in a ring cavity.
\newblock New Journal of Physics. 2016;18(11):113043.

\bibitem{herskind2009realization}
Herskind PF, Dantan A, Marler JP, Albert M, Drewsen M.
\newblock Realization of collective strong coupling with ion Coulomb crystals
  in an optical cavity.
\newblock Nature Physics. 2009;5(7):494-8.

\bibitem{favero2009optomechanics}
Favero I, Karrai K.
\newblock Optomechanics of deformable optical cavities.
\newblock Nature Photonics. 2009;3(4):201-5.

\bibitem{plum2015chiral}
Plum E, Zheludev NI.
\newblock Chiral mirrors.
\newblock Applied Physics Letters. 2015;106(22):221901.

\bibitem{viviescas2003field}
Viviescas C, Hackenbroich G.
\newblock Field quantization for open optical cavities.
\newblock Physical Review A. 2003;67(1):013805.

\bibitem{maier2006plasmonics}
Maier SA.
\newblock Plasmonics: Metal nanostructures for subwavelength photonic devices.
\newblock IEEE Journal of selected topics in quantum electronics.
  2006;12(6):1214-20.

\bibitem{lee2010review}
Lee B, Lee IM, Kim S, Oh DH, Hesselink L.
\newblock Review on subwavelength confinement of light with plasmonics.
\newblock Journal of Modern Optics. 2010;57(16):1479-97.

\bibitem{benz2013strong}
Benz A, Campione S, Liu S, Montano I, Klem J, Allerman A, et~al.
\newblock Strong coupling in the sub-wavelength limit using metamaterial
  nanocavities.
\newblock Nature communications. 2013;4(1):2882.

\bibitem{dintinger2005strong}
Dintinger J, Klein S, Bustos F, Barnes WL, Ebbesen T.
\newblock Strong coupling between surface plasmon-polaritons and organic
  molecules in subwavelength hole arrays.
\newblock Physical Review B. 2005;71(3):035424.

\bibitem{ballarini2019polaritonics}
Ballarini D, De~Liberato S.
\newblock Polaritonics: from microcavities to sub-wavelength confinement.
\newblock Nanophotonics. 2019;8(4):641-54.

\bibitem{todorov2009strong}
Todorov Y, Andrews A, Sagnes I, Colombelli R, Klang P, Strasser G, et~al.
\newblock Strong light-matter coupling in subwavelength metal-dielectric
  microcavities at terahertz frequencies.
\newblock Physical review letters. 2009;102(18):186402.

\bibitem{xiao2018theoretical}
Xiao X, Li X, Caldwell JD, Maier SA, Giannini V.
\newblock Theoretical analysis of graphene plasmon cavities.
\newblock Applied Materials Today. 2018;12:283-93.

\bibitem{li2022strong}
Li M, Liu C, Ruan B, Zhang B, Gao E, Zhang Z, et~al.
\newblock Strong coupling of plasmonic waves in graphene for light confinement.
\newblock Journal of Luminescence. 2022;252:119332.

\bibitem{qing2022strong}
Qing YM, Ren Y, Lei D, Ma HF, Cui TJ.
\newblock Strong coupling in two-dimensional materials-based nanostructures: a
  review.
\newblock Journal of Optics. 2022;24(2):024009.

\bibitem{li2017graphene}
Li K, Fitzgerald JM, Xiao X, Caldwell JD, Zhang C, Maier SA, et~al.
\newblock Graphene plasmon cavities made with silicon carbide.
\newblock ACS omega. 2017;2(7):3640-6.

\bibitem{gan2012strong}
Gan X, Mak KF, Gao Y, You Y, Hatami F, Hone J, et~al.
\newblock Strong enhancement of light--matter interaction in graphene coupled
  to a photonic crystal nanocavity.
\newblock Nano letters. 2012;12(11):5626-31.

\bibitem{koppens2011graphene}
Koppens FH, Chang DE, Garc{\'\i}a~de Abajo FJ.
\newblock Graphene plasmonics: a platform for strong light--matter
  interactions.
\newblock Nano letters. 2011;11(8):3370-7.

\bibitem{maier2006effective}
Maier SA.
\newblock Effective mode volume of nanoscale plasmon cavities.
\newblock Optical and Quantum Electronics. 2006;38:257-67.

\bibitem{hugall2018plasmonic}
Hugall JT, Singh A, van Hulst NF.
\newblock Plasmonic cavity coupling.
\newblock Acs Photonics. 2018;5(1):43-53.

\bibitem{mondal2022strong}
Mondal M, Semenov A, Ochoa MA, Nitzan A.
\newblock Strong Coupling in Infrared Plasmonic Cavities.
\newblock The Journal of Physical Chemistry Letters. 2022;13(41):9673-8.

\bibitem{liang2020fine}
Liang K, Guo J, Huang Y, Yu L.
\newblock Fine-tuning of polariton energies in a tailored plasmon cavity and
  J-aggregates hybrid system.
\newblock Nanoscale. 2020;12(45):23069-76.

\bibitem{zhang2020hybrid}
Zhang H, Liu YC, Wang C, Zhang N, Lu C.
\newblock Hybrid photonic-plasmonic nano-cavity with ultra-high Q/V.
\newblock Optics Letters. 2020;45(17):4794-7.

\bibitem{zhang2019distributed}
Zhang C, ElAfandy R, Han J.
\newblock Distributed Bragg reflectors for GaN-based vertical-cavity
  surface-emitting lasers.
\newblock Applied Sciences. 2019;9(8):1593.

\bibitem{emsley2002silicon}
Emsley MK, Dosunmu O, Unlu M.
\newblock Silicon substrates with buried distributed Bragg reflectors for
  resonant cavity-enhanced optoelectronics.
\newblock IEEE Journal of Selected Topics in Quantum Electronics.
  2002;8(4):948-55.

\bibitem{menghrajani2020strong}
Menghrajani KS, Barnes WL.
\newblock Strong coupling beyond the light-line.
\newblock ACS photonics. 2020;7(9):2448-59.

\bibitem{tao2015strong}
Tao R, Arita M, Kako S, Kamide K, Arakawa Y.
\newblock Strong coupling in non-polar GaN/AlGaN microcavities with
  air-gap/III-nitride distributed Bragg reflectors.
\newblock Applied Physics Letters. 2015;107(10):101102.

\bibitem{butte2005recent}
Butt{\'e} R, Feltin E, Dorsaz J, Christmann G, Carlin JF, Grandjean N, et~al.
\newblock Recent progress in the growth of highly reflective nitride-based
  distributed Bragg reflectors and their use in microcavities.
\newblock Japanese journal of applied physics. 2005;44(10R):7207.

\bibitem{hu2021strong}
Hu ML, Yang ZJ, Du XJ, Ma L, He J.
\newblock Strong couplings between magnetic quantum emitters and subwavelength
  all-dielectric resonators with whispering gallery modes.
\newblock Optics Express. 2021;29(16):26028-38.

\bibitem{farr2014strong}
Farr WG, Goryachev M, Creedon DL, Tobar ME.
\newblock Strong coupling between whispering gallery modes and chromium ions in
  ruby.
\newblock Physical Review B. 2014;90(5):054409.

\bibitem{gupta1995strong}
Gupta SD, Agarwal GS.
\newblock Strong coupling cavity physics in microspheres with whispering
  gallery modes.
\newblock Optics communications. 1995;115(5-6):597-605.

\bibitem{o2011all}
O’shea D, Junge C, P{\"o}llinger M, Vogler A, Rauschenbeutel A.
\newblock All-optical switching and strong coupling using tunable
  whispering-gallery-mode microresonators.
\newblock Applied Physics B. 2011;105:129-48.

\bibitem{matsko2006optical}
Matsko AB, Ilchenko VS.
\newblock Optical resonators with whispering-gallery modes-part I: basics.
\newblock IEEE Journal of selected topics in quantum electronics.
  2006;12(1):3-14.

\bibitem{strekalov2016nonlinear}
Strekalov DV, Marquardt C, Matsko AB, Schwefel HG, Leuchs G.
\newblock Nonlinear and quantum optics with whispering gallery resonators.
\newblock Journal of Optics. 2016;18(12):123002.

\bibitem{matsko2005review}
Matsko A, Savchenkov A, Strekalov D, Ilchenko V, Maleki L.
\newblock Review of applications of whispering-gallery mode resonators in
  photonics and nonlinear optics.
\newblock IPN Progress Report. 2005;42(162):1-51.

\bibitem{kaliteevski2007whispering}
Kaliteevski M, Brand S, Abram R, Kavokin A, Dang LS.
\newblock Whispering gallery polaritons in cylindrical cavities.
\newblock Physical Review B. 2007;75(23):233309.

\bibitem{gautier2022planar}
Gautier J, Li M, Ebbesen TW, Genet C.
\newblock Planar chirality and optical spin--orbit coupling for chiral
  fabry--perot cavities.
\newblock ACS photonics. 2022;9(3):778-83.

\bibitem{voronin2022single}
Voronin K, Taradin AS, Gorkunov MV, Baranov DG.
\newblock Single-handedness chiral optical cavities.
\newblock ACS Photonics. 2022;9(8):2652-9.

\bibitem{feis2020helicity}
Feis J, Beutel D, K{\"o}pfler J, Garcia-Santiago X, Rockstuhl C, Wegener M,
  et~al.
\newblock Helicity-preserving optical cavity modes for enhanced sensing of
  chiral molecules.
\newblock Physical review letters. 2020;124(3):033201.

\bibitem{beutel2021enhancing}
Beutel D, Scott P, Wegener M, Rockstuhl C, Fernandez-Corbaton I.
\newblock Enhancing the optical rotation of chiral molecules using helicity
  preserving all-dielectric metasurfaces.
\newblock Applied Physics Letters. 2021;118(22):221108.

\bibitem{scott2020enhanced}
Scott P, Garcia-Santiago X, Beutel D, Rockstuhl C, Wegener M,
  Fernandez-Corbaton I.
\newblock On enhanced sensing of chiral molecules in optical cavities.
\newblock Applied Physics Reviews. 2020;7(4):041413.

\bibitem{yoo2015chiral}
Yoo S, Park QH.
\newblock Chiral light-matter interaction in optical resonators.
\newblock Physical review letters. 2015;114(20):203003.

\bibitem{liu2020switchable}
Liu M, Plum E, Li H, Duan S, Li S, Xu Q, et~al.
\newblock Switchable chiral mirrors.
\newblock Advanced Optical Materials. 2020;8(15):2000247.

\bibitem{sofikitis2014evanescent}
Sofikitis D, Bougas L, Katsoprinakis GE, Spiliotis AK, Loppinet B, Rakitzis TP.
\newblock Evanescent-wave and ambient chiral sensing by signal-reversing cavity
  ringdown polarimetry.
\newblock Nature. 2014;514(7520):76-9.

\bibitem{hodgkinson2000vacuum}
Hodgkinson I, hong Wu Q, Knight B, Lakhtakia A, Robbie K.
\newblock Vacuum deposition of chiral sculptured thin films with high optical
  activity.
\newblock Applied Optics. 2000;39(4):642-9.

\bibitem{graf2019achiral}
Graf F, Feis J, Garcia-Santiago X, Wegener M, Rockstuhl C, Fernandez-Corbaton
  I.
\newblock Achiral, helicity preserving, and resonant structures for enhanced
  sensing of chiral molecules.
\newblock ACS Photonics. 2019;6(2):482-91.

\bibitem{hentschel2017chiral}
Hentschel M, Sch{\"a}ferling M, Duan X, Giessen H, Liu N.
\newblock Chiral plasmonics.
\newblock Science advances. 2017;3(5):e1602735.

\bibitem{zheng2021discrete}
Zheng G, He J, Kumar V, Wang S, Pastoriza-Santos I, P{\'e}rez-Juste J, et~al.
\newblock Discrete metal nanoparticles with plasmonic chirality.
\newblock Chemical Society Reviews. 2021;50(6):3738-54.

\bibitem{wang2023excitation}
Wang J, Zheng J, Li KH, Wang J, Lin HQ, Shao L.
\newblock Excitation of Chiral Cavity Plasmon Resonances in Film-Coupled Chiral
  Au Nanoparticles.
\newblock Advanced Optical Materials. 2023:2202865.

\bibitem{govorov2012theory}
Govorov AO, Fan Z.
\newblock Theory of chiral plasmonic nanostructures comprising metal
  nanocrystals and chiral molecular media.
\newblock ChemPhysChem. 2012;13(10):2551-60.

\bibitem{lan2016self}
Lan X, Wang Q.
\newblock Self-assembly of chiral plasmonic nanostructures.
\newblock Advanced Materials. 2016;28(47):10499-507.

\bibitem{cohen1997photons}
Cohen-Tannoudji C, Dupont-Roc J, Grynberg G.
\newblock Photons and Atoms-Introduction to Quantum Electrodynamics.
\newblock John Wiley \& Sons; 1997.

\bibitem{craig1998molecular}
Craig DP, Thirunamachandran T.
\newblock Molecular quantum electrodynamics: an introduction to
  radiation-molecule interactions.
\newblock Courier Corporation; 1998.

\bibitem{landau2013electrodynamics}
Landau LD, Bell J, Kearsley M, Pitaevskii L, Lifshitz E, Sykes J.
\newblock Electrodynamics of continuous media. vol.~8.
\newblock elsevier; 2013.

\bibitem{griesemer2001ground}
Griesemer M, Lieb EH, Loss M.
\newblock Ground states in non-relativistic quantum electrodynamics.
\newblock Inventiones mathematicae. 2001;145(3):557-95.

\bibitem{hiroshima2002self}
Hiroshima F.
\newblock Self-adjointness of the Pauli-Fierz Hamiltonian for arbitrary values
  of coupling constants.
\newblock In: Annales Henri Poincar{\'e}. vol.~3. Springer; 2002. p. 171-201.

\bibitem{golenia}
Gol{\'e}nia S.
\newblock Positive commutators, Fermi golden rule and the spectrum of zero
  temperature Pauli--Fierz Hamiltonians.
\newblock Journal of Functional Analysis. 2009;256(8):2587-620.

\bibitem{derezinski}
Derezi{\'n}ski J, Jak{\v{s}}i{\'c} V.
\newblock Spectral theory of Pauli--Fierz operators.
\newblock Journal of Functional analysis. 2001;180(2):243-327.

\bibitem{Bach1995}
Bach V, Fr{\"o}hlich J, Sigal IM.
\newblock Mathematical theory of nonrelativistic matter and radiation.
\newblock Letters in Mathematical Physics. 1995;34(3):183-201.

\bibitem{Bach1999}
Bach V, Fröhlich J, Sigal IM, Hepp K, Hunziker W.
\newblock Spectral analysis for systems of atoms and molecules coupled to the
  quantized radiation field.
\newblock Springer. 1999;207:249-90.

\bibitem{flick2017cavity}
Flick J, Appel H, Ruggenthaler M, Rubio A.
\newblock Cavity Born--Oppenheimer approximation for correlated
  electron--nuclear-photon systems.
\newblock Journal of chemical theory and computation. 2017;13(4):1616-25.

\bibitem{flick2018strong}
Flick J, Rivera N, Narang P.
\newblock Strong light-matter coupling in quantum chemistry and quantum
  photonics.
\newblock Nanophotonics. 2018;7(9):1479-501.

\bibitem{flick2017atoms}
Flick J, Ruggenthaler M, Appel H, Rubio A.
\newblock Atoms and molecules in cavities, from weak to strong coupling in
  quantum-electrodynamics (QED) chemistry.
\newblock Proceedings of the National Academy of Sciences.
  2017;114(12):3026-34.

\bibitem{kowalewski2016cavity}
Kowalewski M, Bennett K, Mukamel S.
\newblock Cavity femtochemistry: Manipulating nonadiabatic dynamics at avoided
  crossings.
\newblock The journal of physical chemistry letters. 2016;7(11):2050-4.

\bibitem{ribeiro2018polariton}
Ribeiro RF, Mart{\'\i}nez-Mart{\'\i}nez LA, Du M, Campos-Gonzalez-Angulo J,
  Yuen-Zhou J.
\newblock Polariton chemistry: controlling molecular dynamics with optical
  cavities.
\newblock Chemical science. 2018;9(30):6325-39.

\bibitem{bennett2016novel}
Bennett K, Kowalewski M, Mukamel S.
\newblock Novel photochemistry of molecular polaritons in optical cavities.
\newblock Faraday discussions. 2016;194:259-82.

\bibitem{vendrell2018collective}
Vendrell O.
\newblock Collective Jahn-Teller interactions through light-matter coupling in
  a cavity.
\newblock Physical review letters. 2018;121(25):253001.

\bibitem{schafer2018ab}
Sch{\"a}fer C, Ruggenthaler M, Rubio A.
\newblock Ab initio nonrelativistic quantum electrodynamics: Bridging quantum
  chemistry and quantum optics from weak to strong coupling.
\newblock Physical Review A. 2018;98(4):043801.

\bibitem{fabri2021born}
F{\'a}bri C, Hal{\'a}sz GJ, Cederbaum LS, Vib{\'o}k {\'A}.
\newblock Born--Oppenheimer approximation in optical cavities: from success to
  breakdown.
\newblock Chemical science. 2021;12(4):1251-8.

\bibitem{power1982quantum}
Power E, Thirunamachandran T.
\newblock Quantum electrodynamics in a cavity.
\newblock Physical Review A. 1982;25(5):2473.

\bibitem{schuler2020vacua}
Schuler M, De~Bernardis D, L{\"a}uchli A, Rabl P.
\newblock The vacua of dipolar cavity quantum electrodynamics.
\newblock SciPost Physics. 2020;9(5):066.

\bibitem{de2018cavity}
De~Bernardis D, Jaako T, Rabl P.
\newblock Cavity quantum electrodynamics in the nonperturbative regime.
\newblock Physical Review A. 2018;97(4):043820.

\bibitem{barut1987quantum}
Barut A, Dowling J.
\newblock Quantum electrodynamics based on self-energy: Spontaneous emission in
  cavities.
\newblock Physical Review A. 1987;36(2):649.

\bibitem{rokaj2022free}
Rokaj V, Ruggenthaler M, Eich FG, Rubio A.
\newblock Free electron gas in cavity quantum electrodynamics.
\newblock Physical Review Research. 2022;4(1):013012.

\bibitem{helgaker2014molecular}
Helgaker T, J{\o}rgensen P, Olsen J.
\newblock Molecular electronic-structure theory.
\newblock John Wiley \& Sons; 2014.

\bibitem{schafer2020relevance}
Sch{\"a}fer C, Ruggenthaler M, Rokaj V, Rubio A.
\newblock Relevance of the quadratic diamagnetic and self-polarization terms in
  cavity quantum electrodynamics.
\newblock ACS photonics. 2020;7(4):975-90.

\bibitem{tokatly2013time}
Tokatly IV.
\newblock Time-dependent density functional theory for many-electron systems
  interacting with cavity photons.
\newblock Physical review letters. 2013;110(23):233001.

\bibitem{rokaj2018light}
Rokaj V, Welakuh DM, Ruggenthaler M, Rubio A.
\newblock Light--matter interaction in the long-wavelength limit: no
  ground-state without dipole self-energy.
\newblock Journal of Physics B: Atomic, Molecular and Optical Physics.
  2018;51(3):034005.

\bibitem{riso2022strong}
Riso RR, Grazioli L, Ronca E, Giovannini T, Koch H.
\newblock Strong coupling in chiral cavities: nonperturbative framework for
  enantiomer discrimination.
\newblock arXiv preprint arXiv:220901987. 2022.

\bibitem{doi:10.1021/acs.jpclett.3c00286}
Sch{\"a}fer C, Baranov DG.
\newblock Chiral Polaritonics: Analytical Solutions, Intuition, and Use.
\newblock The Journal of Physical Chemistry Letters. 2023;14(15):3777-84.

\bibitem{PhysRevA.107.L021501}
Mauro L, Fregoni J, Feist J, Avriller R.
\newblock Chiral discrimination in helicity-preserving Fabry-P\'erot cavities.
\newblock Phys Rev A. 2023 Feb;107:L021501.

\bibitem{list2015beyond}
List NH, Kauczor J, Saue T, Jensen HJA, Norman P.
\newblock Beyond the electric-dipole approximation: A formulation and
  implementation of molecular response theory for the description of absorption
  of electromagnetic field radiation.
\newblock The Journal of chemical physics. 2015;142(24):244111.

\bibitem{bernadotte2012origin}
Bernadotte S, Atkins AJ, Jacob CR.
\newblock Origin-independent calculation of quadrupole intensities in X-ray
  spectroscopy.
\newblock The Journal of chemical physics. 2012;137(20):204106.

\bibitem{list2020beyond}
List NH, Melin TRL, van Horn M, Saue T.
\newblock Beyond the electric-dipole approximation in simulations of x-ray
  absorption spectroscopy: Lessons from relativistic theory.
\newblock The Journal of Chemical Physics. 2020;152(18):184110.

\bibitem{lestrange2015consequences}
Lestrange PJ, Egidi F, Li X.
\newblock The consequences of improperly describing oscillator strengths beyond
  the electric dipole approximation.
\newblock The Journal of Chemical Physics. 2015;143(23):234103.

\bibitem{list2017rotationally}
List NH, Saue T, Norman P.
\newblock Rotationally averaged linear absorption spectra beyond the
  electric-dipole approximation.
\newblock Molecular Physics. 2017;115(1-2):63-74.

\bibitem{rokaj2019quantum}
Rokaj V, Penz M, Sentef MA, Ruggenthaler M, Rubio A.
\newblock Quantum electrodynamical Bloch theory with homogeneous magnetic
  fields.
\newblock Physical review letters. 2019;123(4):047202.

\bibitem{rokaj2022polaritonic}
Rokaj V, Penz M, Sentef MA, Ruggenthaler M, Rubio A.
\newblock Polaritonic Hofstadter butterfly and cavity control of the quantized
  Hall conductance.
\newblock Physical Review B. 2022;105(20):205424.

\bibitem{schafer2021making}
Sch{\"a}fer C, Buchholz F, Penz M, Ruggenthaler M, Rubio A.
\newblock Making ab initio QED functional (s): Nonperturbative and photon-free
  effective frameworks for strong light--matter coupling.
\newblock Proceedings of the National Academy of Sciences.
  2021;118(41):e2110464118.

\bibitem{schafer2022shortcut}
Sch{\"a}fer C, Johansson G.
\newblock Shortcut to self-consistent light-matter interaction and realistic
  spectra from first principles.
\newblock Physical Review Letters. 2022;128(15):156402.

\bibitem{schafer2022polaritonic}
Sch{\"a}fer C.
\newblock Polaritonic chemistry from first principles via embedding radiation
  reaction.
\newblock The Journal of Physical Chemistry Letters. 2022;13(30):6905-11.

\bibitem{ehrenfest1927bemerkung}
Ehrenfest P.
\newblock Bemerkung {\"u}ber die angen{\"a}herte G{\"u}ltigkeit der klassischen
  Mechanik innerhalb der Quantenmechanik.
\newblock Zeitschrift f{\"u}r physik. 1927;45(7):455-7.

\bibitem{RSpertutheory}
Schrödinger E.
\newblock Quantisierung als Eigenwertproblem.
\newblock Annalen der Physik. 1926;386(18):109-39.

\bibitem{hellmann1937}
Hellmann J.
\newblock Einf{\"u}hrung in die Quantenchemie.
\newblock Leipzig: Deuticke; 1937.

\bibitem{feynman1939}
Feynman R.
\newblock Forces in molecules.
\newblock Phys Rev. 1939;56:340-3.

\bibitem{barron2009molecular}
Barron LD.
\newblock Molecular light scattering and optical activity.
\newblock Cambridge University Press; 2009.

\bibitem{houdre1996vacuum}
Houdr{\'e} R, Stanley R, Ilegems M.
\newblock Vacuum-field Rabi splitting in the presence of inhomogeneous
  broadening: Resolution of a homogeneous linewidth in an inhomogeneously
  broadened system.
\newblock Physical Review A. 1996;53(4):2711.

\bibitem{sidler2020polaritonic}
Sidler D, Sch{\"a}fer C, Ruggenthaler M, Rubio A.
\newblock Polaritonic chemistry: Collective strong coupling implies strong
  local modification of chemical properties.
\newblock The journal of physical chemistry letters. 2020;12(1):508-16.

\bibitem{pavovsevic2022wavefunction}
Pavo{\v{s}}evi{\'c} F, Rubio A.
\newblock Wavefunction embedding for molecular polaritons.
\newblock The Journal of Chemical Physics. 2022;157(9):094101.

\bibitem{li2022energy}
Li TE, Nitzan A, Subotnik JE.
\newblock Energy-efficient pathway for selectively exciting solute molecules to
  high vibrational states via solvent vibration-polariton pumping.
\newblock Nature Communications. 2022;13(1):4203.

\bibitem{wang2021defect}
Wang DS, Yelin SF, Flick J.
\newblock Defect polaritons from first principles.
\newblock ACS nano. 2021;15(9):15142-52.

\bibitem{hubener2021engineering}
H{\"u}bener H, De~Giovannini U, Sch{\"a}fer C, Andberger J, Ruggenthaler M,
  Faist J, et~al.
\newblock Engineering quantum materials with chiral optical cavities.
\newblock Nature materials. 2021;20(4):438-42.

\bibitem{li2023strong}
Li M, Nizar S, Saha S, Thomas A, Azzini S, Ebbesen TW, et~al.
\newblock Strong coupling of chiral Frenkel exciton for intense, bisignate
  circularly polarized luminescence.
\newblock Angewandte Chemie International Edition. 2023;62(6):e202212724.

\bibitem{sun2022polariton}
Sun S, Gu B, Mukamel S.
\newblock Polariton ring currents and circular dichroism of Mg-porphyrin in a
  chiral cavity.
\newblock Chemical science. 2022;13(4):1037-48.

\bibitem{allenmark2003induced}
Allenmark S.
\newblock Induced circular dichroism by chiral molecular interaction.
\newblock Chirality: The Pharmacological, Biological, and Chemical Consequences
  of Molecular Asymmetry. 2003;15(5):409-22.

\bibitem{saeva1971induced}
Saeva F, Wysocki J.
\newblock Induced circular dichroism in cholesteric liquid crystals.
\newblock Journal of the American Chemical Society. 1971;93(22):5928-9.

\bibitem{gawronski2003significance}
Gawro{\'n}ski J, Grajewski J.
\newblock The significance of induced circular dichroism.
\newblock Organic letters. 2003;5(18):3301-3.

\bibitem{craig1976dynamic}
Craig DP, Power EA, Thirunamachandran T.
\newblock The dynamic terms in induced circular dichroism.
\newblock Proceedings of the Royal Society of London A Mathematical and
  Physical Sciences. 1976;348(1652):19-38.

\bibitem{bak1992first}
Bak KL, J{\o}rgensen P, Jensen HJA, Olsen J, Helgaker T.
\newblock First-order nonadiabatic coupling matrix elements from
  multiconfigurational self-consistent-field response theory.
\newblock The Journal of chemical physics. 1992;97(10):7573-84.

\bibitem{ruud1993hartree}
Ruud K, Helgaker T, Bak KL, J{\o}rgensen P, Jensen HJA.
\newblock Hartree--Fock limit magnetizabilities from London orbitals.
\newblock The Journal of chemical physics. 1993;99(5):3847-59.

\bibitem{ruud1994theoretical}
Ruud K, Helgaker T, J{\o}rgensen P, Bak KL.
\newblock Theoretical calculations of the magnetizability of some small
  fluorine-containing molecules using London atomic orbitals.
\newblock Chemical physics letters. 1994;223(1-2):12-8.

\bibitem{ruud1998hartree}
Ruud K, {\AA}gren H, Helgaker T, Dahle P, Koch H, Taylor PR.
\newblock The Hartree--Fock magnetizability of C60.
\newblock Chemical physics letters. 1998;285(3-4):205-9.

\bibitem{aastrand1996magnetizabilities}
{\AA}strand PO, Mikkelsen KV, Ruud K, Helgaker T.
\newblock Magnetizabilities and nuclear shielding constants of the
  fluoromethanes in the gas phase and solution.
\newblock The Journal of Physical Chemistry. 1996;100(51):19771-82.

\bibitem{gauss2007gauge}
Gauss J, Ruud K, K{\'a}llay M.
\newblock Gauge-origin independent calculation of magnetizabilities and
  rotational g tensors at the coupled-cluster level.
\newblock The Journal of chemical physics. 2007;127(7):074101.

\bibitem{lutnaes2009benchmarking}
Lutn{\ae}s OB, Teale AM, Helgaker T, Tozer DJ, Ruud K, Gauss J.
\newblock Benchmarking density-functional-theory calculations of rotational g
  tensors and magnetizabilities using accurate coupled-cluster calculations.
\newblock The Journal of chemical physics. 2009;131(14):144104.

\bibitem{london1937}
London F.
\newblock {Th{\'e}orie quantique des courants interatomiques dans les
  combinaisons aromatiques}.
\newblock {J Phys Radium}. 1937;8(10):397-409.

\bibitem{dalgaard1980time}
Dalgaard E.
\newblock Time-dependent multiconfigurational Hartree--Fock theory.
\newblock The Journal of Chemical Physics. 1980;72(2):816-23.

\bibitem{mclachlan1964time}
McLachlan A, Ball M.
\newblock Time-dependent hartree—fock theory for molecules.
\newblock Reviews of Modern Physics. 1964;36(3):844.

\bibitem{di2000hellmann}
Di~Ventra M, Pantelides ST.
\newblock Hellmann-Feynman theorem and the definition of forces in quantum
  time-dependent and transport problems.
\newblock Physical Review B. 2000;61(23):16207.

\bibitem{casida2009time}
Casida ME.
\newblock Time-dependent density-functional theory for molecules and molecular
  solids.
\newblock Journal of Molecular Structure: THEOCHEM. 2009;914(1-3):3-18.

\bibitem{casida1995time}
Casida ME.
\newblock Time-dependent density functional response theory for molecules.
\newblock In: Recent Advances In Density Functional Methods: (Part I). World
  Scientific; 1995. p. 155-92.

\bibitem{kjaergaard2008hartree}
Kj{\ae}rgaard T, J{\o}rgensen P, Olsen J, Coriani S, Helgaker T.
\newblock Hartree-Fock and Kohn-Sham time-dependent response theory in a
  second-quantization atomic-orbital formalism suitable for linear scaling.
\newblock The Journal of chemical physics. 2008;129(5):054106.

\bibitem{dalgaard1982quadratic}
Dalgaard E.
\newblock Quadratic response functions within the time-dependent Hartree-Fock
  approximation.
\newblock Physical Review A. 1982;26(1):42.

\bibitem{pedersen1997coupled}
Pedersen TB, Koch H.
\newblock Coupled cluster response functions revisited.
\newblock The Journal of chemical physics. 1997;106(19):8059-72.

\bibitem{koch1990coupled}
Koch H, J{\o}rgensen P.
\newblock Coupled cluster response functions.
\newblock The Journal of chemical physics. 1990;93(5):3333-44.

\bibitem{stanton1993equation}
Stanton JF, Bartlett RJ.
\newblock The equation of motion coupled-cluster method. A systematic
  biorthogonal approach to molecular excitation energies, transition
  probabilities, and excited state properties.
\newblock The Journal of chemical physics. 1993;98(9):7029-39.

\end{thebibliography}
\bibliographystyle{vancouver}

\end{document}